\documentclass[thmsa,a4paper]{article}
\usepackage{amsfonts}

\usepackage{amsmath}

\input{tcilatex}

\begin{document}

\setcounter{page}{0} \topmargin0pt \oddsidemargin5mm \renewcommand{%
\thefootnote}{\fnsymbol{footnote}} \newpage \setcounter{page}{0} 
\begin{titlepage}
\begin{flushright}
EMPG-03-03 \\
\end{flushright}
\vspace{0.5cm}
\begin{center}
{\Large {\bf Auxiliary matrices for the six-vertex model  at $q^N=1$
and a geometric interpretation of its symmetries} }

\vspace{0.8cm}
{ \large Christian Korff}

\vspace{0.5cm}
{\em School of Mathematics, University of Edinburgh\\
Mayfield Road, Edinburgh EH9 3JZ, U.K.}
\end{center}
\vspace{0.2cm}
 
\renewcommand{\thefootnote}{\arabic{footnote}}
\setcounter{footnote}{0}

\begin{abstract}
The construction of auxiliary matrices for the six-vertex model at a root of unity is investigated 
from a quantum group theoretic point of view. Employing the concept of intertwiners 
associated with the quantum loop algebra $U_q(\tilde{sl}_2)$ at $q^N=1$ a three parameter 
family of auxiliary matrices is constructed. The elements of this family satisfy a 
functional relation with the transfer matrix allowing one to solve the eigenvalue problem 
of the model and to derive the Bethe ansatz equations.
This functional relation is obtained from the decomposition of a tensor product 
of evaluation representations and involves auxiliary matrices with different parameters. 
Because of this dependence on additional parameters the auxiliary matrices break in 
general the finite symmetries of the six-vertex model, such as spin-reversal or 
spin conservation. More importantly, they also lift the extra degeneracies of the transfer matrix 
due to the loop symmetry present at rational coupling values. The extra parameters in 
the auxiliary matrices are shown to be directly related to the elements in the enlarged 
center $Z$ of the algebra $U_q(\tilde{sl}_2)$ at $q^N=1$. This connection provides a 
geometric interpretation of the enhanced symmetry of the six-vertex model at rational 
coupling. The parameters labelling the auxiliary matrices can be interpreted as coordinates 
on a hypersurface Spec$\, Z\subset\mathbb{C}^4$ which remains invariant under the action 
of an infinite-dimensional group $G$ of analytic transformations, called the quantum 
coadjoint action. 
\medskip
\par\noindent
\end{abstract}
\vfill{ \hspace*{-9mm}
\begin{tabular}{l}
\rule{6 cm}{0.05 mm}\\
C.Korff@ed.ac.uk 
\end{tabular}}
\end{titlepage}
\newpage

\section{Introduction}

Almost forty years ago Lieb \cite{Lb67a,Lb67b,Lb67c} and Sutherland \cite
{St67} solved the six-vertex or XXZ model associated with the following
quantum spin-chain Hamiltonian, 
\begin{equation}
H=\sum_{m=1}^{M}\sigma _{m}^{x}\sigma _{m+1}^{x}+\sigma _{m}^{y}\sigma
_{m+1}^{y}+\frac{q+q^{-1}}{2}\,\left( \sigma _{m}^{z}\sigma
_{m+1}^{z}-1\right) ,\;\sigma _{M+1}^{x,y,z}\equiv \sigma _{1}^{x,y,z}\;.
\label{H}
\end{equation}
Here $\sigma _{m}^{x},\sigma _{m}^{y},\sigma _{m}^{z}$ are the Pauli
matrices acting on the $m^{\text{th}}$ site of the spin-chain. Applying the
coordinate space Bethe ansatz \cite{Bethe} the eigenvalues of the
Hamiltonian (\ref{H}) and the associated transfer matrix (see equation (\ref
{T}) below) can be determined by the solutions of the following set of
equations, 
\begin{equation}
\left\{ \frac{\sinh \frac{1}{2}(u_{j}^{B}+i\gamma )}{\sinh \frac{1}{2}%
(u_{j}^{B}-i\gamma )}\right\} ^{M}=\prod_{\substack{ l=1  \\ l\neq j}}%
^{n_{B}}\frac{\sinh \frac{1}{2}(u_{j}^{B}-u_{l}^{B}+2i\gamma )}{\sinh \frac{1%
}{2}(u_{j}^{B}-u_{l}^{B}-2i\gamma )},\quad q=e^{i\gamma },\;\gamma \in 
\mathbb{R\,}.  \label{BE}
\end{equation}
The integer $n_{B}$ is related to the total spin of the associated
eigenvector. Throughout this article I shall refer to the finite solutions $%
u_{j}^{B}$ as Bethe roots and to (\ref{BE}) as the Bethe ansatz
equations.\smallskip

In recent years the Bethe ansatz equations and the algebraic structure of
the six-vertex model at roots of unity,$\;q^{N}=1$, have again been the
focus of discussion, see e.g. \cite{DFM,FM01a,BA,FM01b,Bx02,D02}. These new
developments address the nature of the extra degeneracies in the spectrum of
the six-vertex transfer matrix and (\ref{H}) at a root of unity. The
discovery of these degeneracies originated in Baxter's papers on the
eight-vertex model \cite{Bx73a,Bx73b,Bx73c}, where the six-vertex limit can
be taken. Thirty years later Deguchi, Fabricius and McCoy showed that there
is an infinite-dimensional symmetry underlying these degeneracies \cite{DFM}%
. In the commensurate sectors where the total spin is a multiple of the
order $N$ the transfer matrix and thus the Hamiltonian (\ref{H}) can be
shown to be invariant under the action of the loop algebra $\widetilde{sl}%
_{2}=sl_{2}\otimes \mathbb{C}[t,t^{-1}]$. Although the present discussion is
limited to the six-vertex model with spin one-half, the occurrence of the
loop symmetry at roots of unity is a general phenomenon. See \cite{KM01,KR02}
for generalizations to models with higher spin and higher rank.\smallskip

Fabricius and McCoy pointed out \cite{FM01a,FM01b} that in order to
understand the structure of the degenerate eigenspaces and the enhanced
symmetry at rational coupling it is not sufficient to look at the solutions
of the Bethe ansatz equations (\ref{BE}) alone. In fact, the loop symmetry
shows that the degenerate eigenspaces of (\ref{H}) at $q^{N}=1$ contain
linear combinations of states whose total spin differs by multiples of $N$.
Hence, the coordinate space Bethe ansatz, which starts from the assumption
of spin-conservation, is not suitable for analyzing the full symmetry
present at roots of unity. An alternative approach which solves the
eigenvalue problem of integrable models such as (\ref{H}) and does not rely
on the conservation of the total spin is the concept of auxiliary matrices.
This technique was first introduced by Baxter in connection with his
solution to the eight-vertex model \cite{Bx71,Bx72,Bx73a,Bx73b,Bx73c}. His
method will be briefly described in Section 1.2 below.\smallskip

In this article auxiliary matrices for the six-vertex model at roots of
unity will be constructed which differ from Baxter's. The differences will
be explained in detail in Section 1.3. The construction procedure and the
assumptions made on the form of the auxiliary matrices in this work are
motivated by the known results on root-of-unity representations of the
quantum loop algebra $U_{q}(\widetilde{sl}_{2})$ and its finite counterpart $%
U_{q}(sl_{2})$ \cite{CK,CKP,BK}.

The main result of this work is the derivation of a functional relation (see
equation (\ref{TQ0}) below) between the auxiliary matrices and the
six-vertex transfer matrix from representation theory. All operators in this
functional equation are proven to commute with each other which allows one
to derive the Bethe ansatz equations (\ref{BE}) and to determine the
spectrum of the transfer matrix. This will be explicitly demonstrated for
two examples. It is important to note that this functional equation is
different from the one considered by Baxter (see equation (\ref{BQ})). The
functional equation derived in this article involves three different
auxiliary matrices instead of only a single one.

In addition, the auxiliary matrices are shown to break the finite and the
infinite-dimensional symmetries of the six-vertex model at roots of unity
due to the dependence on special parameters. These parameters encode a rich
geometric structure which makes the auxiliary matrices mathematically
interesting objects for further studies. At the end of this paper the action
of an infinite-dimensional automorphism group will be defined on the
auxiliary matrices using the results in \cite{CK,CKP}. Since all auxiliary
matrices will be shown to commute with the six-vertex transfer matrix, while
they in general do not commute among themselves, this group action manifests
the infinite-dimensional non-abelian symmetry of the six-vertex model at $%
q^{N}=1$. Recall that the loop symmetry has only been established in the
commensurate sectors where the total spin is a multiple of $N$ \cite
{DFM,KM01,KR02}. In contrast the auxiliary matrices will be defined for all
spin-sectors.

The focus of this paper is mainly on the construction of the auxiliary
matrices, their properties and their geometric description. The implications
for the analysis of the degenerate eigenspaces of the six-vertex model will
be subject to future investigations \cite{KP}.\smallskip

As the discussion of the degeneracies at roots of unity and the construction
of auxiliary matrices involves various technical subtleties it is worthwhile
first presenting the definition of the six-vertex model. This will enable us
to make precise statements with regard to the different approaches of
constructing auxiliary matrices. It will also allow us to briefly review
recent developments concerning the degeneracies at roots of unity. The
connection with the literature \cite{Bx73a,BxBook,BS90,AF97,BOU01,RW02}
concerning previous results on auxiliary matrices for the six-vertex model
will be made in Section 1.4.

\subsection{The six-vertex model, definitions and conventions}

Consider an $M\times M^{\prime }$ square lattice with the partition function
written in terms of the transfer matrix $T$ as 
\begin{equation}
Z(z)=\limfunc{tr}_{(\mathbb{C}^{2})^{\otimes M}}T(z)^{M^{\prime }},\quad
T(z)=\limfunc{tr}_{0}R_{0M}(z)R_{0M-1}(z)\cdots R_{01}(z)  \label{T}
\end{equation}
Here $R=R(z,q)$ is a matrix defined over $\mathbb{C}^{2}\otimes \mathbb{C}%
^{2}$ and contains the Boltzmann weights associated with the different
vertex configurations, 
\begin{equation}
R=\tfrac{a+b}{2}1\otimes 1+\tfrac{a-b}{2}\,\sigma ^{z}\otimes \sigma
^{z}+c\,\sigma ^{+}\otimes \sigma ^{-}+c^{\prime }\sigma ^{-}\otimes \sigma
^{+}=\left( 
\begin{array}{cccc}
a &  &  &  \\ 
& b & c &  \\ 
& c^{\prime } & b &  \\ 
&  &  & a
\end{array}
\right) \,.  \label{R}
\end{equation}
The above matrices are defined as $\sigma ^{+}=(\sigma ^{x}+i\sigma
^{y})/2,\sigma ^{-}=(\sigma ^{x}-i\sigma ^{y})/2$ and the lower indices in (%
\ref{T}) indicate on which pair of spaces the $R$-matrix acts in the $(M+1)$%
-fold tensor product of $\mathbb{C}^{2}$. The Boltzmann weights of the six
allowed vertices can be parametrized as follows, 
\begin{equation}
a=\rho ,\;b=\rho \,\frac{\left( 1-z\right) q}{1-zq^{2}},\;c=\rho \,\frac{%
1-q^{2}}{1-zq^{2}},\;c^{\prime }=c\,z,\quad z=e^{u}q^{-1}\in \mathbb{C}%
^{\times }\;.  \label{h}
\end{equation}
The function $\rho =\rho (z,q)$ is an arbitrary normalization factor which
might depend on the spectral parameter $z$ and the deformation parameter $q$%
. Setting $\rho (z=1,q)=1$ the matrix (\ref{R}) becomes the permutation
operator at $z=1$ and the transfer matrix reduces to the shift operator. The
associated XXZ spin-chain Hamiltonian (\ref{H}) is obtained by taking the
logarithmic derivative of the transfer matrix (\ref{T}) with respect to the
spectral parameter. As is well known the six-vertex model is integrable as
the transfer matrix evaluated at different spectral parameters commutes with
itself, $[T(z),T(w)]=0$. Further, symmetries are given by the conservation
of the total spin 
\begin{equation}
\lbrack T(z),S^{z}]=0,\quad S^{z}=\frac{1}{2}\sum_{m=1}^{M}\sigma _{m}^{z}\;
\label{Sz}
\end{equation}
and by the commutation of the transfer matrix with the two idempotent
operators 
\begin{equation}
\frak{R}=\sigma ^{x}\otimes \cdots \otimes \sigma ^{x},\text{\quad }\frak{S}%
=\sigma ^{z}\otimes \cdots \otimes \sigma ^{z}=(-1)^{M/2-|S^{z}|}\;.
\label{R&S}
\end{equation}
The first operator invokes spin-reversal while the second has eigenvalue $+1$
or $-1$ depending whether the number of down spins $n$ in a state is even or
odd. The operators $\frak{R},\frak{S}$ can be used to derive for spin-chains
of even length the useful relations, 
\begin{equation}
M\text{ even}:T(z,q^{-1})=T(z^{-1},q)\quad \text{and\quad }T(z,-q)=\frak{S}%
T(z,q)=T(z,q)\frak{S}.  \label{RST}
\end{equation}
Because of the degeneracies at roots of unity we will not analyze the
structure of the six-vertex model via the coordinate space Bethe ansatz (see
e.g. \cite{Bx02} for a recent discussion). Instead we employ the method of
auxiliary matrices and functional equations, which is described next.

\subsection{Baxter's auxiliary matrix and functional equation}

In 1971 Baxter noted that the Bethe ansatz equations (\ref{BE}) for the
six-vertex model ensure the existence of an auxiliary matrix $Q$ subject to
the following functional relation with the transfer matrix \cite{Bx71}, 
\begin{equation}
T(z)Q(z)=Q(z)T(z)=b(z)^{M}Q(zq^{2})+a(z)^{M}Q(zq^{-2})\;.  \label{BTQ}
\end{equation}
In addition, Baxter postulated (see Section 9.5, page 184 in \cite{BxBook})
that an auxiliary matrix ought to obey the following commutation relations , 
\begin{eqnarray}
\lbrack T(z),Q(w)] &=&0,  \label{BQC} \\
\lbrack Q(z),Q(w)] &=&0,  \label{BQC1} \\
\lbrack Q(z),\frak{S}] &=&0,\quad z,w\in \mathbb{C}\;.  \label{BQC2}
\end{eqnarray}
For $w=z,zq^{\pm 2}$ the first two commutators imply that all matrices in
the functional equation (\ref{BTQ}) can be simultaneously diagonalized.
Hence, all the eigenvalues of the transfer matrix (\ref{T}) can be expressed
in terms of those of an auxiliary matrix $Q(z)$ provided the latter is
non-singular. Imposing the commutation relations (\ref{BQC}) and (\ref{BQC1}%
) for arbitrary values of $w$ guarantees that the eigenvectors can be chosen
independent of the spectral parameter $z$. The last commutator (\ref{BQC2})
imposes invariance under the transformation $q\rightarrow -q$, cf equation (%
\ref{RST}). There is no mentioning made in \cite{BxBook} with respect to the
behaviour of the auxiliary matrix under spin-reversal.

The Bethe ansatz equations (\ref{BE}) are recovered whenever the eigenvalue
of $Q(z)$ vanishes for some value $z=z_{j}^{B}$ while the corresponding
eigenvalues of $Q(z_{j}^{B}q^{2}),Q(z_{j}^{B}q^{-2})$ are nonzero, 
\begin{equation}
0=Q(z_{j}^{B})T(z_{j}^{B})=a(z_{j}^{B})^{M}Q(z_{j}^{B}q^{-2})+b(z_{j}^{B})^{M}Q(z_{j}^{B}q^{2})\;.
\label{BEQ}
\end{equation}
Here the eigenvalues and the corresponding matrices are denoted by the same
symbol. The zeroes $z_{j}^{B}=e^{u_{j}^{B}}q^{-1}$ coincide with the Bethe
roots in (\ref{BE}). Note that there might be further zeroes $z_{j}$ for
which all three eigenvalues in (\ref{BEQ}) simultaneously vanish. This is of
crucial importance for the case when $q$ is a root of unity.

\subsubsection{Degeneracies and complete $N$ strings at $q^{N}=1$}

Suppose we are given a solution $Q(z)$ to Baxter's equation (\ref{BTQ})
which satisfies (\ref{BQC}), (\ref{BQC1}), (\ref{BQC2}) and lifts the
degeneracy in the eigenspaces of the six-vertex transfer matrix. The
functional equation then implies for even roots of unity, $N=2N^{\prime }$,
that the eigenvalues of $Q(z)$ must contain additional factors of the form 
\cite{Bx73c,FM01a} 
\begin{equation}
Q_{N^{\prime }}(z,z_{o})=\prod_{\ell =0}^{N^{\prime }-1}\left(
z-z_{o}q^{2\ell }\right) =z^{N^{\prime }}-z_{o}^{N^{\prime }},\quad
q^{2N^{\prime }}=1\;.  \label{Nstring}
\end{equation}
These factors amount to the existence of complete $N^{\prime }$-strings\ $%
(z_{o},z_{o}q^{2},...,z_{o}q^{2N^{\prime }-2})$ first observed in the
context of the eight-vertex model \cite{Bx73c}. As has been argued by
Fabricius and McCoy \cite{FM01a} the string center $z_{o}$ is not fixed by
the Bethe ansatz equations (\ref{BE}) since the factors (\ref{Nstring}) drop
out of the functional equation (\ref{BTQ}) due to the obvious periodicity 
\begin{equation}
Q_{N^{\prime }}(z,z_{o})=Q_{N^{\prime }}(zq^{2},z_{o})\;.  \label{period}
\end{equation}
Note that the string center $z_{o}$ is in fact only determined up to
multiplication by $q^{2}$, whence the factors (\ref{Nstring}) depend on $%
z_{o}^{N^{\prime }}$ rather than $z_{o}$. For odd roots of unity the period
of the complete strings is given by $q$ rather than $q^{2}$.\smallskip

Consequently, the Bethe ansatz equations (\ref{BE}) alone are not sufficient
to describe the degenerate eigenspaces of the transfer matrix \cite{FM01a}.
One ought to construct a consistent auxiliary matrix which lifts the
degeneracy of the six-vertex transfer matrix at roots of unity by fixing the
string centers $z_{o}$. In \cite{Bx02} (see the comment on page 25, after
equation (94)) Baxter argued that the arbitrariness in choosing $z_{o}$
should allow for the existence of a one-parameter family of auxiliary
matrices at a root of unity.\smallskip

We shall see in Section 5 of this article that for $N$ odd there even exists
a three-parameter family provided the conditions (\ref{BQC1}) and (\ref{BQC2}%
) are dropped.

\subsubsection{Baxter's construction of auxiliary matrices}

In Baxter's approach to the eight \cite{Bx72,Bx73a,Bx73b,Bx73c,BxBook} and
six-vertex model \cite{Bx73a,BxBook} two ``preliminary'' auxiliary matrices $%
Q_{\text{R,L}}(z)$ are introduced both of which satisfy the functional
equation (\ref{BTQ}) and are of the following form, 
\begin{equation}
Q(z)=\limfunc{tr}_{_{0}}L_{0M}(z/\mu )L_{0M-1}(z/\mu )\cdots L_{01}(z/\mu
),\quad L_{0m}\in \limfunc{End}(V_{0}\otimes V_{m})\;.  \label{Q}
\end{equation}
Here $\mu \in \mathbb{C}$ is a possible scaling factor, $V_{0}$ denotes the
auxiliary space and the tensor product of the vector spaces $V_{m}\cong 
\mathbb{C}^{2},\;1\leq m\leq M$ forms the quantum spin-chain. Neither of the
matrices $Q_{\text{R}},Q_{\text{L}}$ commutes with the transfer matrix. The
final auxiliary matrix which commutes with the transfer matrix is given by 
\begin{equation}
Q_{\text{B}}(z)=Q_{\text{R}}(z)Q_{\text{R}}(z_{\text{R}})^{-1}=Q_{\text{L}%
}(z_{\text{L}})^{-1}Q_{\text{L}}(z)\;.  \label{QB}
\end{equation}
Here $z_{\text{R,L}}$ are some arbitrary reference points at which the
matrices $Q_{\text{R,L}}$ are supposed to be non-singular. In general the
auxiliary matrix is therefore not of the simple form (\ref{Q}) which makes
it unwieldy in light of algebraic manipulations.

In the case of the six-vertex model an explicit formula for an auxiliary
matrix was given by Baxter only for the sectors of vanishing total spin \cite
{Bx73a}, 
\begin{equation}
S^{z}=0:\quad Q(z)_{\alpha _{1}\cdots \alpha _{M}}^{\beta _{1}\cdots \beta
_{M}}=\mathcal{N}_{\infty }\exp \left( \tfrac{1}{4}i\gamma
\tsum_{m=1}^{M}\tsum_{n=1}^{m-1}\left( \alpha _{n}\beta _{m}-\alpha
_{m}\beta _{n}\right) +\tfrac{1}{4}u\tsum_{m=1}^{M}\alpha _{m}\beta
_{m}\right) .  \label{BQ}
\end{equation}
Here $\alpha _{m},\beta _{m}=\pm 1$ are the eigenvalues of $\sigma ^{z}$ at
the $m^{\text{th}}$ site. This expression ought to hold for all values of $%
\gamma \in \mathbb{R}$ and spin chains of even length.

\subsection{Construction of auxiliary matrices via quantum groups}

In this article a different approach will be used which requires that the
final auxiliary matrix be of the simple form (\ref{Q}); see also e.g. \cite
{AF97,RW02}. This assumption has several consequences for the choice of the $%
L$-matrix used in (\ref{Q}) and for the form of the functional equation with
the transfer matrix which will turn out to be different from Baxter's
equation (\ref{BTQ}).

In order to satisfy (\ref{BQC}) one now demands that the $R$ and $L$-matrix
obey the Yang-Baxter equation \cite{Y,Bx70} 
\begin{equation}
L_{12}(w/z)L_{13}(w)R_{23}(z)=R_{23}(z)L_{13}(w)L_{12}(w/z)\;.  \label{RLL}
\end{equation}
In the context of trigonometric integrable vertex models the solutions to
this equation can be classified through intertwiners associated with quantum
groups. The latter are non-cocommutative Hopf algebras introduced by
Drinfel'd \cite{Drin} and Jimbo \cite{Jimbo}. The algebraic structure of
quantum groups is intimately linked with the quantum inverse scattering
method of the Faddeev school \cite{QISM,QISM1,FRT}.

The simplest example of an intertwiner is the six-vertex $R$-matrix (\ref{R}%
), 
\begin{equation}
R(z)(\pi _{z}^{0}\otimes \pi _{1}^{0})\Delta (x)=\left[ \left( \pi
_{z}^{0}\otimes \pi _{1}^{0}\right) \Delta ^{\text{op}}(x)\right] R(z),\quad
x\in U_{q}(\widetilde{sl}_{2})\;.  \label{R2}
\end{equation}
Here $\pi _{z}^{0}$ denotes the two-dimensional evaluation representation of
the quantum loop algebra $U_{q}(\widetilde{sl}_{2})$, 
\begin{eqnarray}
\pi _{z}^{0}(e_{0}) &=&z\sigma ^{-},\;\pi _{z}^{0}(f_{0})=z^{-1}\sigma
^{+},\;\pi _{z}^{0}(k_{0})=q^{-\sigma ^{z}},  \notag \\
\pi _{z}^{0}(e_{1}) &=&\sigma ^{+},\;\pi _{z}^{0}(f_{1})=\sigma ^{-},\;\pi
_{z}^{0}(k_{1})=q^{\sigma ^{z}}\;.  \label{pi0}
\end{eqnarray}
The symbols $\Delta ,\Delta ^{\text{op}}$ stand for the coproduct and
opposite coproduct whose definition will be given in the text (see equations
(\ref{cop}) and (\ref{AQG})). The symbols $\{e_{i},f_{i},k_{i}\}_{i=0,1}$
denote the Chevalley-Serre generators of $U_{q}(\widetilde{sl}_{2})$%
.\smallskip

In the present construction of auxiliary matrices the $L$-operator will also
be defined as an intertwiner but with the representation (\ref{pi0}) in the
first factor replaced by a more general representation of the quantum loop
algebra at a primitive root of unity, $q^{N}=1$ with $N\geq 3$. That is, the 
$L$-matrix in (\ref{Q}) has to obey the relation 
\begin{equation}
L^{p}(w/z)(\pi _{w}^{p}\otimes \pi _{z}^{0})\Delta (x)=\left[ \left( \pi
_{w}^{p}\otimes \pi _{z}^{0}\right) \Delta ^{\text{op}}(x)\right]
L^{p}(w/z),\quad x\in U_{q}(\widetilde{sl}_{2}),  \label{L}
\end{equation}
where $\pi _{w}^{p}:U_{q}(\widetilde{sl}_{2})\rightarrow \limfunc{End}%
V_{0}\cong \limfunc{End}\mathbb{C}^{N}$ is some finite dimensional
irreducible evaluation representation. We will see explicit examples in
Section 2; see definition (\ref{basis}). However, since the trace is taken
in (\ref{Q}) the explicit form of the representation $\pi _{w}^{p}$ only
matters up to isomorphism. As will be explained in the text one may use the
results in \cite{CK,CKP} to show that all isomorphic representations can be
labelled in terms of points $p$ on a three-dimensional complex hypersurface $%
\limfunc{Spec}Z\subset \mathbb{C}^{4}$. For example when $N$ is odd the
points on the hypersurface obey 
\begin{equation}
p=(\mathbf{x},\mathbf{y},\mathbf{z},\mathbf{c}=\mu +\mu ^{-1})\in \mathbb{C}%
^{4},\quad \mathbf{x\,y}+\mathbf{z+z}^{-1}=\mu ^{N}+\mu ^{-N}\;.  \label{p0}
\end{equation}
The structure of $\limfunc{Spec}Z$ is connected with the algebraic
properties of the center $Z$ of the algebra $U_{q}(\widetilde{sl}_{2})$; see
definition (\ref{SZ}) in Section 2 of this article. The parameter $\mu \neq
1 $ is identical with the scaling factor in (\ref{Q}).\smallskip

Thus, while there are many different ways of writing down solutions to (\ref
{RLL}) respectively (\ref{L}) the final auxiliary matrices $\{Q_{p}(z)\}$
(defined in Section 5, equation (\ref{Q0})) only depend on the point $p\in 
\limfunc{Spec}Z$. The choice of a representation $\pi _{w}^{p}$ corresponds
to a choice of coordinates on the hypersurface and does not effect the final
form of the auxiliary matrix $Q_{p}(z)$.\smallskip

The second step in the construction is to find a functional relation with
the transfer matrix (\ref{T}) analogous to (\ref{BTQ}). In fact, we will see
that due to the different assumptions made in comparison to \cite{BxBook}
the family of the auxiliary matrices obeys a functional equation of a more
general form than (\ref{BTQ}), namely, 
\begin{equation}
Q_{p}(z)T(z)=b(z)^{M}Q_{p^{\prime }}(zq^{2})+a(z)^{M}Q_{p^{\prime \prime
}}(zq^{-2})\;.  \label{TQ0}
\end{equation}
The points $p^{\prime }$ and $p^{\prime \prime }$ are determined via an
exact sequence which describes the decomposition of the tensor product $\pi
_{w}^{p}\otimes \pi _{z}^{0}$ when the ratio $z/w$ is fixed to the value $%
\mu $ in (\ref{p0}), 
\begin{equation}
0\rightarrow \pi _{w^{\prime }}^{p^{\prime }}\hookrightarrow \pi
_{w}^{p}\otimes \pi _{z}^{0}\rightarrow \pi _{w^{\prime \prime }}^{p^{\prime
\prime }}\rightarrow 0\;.  \label{seq}
\end{equation}
Here $\pi _{w^{\prime }}^{p^{\prime }},\pi _{w^{\prime \prime }}^{p^{\prime
\prime }}$ denote some other irreducible root-of-unity representations of $%
U_{q}(\widetilde{sl}_{2})$ which will be explicitly calculated. \smallskip

In contrast to (\ref{BTQ}) the three auxiliary matrices appearing in (\ref
{TQ0}) have different spectra even when they are evaluated at the same value
of the spectral variable $z$. That is, the eigenvalues of $%
Q_{p}(z),Q_{p^{\prime }}(z),Q_{p^{\prime \prime }}(z)$ do in general not
coincide. This is the major difference with expression (\ref{BQ}) and
Baxter's construction. Now one has to allow for a shift in the additional
parameters. Furthermore, for generic points $p\in \limfunc{Spec}Z$ the
auxiliary matrices $Q_{p}$ constructed in this article violate (\ref{BQC1})
and (\ref{BQC2}). However, those restrictions are in general too strong: the
minimal requirement to derive the eigenvalues of the transfer matrix (\ref{T}%
) and the Bethe ansatz equations (\ref{BE}) from (\ref{TQ0}) is to demand
that all matrices in the functional equation commute with each other.

Note that by defining the $L$-matrix as a solution to (\ref{RLL}) we have
also imposed the overly restrictive condition (\ref{BQC}). The motivation
for this definition is the connection with representation theory via (\ref{L}%
).\smallskip

Despite the described differences between the two approaches of constructing
auxiliary matrices, we will see in Section 5 that solutions to Baxter's
functional equation (\ref{BTQ}) can be obtained by taking a finite sum over
the solutions to (\ref{TQ0}); cf equations (\ref{Qel}) and (\ref{Qfiber}) in
the text. It can happen that these solutions sum up to zero in certain
spin-sectors. However, in Section 6 we will explicitly calculate the
eigenvalues of these solutions for the spin-zero sector of the four chain
and verify that they are non-vanishing. While they yield the correct
eigenvalues of the transfer matrix and give the correct Bethe roots they are
shown to be different from Baxter's expression (\ref{BQ}).

\subsubsection{Infinite dimensional symmetries of the six-vertex model}

The occurrence of parameter dependent auxiliary matrices in (\ref{TQ0}) does
not pose a problem for solving the eigenvalue problem of the transfer matrix
(\ref{T}). As we will see in explicit examples the dependence on the
additional parameters in $p$ drops out of the functional relation (\ref{TQ0}%
) when it is written in terms of eigenvalues. This is due to the fact that
the dependence on $p$ enters either through common normalization factors or
the complete $N^{\prime }$-strings (\ref{Nstring}).

Since different auxiliary matrices occur in (\ref{TQ0}) there is a third
possibility. Because of the simultaneous shift in the spectral variable $z$
and the points $p$ single factors in the eigenvalues of the respective
auxiliary matrices can cancel on both sides of the functional relation.
Again we will see this realized in two concrete examples in Section 6.

As the dependence of the auxiliary matrices $Q_{p}$ on the point $p$ lifts
the degeneracies of the transfer matrix, the cancellation of the additional
parameters in $p$ when the operators in (\ref{TQ0}) are diagonalized
manifests the infinite-dimensional non-abelian symmetry of the six-vertex
model at roots of unity. The point $p\in \limfunc{Spec}Z$ can be chosen
arbitrarily, therefore one may allow for analytic transformations leaving
the complex hypersurface $\limfunc{Spec}Z$ invariant. This set of
transformations is induced on $\limfunc{Spec}Z$ by the action of a
non-abelian infinite-dimensional automorphism group $G$, called the quantum
coadjoint action \cite{CK,CKP}. See definition (\ref{G}) in Section 2. This
provides a geometric interpretation of the infinite-dimensional symmetry of
the six-vertex model at $q^{N}=1$ for all spin-sectors.

\subsection{Previous results on auxiliary matrices in the literature}

In Section 1.2 Baxter's result for the six-vertex auxiliary matrix (\ref{BQ}%
) limited to the spin-sectors $S^{z}=0$ has already been mentioned. His
result applies to all values of $q$.

\subsubsection{Results for $q^{N}\neq 1$}

Away from a root of unity auxiliary matrices for the six-vertex model have
been investigated in \cite{AF97,RW02} by considering infinite-dimensional
representations of the upper triangular Borel subalgebra of $U_{q}(%
\widetilde{sl}_{2})$. An extension of the expression (\ref{BQ}) to all
spin-sectors was recently derived in \cite{RW02}. In order to solve (\ref
{BTQ}) within the framework of quantum group theory the choice of an
infinite-dimensional auxiliary space seems to be necessary. This leads,
however, to technical subtleties as one has to introduce the formal power
series 
\begin{equation}
\mathcal{N}_{\infty }(q)\propto \sum_{n\in \mathbb{Z}}q^{nS^{z}}  \label{N8}
\end{equation}
in the normalization ``constant'' in (\ref{BQ}) \cite{RW02}. For $q^{N}\neq
1 $ the properties of this series are needed to satisfy the functional
equation (\ref{BTQ}) in the sectors $S^{z}\neq 0$, while $\mathcal{N}%
_{\infty }$ has to be removed in the sectors $S^{z}=0$ where it becomes
ill-defined. When the root of unity limit is taken, $q^{N}\rightarrow 1$,
the expression (\ref{BQ}) without the factor (\ref{N8}) solves the
functional relation (\ref{BTQ}) only in the spin-sectors $S^{z}=0\func{mod}N$
\cite{RW02}.

\subsubsection{Results for $q^{N}=1$}

The Yang-Baxter algebra (\ref{RLL}) of the six-vertex model at roots of
unity has been the starting point for previous investigations in the
literature \cite{BS90,DJMM90,T92,BOU01}, in particular with regard to the
chiral Potts model \cite{AYMCPTY,BPAY}. For a certain choice of a root of
unity representation the intertwiner (\ref{L}) is contained in the solutions
discussed in \cite{BS90,T92,BOU01}. (The relation will be described at the
end of Section 5 in this article.) However, in this representation one
cannot take the limit from cyclic to nilpotent representations; see Section
2 for an explanation of these terms. This limit is needed for discussing
even roots of unity which have been excluded in \cite{BS90,T92,BOU01}.

As explained above one of the main results in this work is the connection
with representation theory and the results of \cite{CK,CKP}. For this the
solution of (\ref{L}) for any root of unity representation is needed which
is given in Section 3 of this work, cf (\ref{Lsol}) and (\ref{Lodd}).

Moreover, in \cite{BS90,BOU01} five parameter families of auxiliary matrices
at roots of unity are introduced. In contrast to the construction in this
article the authors of \cite{BS90,BOU01} discuss solutions to the functional
equation (\ref{BTQ}) in Baxter's approach.

The equation (\ref{TQ0}) and the exact sequence (\ref{seq}) are new results.
Again the precise relation between the outcome of \cite{BS90,BOU01} and the
results presented here is explained at the end of Section 5.

In \cite{DJMM90} the connection between representations of $U_{q}(\widetilde{%
sl}_{2,3})$ and the chiral Potts model has been considered. The construction
of auxiliary matrices and the decomposition of the tensor product via (\ref
{seq}) have not been analyzed.

\subsection{Outline of the article}

In Section 2 the representation theoretic results on the quantum groups $%
U_{q}(\widetilde{sl}_{2})$ and $U_{q}(sl_{2})$ we shall need are briefly
summarized. In particular, the enlarged center of the algebras, the concept
of evaluation representations and the existence criteria for intertwiners
are reviewed. Also the hypersurface whose points will label the auxiliary
matrices and the infinite-dimensional automorphism group $G$ are introduced.

Section 3 gives a concrete solution for the intertwiner (\ref{L}) which is
the basic constituent for the construction of the auxiliary matrices. Its
transformation properties under spin-reversal and the difference between
principal and homogeneous gradation are discussed.

In Section 4 the exact sequence (\ref{seq}) is described. In particular, it
is stated in terms of representation theory how the points $p^{\prime
},p^{\prime \prime }$ are related to $p$.

Section 5 contains the definition of the auxiliary matrices and the proof of
the functional equation (\ref{TQ0}). In addition, the transformation
properties of the auxiliary matrices under the finite symmetries (\ref{Sz}),
(\ref{R&S}) and the commutation relations of the operators in (\ref{TQ0})
are discussed.

In Section 6 the construction procedure of the auxiliary matrix is
illustrated for the two simple examples $N=3,M=3,4$. The eigenvalues of the
transfer matrix are calculated from the ones of the auxiliary matrix and the
Bethe roots among the zeroes of the auxiliary matrix are identified. As
expected they satisfy the Bethe ansatz equations (\ref{BE}). One example of
a complete $N$-string is given and it is shown that its center is determined
via the central elements of the quantum group. Furthermore, we will see that
the auxiliary matrices constructed in this article do not coincide with the
expression (\ref{BQ}).

Section 7 summarizes the results and gives the conclusions. It is also
explained how the auxiliary matrices can be equipped with the quantum
coadjoint action.

\section{Quantum groups at roots of unity - a reminder}

In this section the known results about representations of $U_{q}(\widetilde{%
sl}_{2}),U_{q}(sl_{2})$ at a root of unity are briefly reviewed focussing
only on those facts which are necessary for our discussion. The original
references for the results stated are \cite{CK,CKP,BK}. The material may
also be found in several textbooks \cite{CP,JmbNotes}.

\subsection{The finite quantum group $U_{q}(sl_{2})$}

For simplicity let us start with the finite quantum group $U_{q}(sl_{2})$
defined in terms of the Chevalley generators $\{e,f,k\}$ obeying the
algebraic relations 
\begin{equation}
kek^{-1}=q^{2}e,\quad kfk^{-1}=q^{-2}f,\quad \lbrack e,f]=\frac{k-k^{-1}}{%
q-q^{-1}}\;.  \label{QG}
\end{equation}
There exists a unique Hopf algebra structure on $U_{q}(sl_{2})$ with
comultiplication $\Delta $, counit $\varepsilon $ and antipode $\Gamma $
such that 
\begin{eqnarray}
\Delta (e) &=&e\otimes 1+k\otimes e,\;\Delta (f)=f\otimes k^{-1}+1\otimes
f,\;\Delta (k)=k\otimes k,  \label{cop} \\
\Gamma (k) &=&k^{-1},\;\Gamma (e)=-k^{-1}e,\,\Gamma (f)=-fk,\;\varepsilon
(e)=\varepsilon (f)=0,\;\varepsilon (k)=1\;.  \notag
\end{eqnarray}
The opposite coproduct $\Delta ^{\text{op}}$ is obtained by permuting the
two factors. There is an alternative variant of this quantum algebra which
occurs in the literature and will be important for our discussion. Suppose
that $q^{4}\neq 0,1$. Then we denote by $\breve{U}_{q}(sl_{2})$ the algebra
generated by $\{\breve{e},\breve{f},t\}$ subject to the relations 
\begin{equation}
t\breve{e}t^{-1}=q\breve{e},\;t\breve{f}t^{-1}=q^{-1}\breve{f},\;[\breve{e},%
\breve{f}]=\frac{t^{2}-t^{-2}}{q-q^{-1}}\;.  \label{QG2}
\end{equation}
The Hopf algebra structure is now given by 
\begin{eqnarray}
\Delta (\breve{e}) &=&\breve{e}\otimes t^{-1}+t\otimes \breve{e},\;\Delta (%
\breve{f})=\breve{f}\otimes t^{-1}+t\otimes \breve{f},\;\Delta (t)=t\otimes
t,  \label{cop2} \\
\Gamma (\breve{e}) &=&-q\,\breve{e},\,\Gamma (\breve{f})=-q^{-1}\breve{f}%
,\,\Gamma (t)=t^{-1},\;\varepsilon (\breve{e})=\varepsilon (\breve{f}%
)=0,\;\varepsilon (t)=1\;.  \notag
\end{eqnarray}
The algebras $U_{q}(sl_{2})$ and $\breve{U}_{q}(sl_{2})$ are not isomorphic.
But there exists an injective Hopf algebra homomorphism $U_{q}(sl_{2})%
\hookrightarrow \breve{U}_{q}(sl_{2})$ via the embedding 
\begin{equation}
e\hookrightarrow \breve{e}\,t,\;f\hookrightarrow t^{-1}\breve{f}%
,\;k\hookrightarrow t^{2}\;.  \label{em}
\end{equation}
Thus, we can view $U_{q}(sl_{2})$ as an Hopf subalgebra of $\breve{U}%
_{q}(sl_{2})$. For the moment we concentrate on $U_{q}(sl_{2})$ but we will
return to the algebra $\breve{U}_{q}(sl_{2})$ when discussing the
intertwiner (\ref{L}).

For values of the deformation parameter $q$ different from a root of unity
the center of the algebra $U_{q}(sl_{2})$ is generated by the Casimir
element, 
\begin{equation}
\mathbf{c}=qk+q^{-1}k^{-1}+(q-q^{-1})^{2}fe\;.  \label{c}
\end{equation}

Henceforth let $q^{N}=1,N\geq 3$ with $q$ being primitive. Then the center
of the quantum group is enlarged by the additional central elements, 
\begin{equation}
\mathbf{x}=\left( (q-q^{-1})e\right) ^{N^{\prime }},\mathbf{y}=\left(
(q-q^{-1})f\right) ^{N^{\prime }},\mathbf{z}^{\pm 1}=k^{\pm N^{\prime
}},\;N^{\prime }=\left\{ 
\begin{array}{cc}
N\,, & N\text{ odd} \\ 
N/2\,, & N\text{ even}
\end{array}
\right. .  \label{Z0}
\end{equation}
The presence of additional central elements at $q^{N}=1$ compared with $%
q^{N}\neq 1$ considerably enriches the representation theory and will
reflect on an algebraic level the additional symmetry encountered in the
six-vertex model.

In the following we denote by $Z_{0}$ the commutative subalgebra generated
by the elements (\ref{Z0}) and $Z=Z_{0}\cup \{\mathbf{c}\}$ the center of $%
U_{q}(sl_{2})$. It is important to note that $\mathbf{x},\mathbf{y},\mathbf{z%
}$ are algebraically independent, while the Casimir element is algebraic
over $Z_{0}$. To express this algebraic dependence we define 
\begin{equation}
F_{N}(x)=\left\{ 
\begin{array}{cc}
\prod\limits_{\ell =0}^{N-1}\left( x+q^{\ell }+q^{-\ell }\right) -2\,, & N%
\text{ odd} \\ 
\prod\limits_{\ell =0,\text{even}}^{N-1}\left( x-q^{\ell +1}-q^{-\ell
-1}\right) -2\,, & N\text{ even}
\end{array}
\right. \;.  \label{FN}
\end{equation}
Then the algebraic relation between the elements (\ref{Z0}) and the Casimir
operator is given by \cite{CK}, 
\begin{equation}
\mathbf{xy}+(-1)^{N+1}\left( \mathbf{z}+\mathbf{z}^{-1}\right) =F_{N}(%
\mathbf{c})\;.  \label{cxyz}
\end{equation}
According to Schur's lemma the elements in the center act as scalars in any
finite-dimensional irreducible representation $\pi $. They can be therefore
treated as ordinary complex numbers and by abuse of notation $\mathbf{x},%
\mathbf{y},\mathbf{z},\mathbf{c}$ will sometimes stand for the algebraic
elements (\ref{Z0}),(\ref{c}) and sometimes for their numerical values in
some (unspecified) representation. Denote by Rep$\,U_{q}(sl_{2})$ the set of
equivalence classes $[\pi ]$ of finite-dimensional irreducible
representations $\pi $\ and define the following hypersurface in $\mathbb{C}%
^{4}$, 
\begin{equation}
\text{$\limfunc{Spec}$}\,Z=\left\{ p=(\mathbf{x},\mathbf{y},\mathbf{z},%
\mathbf{c})\,\left| \,\mathbf{xy}+(-1)^{N+1}(\mathbf{z}+\mathbf{z}%
^{-1})=F_{N}(\mathbf{c})\right. \right\} .  \label{SZ}
\end{equation}
Moreover, there exists the following sequence of surjective maps \cite{CK}, 
\begin{equation}
\text{Rep}\,U_{q}(sl_{2})\overset{\frak{X}}{\rightarrow }\text{$\limfunc{Spec%
}$}\,Z\overset{}{\rightarrow }\text{$\limfunc{Spec}$\thinspace }Z_{0}=%
\mathbb{C}^{2}\times \mathbb{C}^{\times },  \label{Zseq}
\end{equation}
where the map Rep\thinspace $U_{q}(sl_{2})\rightarrow \limfunc{Spec}\,Z$
assigns to each equivalence class its values of the central elements, 
\begin{equation}
\frak{X}:[\pi ]\rightarrow p=(\pi (\mathbf{x}),\pi (\mathbf{y}),\pi (\mathbf{%
z}),\pi (\mathbf{c}))\;.
\end{equation}
Away from the singular points 
\begin{equation}
D=\left\{ 
\begin{array}{cc}
\left\{ (0,0,\pm 1,\pm (q^{\ell }+q^{-\ell }))\,|\,1\leq \ell \leq
N-1\right\} , & N\text{ odd} \\ 
\left\{ (0,0,(-1)^{\ell -1},(q^{\ell }+q^{-\ell }))\,|\,1\leq \ell \leq
N-1,\ell \neq N^{\prime }\right\} , & N\text{ even}
\end{array}
\right. .  \label{D}
\end{equation}
$\frak{X}$ is not only surjective but also injective. In other words the
values of the central elements $(\mathbf{x},\mathbf{y},\mathbf{z},\mathbf{c}%
) $ fix the representation up to isomorphism provided $\frak{X}\left[ \pi 
\right] \notin D$. The projection $\limfunc{Spec}$\thinspace $Z\rightarrow 
\limfunc{Spec}\,Z_{0},$%
\begin{equation}
p=(\pi (\mathbf{x}),\pi (\mathbf{y}),\pi (\mathbf{z}),\pi (\mathbf{c}%
))\rightarrow p_{\ast }=(\pi (\mathbf{x}),\pi (\mathbf{y}),\pi (\mathbf{z})),
\end{equation}
however, is not injective, its inverse image consists in general of $%
N^{\prime }$ points. Away from the discriminant set $D$ the fiber in $%
\limfunc{Spec}\,Z$ over the base point $p_{\ast }=(\mathbf{x},\mathbf{y},%
\mathbf{z})$ is given by the set \cite{A94} 
\begin{equation}
p\notin D:\quad \left\{ p_{\ell }=(\mathbf{x},\mathbf{y},\mathbf{z},\mu
q^{\ell }+\mu ^{-1}q^{-\ell })\,|\,\ell \in \mathbb{Z}_{N^{\prime }}\right\}
\;.  \label{fiber}
\end{equation}
(The parameter $\mu $ has been implicitly defined in (\ref{p0}).) This is
immediate to see when employing the identity 
\begin{equation}
\mathbf{xy}+(-1)^{N+1}(\mathbf{z}+\mathbf{z}^{-1})=F_{N}(\mu +\mu ^{-1})=\mu
^{N^{\prime }}+\mu ^{-N^{\prime }}\;.  \label{imu}
\end{equation}
There is a fundamental difference between the irreducible representations
whose image under $\frak{X}$ lies in the discriminant set (\ref{D}) and
those for which it does not. Let $\frak{X}\left[ \pi \right] \in D$ then the
representation $\pi $ can be obtained by taking the limit $q^{N}\rightarrow
1 $ of some standard representations at $q^{N}\neq 1$ with dimension $\dim
\pi <N^{\prime }$. These representations are the quantum analogue of
representations of the non-deformed algebra $sl_{2}$.

Those representations whose image $\frak{X}\left[ \pi \right] $ lies outside
the discriminant set (\ref{D}) are $N^{\prime }$-dimensional and have no
``classical'' counterparts. They depend in general on three parameters, $\pi
=\pi _{N}^{\xi ,\zeta ,\lambda }$ with $\xi ,\zeta ,\lambda \in \mathbb{C}%
^{2}\times \mathbb{C}^{\times }$. Let $v_{n},n=0,1,...,N^{\prime }-1$ denote
the canonical basis in $\mathbb{C}^{N^{\prime }}$ then the representation $%
\pi _{N}^{\xi ,\zeta ,\lambda }$ is defined via the relations \cite{RA89,CK} 
\begin{eqnarray}
\pi _{N}^{\xi ,\zeta ,\lambda }(k)v_{n} &=&\lambda q^{-2n}v_{n}\;,  \notag \\
\pi _{N}^{\xi ,\zeta ,\lambda }(f)v_{n} &=&v_{n+1}\;,\quad  \notag \\
\pi _{N}^{\xi ,\zeta ,\lambda }(f)v_{N^{\prime }-1} &=&\zeta v_{0}\;,  \notag
\\
\pi _{N}^{\xi ,\zeta ,\lambda }(e)v_{n} &=&\left( [\lambda
;n-1]_{q}[n]_{q}+\xi \zeta \right) v_{n-1},\;n>0\;,  \notag \\
\pi _{N}^{\xi ,\zeta ,\lambda }(e)v_{0} &=&\xi v_{N^{\prime }-1}
\label{basis}
\end{eqnarray}
with 
\begin{equation}
\lbrack \lambda ;n]_{q}:=\frac{\lambda q^{-n}-\lambda ^{-1}q^{n}}{q-q^{-1}}%
\;,\text{\quad }[n]_{q}:=\frac{q^{n}-q^{-n}}{q-q^{-1}}\;.
\end{equation}
The values of the central elements (\ref{Z0}) in terms of the parameter set $%
(\xi ,\zeta ,\lambda )$ are given by 
\begin{eqnarray}
\pi _{N}^{\xi ,\zeta ,\lambda }(\mathbf{x}) &=&\xi (q-q^{-1})^{N^{\prime
}}\prod_{n=1}^{N^{\prime }-1}\left( [\lambda ;n-1]_{q}[n]_{q}+\xi \zeta
\right) =:(q-q^{-1})^{N^{\prime }}\eta ,  \label{eta} \\
\pi _{N}^{\xi ,\zeta ,\lambda }(\mathbf{y}) &=&\zeta (q-q^{-1})^{N^{\prime
}},\;\pi _{N}^{\xi ,\zeta ,\lambda }(\mathbf{z})=\lambda ^{N^{\prime }}, 
\notag
\end{eqnarray}
and 
\begin{equation}
\pi _{N}^{\xi ,\zeta ,\lambda }(\mathbf{c})=q\lambda +q^{-1}\lambda
^{-1}+(q-q^{-1})^{2}\xi \zeta \;.  \label{cpi}
\end{equation}
Note that the representations (\ref{basis}) provide coordinates on the
hypersurface (\ref{SZ}) via (\ref{eta}) and (\ref{cpi}), 
\begin{equation}
\varphi :\mathbb{C}\times \mathbb{C}\times \mathbb{C}^{\times }\rightarrow 
\text{$\limfunc{Spec}$\thinspace }Z\backslash D,\;(\xi ,\zeta ,\lambda
)\rightarrow \varphi (\xi ,\zeta ,\lambda )=\frak{X}[\pi _{N}^{\xi ,\zeta
,\lambda }]\;.  \label{phi}
\end{equation}
In the subsequent sections this coordinate map will be often used to perform
calculations in a concrete representation. However, the results obtained
will not depend on this particular choice of ``coordinates''.

Throughout this article it will be important to distinguish between certain
subvarieties of representations in $\limfunc{Spec}$\thinspace $Z_{0},%
\limfunc{Spec}$\thinspace $Z$. We will use the nomenclature presented in the
table below.\bigskip

\begin{center}
\begin{tabular}{|c|c|c|c|}
\hline\hline
representation & $\pi (\mathbf{x})$ & $\pi (\mathbf{y})$ & $\pi (\mathbf{z})$
\\ \hline\hline
nilpotent & $0$ & $0$ & $\mathbb{C}^{\times }$ \\ \hline\hline
semi-cyclic & $0\;(\mathbb{C}^{\times })$ & $\mathbb{C}^{\times }\;(0)$ & $%
\mathbb{C}^{\times }$ \\ \hline\hline
cyclic & $\mathbb{C}^{\times }$ & $\mathbb{C}^{\times }$ & $\mathbb{C}%
^{\times }$ \\ \hline\hline
\end{tabular}
\medskip

{\small Table 2.1. The different types of representations of }$U_{q}(sl_{2})$%
{\small \ at a root of unity.}\bigskip
\end{center}

\noindent Note that nilpotent and semi-cyclic representations possess a
highest or lowest weight vector, while cyclic representations do not. This
can be explicitly seen in the concrete representation (\ref{basis}).
Applying the generator $e$ or $f$ to a basis vector $N^{\prime }$ times
yields the same vector again, hence the name cyclic representation.

\subsection{The quantum coadjoint action}

Following \cite{CK,CKP} one can define for $q^{N}\neq 1$ the following
infinitesimal automorphisms on $U_{q}(sl_{2})$, 
\begin{equation}
\underline{e}(x)=[e^{N^{\prime }}/[N^{\prime }]_{q}!,x],\;\underline{f}%
(x)=[f^{N^{\prime }}/[N^{\prime }]_{q}!,x],\;\underline{k}(x)=[k^{N^{\prime
}}/[N^{\prime }]_{q}!,x]
\end{equation}
with $[n]_{q}!=[n]_{q}[n-1]_{q}\cdots \lbrack 1]_{q}$. Noteworthy, the above
derivations stay well-defined in the root of unity limit $q^{N}\rightarrow 1$
where their action on the Chevalley-Serre generators reads 
\begin{eqnarray}
\underline{e}(e) &=&0,\;\underline{e}(f)=\frac{kq-k^{-1}q^{-1}}{q-q^{-1}}\,%
\frac{e^{N-1}}{[N-1]!},\;\underline{e}(k^{\pm 1})=\mp N^{-1}\mathbf{x}\frak{%
\,}k^{\pm 1}  \notag \\
\underline{f}(e) &=&-\frac{f^{N-1}}{[N-1]!}\frac{kq-k^{-1}q^{-1}}{q-q^{-1}}%
,\;\underline{f}(f)=0\,,\;\underline{f}(k^{\pm 1})=\pm N^{-1}\mathbf{y}\frak{%
\,}k^{\pm 1}\;.
\end{eqnarray}
Of particular interest is their action on the central elements (\ref{Z0}), 
\begin{eqnarray*}
\underline{e}(\mathbf{x}) &=&0,\quad \underline{e}(\mathbf{y})=\mathbf{z}-%
\mathbf{z}^{-1},\quad \underline{e}(\mathbf{z}^{\pm 1})=\mp \mathbf{x}\frak{%
\,}\mathbf{z}^{\pm 1}, \\
\underline{f}(\mathbf{y}) &=&0,\;\underline{f}(\mathbf{x})=\mathbf{z}^{-1}-%
\mathbf{z},\;\underline{f}(\mathbf{z}^{\pm 1})=\pm \mathbf{y}\frak{\,}%
\mathbf{z}^{\pm 1},\quad \quad \underline{k}(\mathbf{x})=\mathbf{x}\frak{\,}%
\mathbf{z},
\end{eqnarray*}
from which one deduces that their exponentials yield analytic
transformations on the hypersurface (\ref{SZ}), \noindent 
\begin{eqnarray}
\exp (t\underline{e})\mathbf{x} &=&\mathbf{x},\;\exp (t\underline{e})\mathbf{%
z}^{\pm 1}=e^{\mp t\mathbf{x}}\mathbf{z}^{\pm 1},\;\exp (t\underline{e})%
\mathbf{y}=\mathbf{y}-\left( \mathbf{z}\tfrac{e^{-t\mathbf{x}}-1}{\mathbf{x}}%
+\mathbf{z}^{-1}\tfrac{e^{t\mathbf{x}}-1}{\mathbf{x}}\right) ,  \notag \\
\exp (t\underline{f})\mathbf{y} &=&\mathbf{y},\;\exp (t\underline{f})\mathbf{%
z}^{\pm 1}=e^{\pm t\mathbf{y}}\mathbf{z}^{\pm 1},\;\exp (t\underline{f})%
\mathbf{x}=\mathbf{x}+\left( \mathbf{z}\tfrac{e^{-t\mathbf{y}}-1}{\mathbf{y}}%
+\mathbf{z}^{-1}\tfrac{e^{t\mathbf{y}}-1}{\mathbf{y}}\right) .  \label{G}
\end{eqnarray}
Here $t\in \mathbb{C}$ is a free parameter. The group $G$ generated by these
automorphisms is infinite-dimensional and its action on the hypersurface is
called the quantum coadjoint action \cite{CK,CKP}. The fixed point set under
this action is given by the discriminant set (\ref{D}). Note further that
the Casimir element remains invariant, i.e. the polynomial $\mathbf{xy}+%
\mathbf{z}+\mathbf{z}^{-1}$ is fixed under the action of $G$.

\subsection{The quantum loop algebra $U_{q}(\widetilde{sl}_{2})$}

In order to make contact with the six-vertex model (\ref{R}) one needs to
consider instead of the finite quantum group $U_{q}(sl_{2})$ the quantum
loop algebra $U_{q}(\widetilde{sl}_{2})$. For its definition we assume
temporarily $q$ to be generic. The quantum loop algebra is defined through
the algebraic relations 
\begin{equation}
k_{i}e_{j}k_{i}^{-1}=q^{A_{ij}}e_{j},\quad
k_{i}f_{j}k_{i}^{-1}=q^{-A_{ij}}f_{j},\quad k_{i}k_{j}=k_{j}k_{i},\quad
i,j=0,1\;,  \label{AQG}
\end{equation}
where the Cartan matrix $A$\ is 
\begin{equation*}
A=\left( 
\begin{array}{cc}
2 & -2 \\ 
-2 & 2
\end{array}
\right) \;.
\end{equation*}
In addition one has to impose the Chevalley-Serre relations, 
\begin{eqnarray}
e_{i}^{3}e_{j}-[3]_{q}e_{i}^{2}e_{j}e_{i}+[3]_{q}e_{i}e_{j}e_{i}^{2}-e_{j}e_{i}^{3} &=&0,
\notag \\
f_{i}^{3}f_{j}-[3]_{q}f_{i}^{2}f_{j}f_{i}+[3]_{q}f_{i}f_{j}f_{i}^{2}-f_{j}f_{i}^{3} &=&0,\quad i\neq j,\;i,j=0,1\;.
\label{CS}
\end{eqnarray}
Similar to the case $U_{q}(sl_{2})$ the quantum loop algebra $U_{q}(%
\widetilde{sl}_{2})$ can be made into a Hopf algebra using the definitions
analogous to (\ref{cop}). Again we may view $U_{q}(\widetilde{sl}_{2})$ as a
subalgebra of the larger Hopf algebra $\breve{U}_{q}(\widetilde{sl}_{2})$
which is the counterpart to $\breve{U}_{q}(sl_{2})$. The quantum loop
algebra at roots of unity is then obtained via the specialization map, see
1.9 on page 398 in \cite{BK}.

A complete classification of the irreducible representations of the
non-restricted algebra $U_{q}(\widetilde{sl}_{2})$ at roots of unity is
presently not known \cite{BK}. In contrast to the $U_{q}(sl_{2})$ the
quantum loop algebra is not finitely generated over its center $\tilde{Z}$
and the structure of $\limfunc{Spec}\tilde{Z}$ is less well understood. In
particular, the canonical map Rep\thinspace $U_{q}(\widetilde{sl}%
_{2})\rightarrow \limfunc{Spec}\tilde{Z}$ is neither surjective or bijective
in general \cite{BK}. It has been demonstrated, however, that all possible
values of the central elements can be obtained by considering irreducible
subquotients of tensor products of evaluation representations \cite{BK}.

An evaluation representation is constructed by composing a representation of
the finite quantum group with the evaluation homomorphism ev$_{w}:U_{q}(%
\widetilde{sl}_{2})\rightarrow U_{q}(sl_{2})$ defined by \cite{Jimbo2} 
\begin{eqnarray}
e_{0} &\rightarrow &w\,f,\;f_{0}\rightarrow w^{-1}e,\;k_{0}\rightarrow
k^{-1},  \notag \\
e_{1} &\rightarrow &e,\;f_{1}\rightarrow f,\;k_{1}\rightarrow k,\quad w\in 
\mathbb{C}\;.  \label{ev}
\end{eqnarray}
In the following I shall restrict discussion to such evaluation
representations and define 
\begin{equation}
\pi _{w}^{p}\equiv \pi ^{p}\circ \text{ev}_{w}\;,\quad p\in \text{$\limfunc{%
Spec}$\thinspace }Z\backslash D,\;[\pi ^{p}]=\frak{X}^{-1}(p)\;.  \label{pia}
\end{equation}
The choice to consider only a single evaluation representation will keep the
subsequent calculations feasible. Obviously, the representation $\pi ^{p}$
is only determined up to isomorphism. For some calculations it will be
necessary to remove this ambiguity. This can be achieved by employing the
coordinate map (\ref{phi}), 
\begin{equation}
\pi ^{p}\equiv \pi _{N^{\prime }}^{\xi ,\zeta ,\lambda },\quad (\xi ,\zeta
,\lambda )=\varphi ^{-1}(p),\;  \label{piphi}
\end{equation}
which gives now a concrete realization according to (\ref{basis}). From the
definition (\ref{pia}) one deduces the six central values 
\begin{eqnarray}
\pi _{w}^{p}(\mathbf{x}_{0})/w^{N^{\prime }} &=&\pi _{w}^{p}(\mathbf{y}%
_{1})=(q-q^{-1})^{N^{\prime }}\eta ,\; \\
\pi _{w}^{p}(\mathbf{y}_{0})w^{N^{\prime }} &=&\pi _{w}^{p}(\mathbf{x}%
_{1})=(q-q^{-1})^{N^{\prime }}\zeta ,\;\pi _{w}^{p}(\mathbf{z}_{0}^{\mp
1})=\pi _{w}^{p}(\mathbf{z}_{1}^{\pm 1})=\lambda ^{\pm N^{\prime }}
\end{eqnarray}
leaving only the four free parameters $w$ and $\varphi ^{-1}(p)=(\xi ,\zeta
,\lambda )$.

\subsection{Necessary existence criteria for intertwiners}

Besides the occurrence of irreducible representations depending on
continuous parameters there is another important characteristic feature of
quantum groups at roots of unity. The quantum group $U_{q}(\widetilde{sl}%
_{2})$ at a root of unity is no longer quasitriangular and the concept of a
universal $R$-matrix present at $q^{N}\neq 1$ is problematic. Now one has to
satisfy certain necessary existence criteria for intertwiners. The argument
is by now standard, see e.g. \cite{JmbNotes, CP}, and it is worthwhile to
repeat it in order to show why one can find an intertwiner (\ref{L}) for
generic representations when $q$ is an odd primitive root of unity but not
for an even one. Taking into account that the central subalgebra (\ref{Z0})
forms a Hopf subalgebra, 
\begin{eqnarray*}
\Delta (\mathbf{x}_{i}) &=&(e_{i}\otimes 1+k_{i}\otimes e_{i})^{N^{\prime }}=%
\mathbf{x}_{i}\otimes 1+\mathbf{z}_{i}\otimes \mathbf{x}_{i} \\
\Delta (\mathbf{y}_{i}) &=&(f_{i}\otimes k_{i}^{-1}+1\otimes
f_{i})^{N^{\prime }}=\mathbf{y}_{i}\otimes \mathbf{z}_{i}^{-1}+1\otimes 
\mathbf{y}_{i} \\
\Delta (\mathbf{z}_{i}) &=&\mathbf{z}_{i}\otimes \mathbf{z}_{i}
\end{eqnarray*}
an intertwiner between two representations $\pi ,\pi ^{\prime }$ at $q^{N}=1$
can only exist if the following equalities hold, 
\begin{eqnarray}
\pi (\mathbf{x}_{i})+\pi (\mathbf{z}_{i})\,\pi ^{\prime }(\mathbf{x}_{i})
&=&\pi ^{\prime }(\mathbf{x}_{i})+\pi (\mathbf{x}_{i})\,\pi ^{\prime }(%
\mathbf{z}_{i})  \notag \\
\pi (\mathbf{y}_{i})\pi ^{\prime }(\mathbf{z}_{i}^{-1})+\pi ^{\prime }(%
\mathbf{y}_{i}) &=&\pi ^{\prime }(\mathbf{y}_{i})\,\pi (\mathbf{z}%
_{i}^{-1})+\pi (\mathbf{y}_{i})\;.  \label{crit}
\end{eqnarray}
As the limit $q^{N}\rightarrow 1$ of the six-vertex model is connected with
the two-dimensional nilpotent representation (\ref{pi0}), 
\begin{equation}
q^{N}\rightarrow 1:\pi _{z}^{0}(\mathbf{x}_{i})=\pi _{z}^{0}(\mathbf{y}%
_{i})=0\,,\;\pi _{z}^{0}(\mathbf{z}_{i})=q^{N^{\prime }}=\pm 1,\;i=0,1
\end{equation}
one arrives at the conditions 
\begin{equation}
\pi _{w}^{p}(\mathbf{x}_{i})=\pi _{w}^{p}(\mathbf{x}_{i})\,\pi _{z}^{0}(%
\mathbf{z}_{i})\quad \text{and}\quad \pi _{w}^{p}(\mathbf{y}_{i})=\pi
_{w}^{p}(\mathbf{y}_{i})\,\pi _{z}^{0}(\mathbf{z}_{i})
\end{equation}
for an intertwiner to exist in the case of the tensor product $\pi
_{w}^{p}\otimes \pi _{z}^{0}$. If we take $N\,$\ to be odd then $\pi
_{z}^{0}(\mathbf{z}_{i})=1$ and there are no restrictions on the existence
of an intertwiner between $\pi _{z}^{0}$ and some cyclic, semi-cyclic or
nilpotent evaluation representation $\pi _{w}^{p}$ at a root of unity. For $%
N $ even $\pi _{z}^{0}(\mathbf{z}_{i})=-1$ and an intertwiner can only exist
for nilpotent representations, i.e. $\pi ^{p}(\mathbf{x})=\pi ^{p}(\mathbf{y}%
)=0$.

\section{Intertwiner for $\protect\pi _{w}^{p}\otimes \protect\pi _{z}^{0}$}

Having investigated the existence criteria for intertwiners at a root of
unity we now need to explicitly construct the operator (\ref{L}) which will
provide us with the basic constituent of the auxiliary matrix (\ref{Q}). For
this purpose it will be necessary to consider the larger Hopf algebra $%
\breve{U}_{q}(\widetilde{sl}_{2})$ respectively $\breve{U}_{q}(sl_{2})$ as
the matrix elements of the intertwiner will lie in these algebras and not in 
$U_{q}(\widetilde{sl}_{2}),$ $U_{q}(sl_{2})$. This becomes important since
there is a crucial difference between the representations of $U_{q}(sl_{2})$
for $N$ odd and $N$ even. Given any point $p=(\mathbf{x},\mathbf{y},\mathbf{z%
},\mathbf{c})\in \limfunc{Spec}Z$ all of the associated representations $\pi
^{p}$ in the class $\frak{X}^{-1}(p)$ can be extended to a representation of 
$\breve{U}_{q}(sl_{2})$ when $q$ is an odd root of unity. For example, let $%
p=\varphi (\xi ,\zeta ,\lambda )$ with respect to the coordinate map (\ref
{phi}). Then setting 
\begin{equation}
\pi _{N}^{\xi ,\zeta ,\lambda }(t):=\pi _{N}^{\xi ,\zeta ,\lambda }(k)^{%
\frac{1}{2}}\quad \text{with}\quad \pi _{N}^{\xi ,\zeta ,\lambda }(k)^{\frac{%
1}{2}}v_{n}=\lambda ^{\frac{1}{2}}q^{-n}v_{n}  \label{t1}
\end{equation}
and 
\begin{equation}
\pi _{N}^{\xi ,\zeta ,\lambda }(\breve{e}):=\pi _{N}^{\xi ,\zeta ,\lambda
}(e)\pi _{N}^{\xi ,\zeta ,\lambda }(k)^{-\frac{1}{2}},\quad \pi _{N}^{\xi
,\zeta ,\lambda }(\breve{f}):=\pi _{N}^{\xi ,\zeta ,\lambda }(k)^{\frac{1}{2}%
}\pi _{N}^{\xi ,\zeta ,\lambda }(f)  \label{t2}
\end{equation}
gives a well-defined representation of $\breve{U}_{q}(sl_{2})$. For even
roots of unity this ceases to be valid, unless the representation is
nilpotent, i.e. $\mathbf{x}=\mathbf{y}=0$. The quantum group relations (\ref
{QG2}) are not satisfied for cyclic or semi-cyclic representations because
of the identities 
\begin{eqnarray}
\pi _{N}^{\xi ,\zeta ,\lambda }(k)^{\frac{1}{2}}\pi _{N}^{\xi ,\zeta
,\lambda }(e)\pi _{N}^{\xi ,\zeta ,\lambda }(k)^{-\frac{1}{2}}v_{0}
&=&q^{N^{\prime }+1}\pi _{N}^{\xi ,\zeta ,\lambda }(e)v_{0},  \notag \\
\pi _{N}^{\xi ,\zeta ,\lambda }(k)^{\frac{1}{2}}\pi _{N}^{\xi ,\zeta
,\lambda }(f)\pi _{N}^{\xi ,\zeta ,\lambda }(k)^{-\frac{1}{2}}v_{N^{\prime
}-1} &=&q^{N^{\prime }-1}\pi _{N}^{\xi ,\zeta ,\lambda }(f)v_{N^{\prime
}-1}\;.
\end{eqnarray}
This is directly related to the fact that an intertwiner only exists for
nilpotent representations $\pi ^{p}$ as the cyclicity of the representation
enforces the above sign change.

In order to unburden the notation let us write temporarily $%
e_{i},f_{i},k_{i} $ for the $N^{\prime }\times N^{\prime }$ matrices $\pi
_{w}^{p}(e_{i}),\pi _{w}^{p}(f_{i}),\pi _{w}^{p}(k_{i})$ of $U_{q}(%
\widetilde{sl}_{2})$ and it will be understood that $\mathbf{x}=\mathbf{y}=0$
for $N$ even. Recall the defining property of the intertwiner, 
\begin{equation*}
L^{p}(w/z)(\pi _{w}^{p}\otimes \pi _{z}^{0})\Delta (x)=\left[ \left( \pi
_{w}^{p}\otimes \pi _{z}^{0}\right) \Delta ^{\text{op}}(x)\right]
L^{p}(w/z),\quad x\in U_{q}(\widetilde{sl}_{2})\;.
\end{equation*}
To solve this set of equations it is helpful to decompose the $L$-matrix
over the second factor as (for the moment let us drop the explicit
dependence on $p$ in the notation) 
\begin{equation}
L=A\otimes \sigma ^{+}\sigma ^{-}+B\otimes \sigma ^{+}+C\otimes \sigma
^{-}+D\otimes \sigma ^{-}\sigma ^{+},\;A,B,C,D\in \limfunc{End}(\mathbb{C}%
^{N^{\prime }})  \label{Ldec}
\end{equation}
and to introduce the $q$-deformed commutator 
\begin{equation*}
\lbrack X,Y]_{q}=XY-qYX\;.
\end{equation*}
One then deduces for the Chevalley generators the following commutation
relations for the $L$-matrix entries $A,B,C,D$, 
\begin{equation}
\lbrack
D,k_{i}]=0,\;[A,k_{i}]=0,\;k_{i}Bk_{i}^{-1}=q^{(-)^{i}2}B,%
\;k_{i}Ck_{i}^{-1}=q^{(-)^{i+1}2}C
\end{equation}
\begin{eqnarray}
B &=&\left[ D,e_{0}\right] _{q}=-k_{1}\left[ f_{1},D\right] _{q}=-\left[
A,e_{0}\right] _{q^{-1}}k_{0}^{-1}=\left[ f_{1},A\right] _{q^{-1}} \\
C &=&\left[ f_{0},D\right] _{q^{-1}}=-[D,e_{1}]_{q^{-1}}k_{1}^{-1}=-k_{0}%
\left[ f_{0},A\right] _{q}=\left[ A,e_{1}\right] _{q}  \notag
\end{eqnarray}
\begin{eqnarray}
\lbrack e_{1},C]_{q} &=&[f_{1},B]_{q}=[e_{0},B]_{q}=[f_{0},C]_{q}=0  \notag
\\
\lbrack C,e_{0}]_{q} &=&A-Dk_{0},\quad \lbrack C,f_{1}]_{q}=q(k_{1}^{-1}A-D)
\notag \\
\left[ B,e_{1}\right] _{q} &=&D-Ak_{1},\quad \left[ B,f_{0}\right]
_{q}=q(k_{0}^{-1}D-A)
\end{eqnarray}
Invoking the quantum group relations (\ref{QG}), (\ref{AQG}) and (\ref{ev}),
(\ref{pia}), (\ref{em}), (\ref{t1}), (\ref{t2}) one verifies by direct
computation that the ansatz 
\begin{eqnarray}
A_{p} &=&\rho _{+}\pi ^{p}(t)-\rho _{-}\pi ^{p}(t)^{-1},\quad B_{p}=\rho
_{+}(q-q^{-1})\pi ^{p}(\breve{f}),\;  \notag  \label{Le} \\
C_{p} &=&\rho _{-}\left( q-q^{-1}\right) \pi ^{p}(\breve{e}),\quad
D_{p}=\rho _{+}\pi ^{p}(t)^{-1}-\rho _{-}\pi ^{p}(t)  \label{Lsol}
\end{eqnarray}
yields a valid solution provided we fix the ratio of the coefficients to the
specific value 
\begin{equation}
\rho _{+}/\rho _{-}=qw/z\;,\quad \rho _{\pm }=\rho _{\pm }(w/z,q)\;.
\label{rho}
\end{equation}
Here the representation $\pi ^{p}$ has been introduced again into the
notation to display the explicit dependence of the quantum group generators
on $p\in \limfunc{Spec}\,Z$. Note that the normalization functions can be
chosen independent of $p$. This will become important when considering the
transformation of the intertwiner under the symmetries of the six-vertex
model. Because the representation $\pi ^{p}$ is only fixed up to isomorphism
(\ref{Lsol}) is defined up to the gauge transformations 
\begin{equation}
L^{p}\rightarrow (\phi \otimes 1)L^{p}(\phi ^{-1}\otimes 1),\quad \lbrack
\phi \pi ^{p}\phi ^{-1}]=[\pi ^{p}]\;.  \label{gauge}
\end{equation}
As pointed out earlier, this ambiguity may be removed by applying the
coordinate map (\ref{piphi}). For the definition of the auxiliary matrix
this ambiguity is unimportant as the trace is taken in (\ref{Q}).

Since the tensor product is indecomposable for generic values of the ratio $%
z/w$ one can conclude that the above solution is the only one up to a
normalization factor. In addition, it follows \cite{DJMM90} that the
intertwiner has to satisfy the Yang-Baxter equation (\ref{RLL}). This can
also be verified by an explicit calculation.

Note that the intertwiner (\ref{L}) can be expressed solely in terms of $%
U_{q}(sl_{2})$ for odd roots of unity upon setting alternatively 
\begin{eqnarray}
A_{p} &=&\rho _{+}\pi ^{p}(k)^{\frac{N+1}{2}}-\rho _{-}\pi ^{p}(k)^{\frac{N-1%
}{2}},\;B_{p}=\rho _{+}\left( q-q^{-1}\right) \pi ^{p}(k)^{\frac{N+1}{2}}\pi
^{p}(f),\;  \notag \\
C_{p} &=&\rho _{-}\left( q-q^{-1}\right) \pi ^{p}(e)\pi ^{p}(k)^{\frac{N-1}{2%
}},\;D_{p}=\rho _{+}\pi ^{p}(k)^{\frac{N-1}{2}}-\rho _{-}\pi ^{p}(k)^{\frac{%
N+1}{2}}\;.  \label{Lodd}
\end{eqnarray}
The normalization functions $\rho _{\pm }$ have again to satisfy (\ref{rho})
and can be chosen independent of the parameters $p$ as before. The advantage
of this solution is that we can now employ the results on irreducible
representations of $U_{q}(sl_{2})$ at roots of unity \cite{CK,CKP} to the
intertwiner. This will allow us to identify auxiliary matrices with points
on the hypersurface (\ref{SZ}) and to obtain a geometric picture for the
symmetries of the six-vertex model.

For a particular root of unity representation the intertwiner (\ref{Lsol})
is contained in the solutions to the Yang-Baxter algebra obtained in \cite
{BS90,T92}. The relation will be explained at the end of Section 5. It is
important to note, however, that the expression (\ref{Lsol}) derived here
gives the intertwiner in terms of an arbitrary representation $\pi ^{p}$ and
that only the quantum group relations have to be used in order to verify (%
\ref{L}). This is important for two reasons. First, the particular
representation used in \cite{BS90} is not suitable for discussing the
nilpotent limit which has to be taken when $N$ is even. Second, the
expression (\ref{Lodd}) displaying the dependence on the quantum group
generators will ultimately allow us to identify equivalent auxiliary
matrices by associating them with points in the hypersurface (\ref{SZ}).

\subsection{Transformation under spin-reversal}

As the six-vertex model is invariant under spin-reversal we need to
investigate the behavior of the constructed intertwiner under this
transformation. From the decomposition (\ref{Ldec}) one deduces the simple
transformation property 
\begin{equation}
\left( 1\otimes \sigma ^{x}\right) L\left( 1\otimes \sigma ^{x}\right)
=\left( 
\begin{array}{cc}
D & C \\ 
B & A
\end{array}
\right) \;.  \label{revL}
\end{equation}
This transformation can be interpreted in terms of representation theory by
noting that the algebraic relations resulting from the intertwiner condition
stay invariant provided we apply simultaneously the $U_{q}(\widetilde{sl}%
_{2})$ algebra automorphism $(e_{i},f_{i},k_{i})\overset{\hat{\omega}}{%
\rightarrow }(f_{i+1},e_{i+1},k_{i+1})$, $i\in \mathbb{Z}_{2}$. Hence,
spin-reversal amounts to the replacement 
\begin{equation*}
\pi _{w}^{p}\rightarrow \pi ^{p}\circ \text{ev}_{w}\circ \hat{\omega}
\end{equation*}
in the intertwiner equation (\ref{L}). In order to obtain again an
evaluation representation of the form (\ref{pia}) it is of advantage to
rewrite this in terms of the $U_{q}(sl_{2})$ algebra automorphism given by 
\begin{equation}
\omega (e)=f,\;\omega (f)=e,\;\omega (k)=k^{-1}\;.  \label{om}
\end{equation}
Introducing on the hypersurface (\ref{SZ}) the map 
\begin{equation}
\text{$\limfunc{Spec}$}\,Z\ni p=(\mathbf{x},\mathbf{y},\mathbf{z},\mathbf{c}%
)\rightarrow \frak{R}\,p:=(\mathbf{y},\mathbf{x},\mathbf{z}^{-1},\mathbf{c})
\end{equation}
and observing that the value of the Casimir element (\ref{c}) stays
invariant under the application of $\omega $, 
\begin{equation}
\mathbf{c}=qk+q^{-1}k^{-1}+(q-q^{-1})^{2}fe=qk^{-1}+q^{-1}k+(q-q^{-1})^{2}ef,
\label{RZ}
\end{equation}
one immediately verifies the following equality of equivalence classes, 
\begin{equation}
\lbrack \pi _{w}^{\frak{R}p}]=[\pi ^{p}\circ \omega \circ \text{ev}_{w}]\;.
\label{Rpi}
\end{equation}
Hence, there exists a non-singular $N^{\prime }\times N^{\prime }$ matrix $%
\phi $ transforming one representation into the other and an elementary
calculation now shows that for the intertwiner solution (\ref{Lodd}) one has 
\begin{eqnarray}
A_{p} &=&\mathbf{z}\frak{\,}\left( \rho _{+}\pi ^{p}(k)^{-\frac{N-1}{2}%
}-\rho _{-}\pi ^{p}(k)^{-\frac{N+1}{2}}\right) =\mathbf{z}\frak{\,}\phi D_{%
\frak{R}p}\,\phi ^{-1}  \notag \\
B_{p} &=&\mathbf{z}\frak{\,}\rho _{-}w\left( q-q^{-1}\right) \pi
^{p}(f)\,\pi ^{p}(k)^{-\frac{N-1}{2}}=\mathbf{z}\frak{\,}w\phi C_{\frak{R}%
p}\,\phi ^{-1}  \notag \\
C_{p} &=&\mathbf{z}\frak{\,}\rho _{+}w^{-1}\left( q-q^{-1}\right) \pi
^{p}(k)^{-\frac{N+1}{2}}e=\mathbf{z}\frak{\,}w^{-1}\phi B_{\frak{R}p}\,\phi
^{-1}  \notag \\
D_{p} &=&\mathbf{z}\frak{\,}\left( \rho _{+}\pi ^{p}(k)^{-\frac{N+1}{2}%
}-\rho _{-}\pi ^{p}(k)^{-\frac{N-1}{2}}\right) =\mathbf{z}\frak{\,}\phi A_{%
\frak{R}p}\,\phi ^{-1}\;.
\end{eqnarray}
For the alternative solution (\ref{Lsol}), whose matrix elements lie in $%
\breve{U}_{q}(sl_{2})$, one obtains an analogous result with the exception
that the factor $\mathbf{z}$ needs to be omitted. In summary, we therefore
obtain the transformation law, 
\begin{equation}
\left( 1\otimes \sigma ^{x}\right) L^{p}(w)\left( 1\otimes \sigma
^{x}\right) =\mathbf{z}\frak{\,}(\phi \otimes w^{-\frac{\sigma ^{z}}{2}})L^{%
\frak{R}p}(w)(\phi ^{-1}\otimes w^{\frac{\sigma ^{z}}{2}})\;.  \label{RL0}
\end{equation}
Again the factor $\mathbf{z}$ is absent for the other solution (\ref{Lsol}).
In terms of the coordinate map (\ref{phi}) spin-reversal then amounts to the
coordinate change 
\begin{equation}
\varphi ^{-1}(p)=(\xi ,\zeta ,\lambda )\rightarrow (\xi ^{\frak{R}},\eta
,\lambda ^{-1}q^{-2})=\varphi ^{-1}(\frak{R}p),  \label{Rphi}
\end{equation}
where $\eta $ is given by (\ref{eta}) and the parameter $\xi ^{\frak{R}}$,
which depends on $(\eta ,\zeta ,\lambda )$, is chosen such that 
\begin{equation}
\zeta =\xi ^{\frak{R}}\prod_{n=0}^{N^{\prime }-1}([\lambda
^{-1}q^{-2};n-1][n]+\xi ^{\frak{R}}\eta )\;.  \label{par}
\end{equation}
On basis of this result we will derive the transformation property of the
auxiliary matrix under spin-reversal. The additional spectral parameter
dependence in (\ref{RL0}) induced by the spin-reversal transformation can be
avoided when changing to the principal gradation.

\subsection{The principal gradation}

When defining the evaluation homomorphism $U_{q}(\widetilde{sl}%
_{2})\rightarrow U_{q}(sl_{2})$ one has the choice between two gradations of
the loop algebra. The definition (\ref{ev}) corresponds to the homogeneous
gradation and is associated with the degree operator 
\begin{equation*}
\lbrack d,e_{i}]=\delta _{i0}e_{i},\;[d,f_{i}]=-\delta
_{i0}f_{i},\;[d,k_{i}]=0,
\end{equation*}
while the principal gradation is induced by 
\begin{equation*}
\lbrack \hat{\rho},e_{i}]=e_{i},\;[\hat{\rho},f_{i}]=-f_{i},\;[\hat{\rho}%
,k_{i}]=0.\;
\end{equation*}
Both degrees are related by the identity $\hat{\rho}=h^{\vee }d+\rho $ with $%
h^{\vee }=2$ being the dual Coxeter number and $\rho $ the Weyl vector.
While the homogeneous gradation will mainly be used throughout this article
for algebraic simplicity, the principal gradation is more natural in order
to discuss the behaviour of the intertwiner (\ref{L}) under spin-reversal.
The evaluation homomorphism then reads 
\begin{eqnarray}
e_{0} &\rightarrow &x\,f,\;f_{0}\rightarrow x^{-1}e,\;k_{0}\rightarrow
k^{-1},  \notag \\
e_{1} &\rightarrow &xe,\;f_{1}\rightarrow x^{-1}f,\;k_{1}\rightarrow k,\quad
x=z^{\frac{1}{2}}\in \mathbb{C}\;.  \label{evp}
\end{eqnarray}
This change from the homogeneous gradation to the principal one is reflected
by the following well known gauge transformation of the six-vertex $R$%
-matrix (\ref{R}), 
\begin{equation}
\mathcal{R}(x)=(x^{\frac{\sigma ^{z}}{2}}\otimes 1)R(x^{2})(x^{-\frac{\sigma
^{z}}{2}}\otimes 1)=(1\otimes x^{-\frac{\sigma ^{z}}{2}})R(x^{2})(1\otimes
x^{\frac{\sigma ^{z}}{2}})\;.  \label{Rp}
\end{equation}
The corresponding Boltzmann weights read in this gauge 
\begin{equation}
a=\varrho ,\;b=\varrho \,\frac{\left( 1-x^{2}\right) q}{1-x^{2}q^{2}}%
,\;c=c^{\prime }=\varrho \,\frac{\left( 1-q^{2}\right) x}{1-x^{2}q^{2}}%
,\;\varrho (x,q)=\rho (x^{2},q)\;.  \label{prin}
\end{equation}
Obviously, the six-vertex R-matrix in the homogenous gauge (\ref{R})
violates spin reversal symmetry while the one in the principal gauge (\ref
{Rp}) is invariant. Clearly, this gauge change does not matter on the level
of the transfer matrix (\ref{T}) (upon identifying $z=x^{2}$) as the gauge
transformation can be invoked in the auxiliary space over which the trace is
taken. However, this argument ceases to be valid in the case of the
auxiliary matrix (\ref{Q}) when cyclic or semi-cyclic representations enter
the definition of the $L$-matrix. Let us define the intertwiner (\ref{L}) in
the principal gradation as 
\begin{equation}
\mathcal{L}^{p}(y)=(1\otimes y^{-\frac{\sigma ^{z}}{2}})L^{p}(y^{2})(1%
\otimes y^{\frac{\sigma ^{z}}{2}})=\left( 
\begin{array}{cc}
A_{p} & B_{p}\,y^{-1} \\ 
C_{p}\,y & D_{p}
\end{array}
\right) \;,  \label{Lp}
\end{equation}
where the coefficients (\ref{rho}) entering the matrices $A,B,C,D$ now have
to obey the relation 
\begin{equation}
\varrho _{+}/\varrho _{-}=qy^{2}\;,\quad \varrho _{\pm }(y,q)=\rho _{\pm
}(y^{2},q)\;.
\end{equation}
For nilpotent representations the gauge transformation may also be cast into
the form 
\begin{equation*}
\mathcal{L}^{p}(y)=(\Gamma _{y}\otimes 1)L^{p}(y^{2})(\Gamma
_{y}^{-1}\otimes 1),\quad (\Gamma _{y})_{mn}:=\delta _{mn}y^{-\frac{n}{2}}
\end{equation*}
because of the crucial commutation relations 
\begin{equation*}
\Gamma _{y}B\Gamma _{y}^{-1}=y^{-1}B\quad \text{and\quad }\Gamma _{y}C\Gamma
_{y}^{-1}=yC\;.
\end{equation*}
This does not hold true for $\mathbf{x}\neq 0$ or $\mathbf{y}\neq 0$ due to
the cyclicity of the associated representation unless $y^{N^{\prime }}=1$.
Thus, there is a genuine difference between the principal and the homogenous
gauge on the level of the auxiliary matrix (\ref{Q}) when the associated
representation is not nilpotent. Returning to the question of the
transformation property under spin-reversal one now calculates 
\begin{eqnarray*}
\left( 1\otimes \sigma ^{x}\right) \mathcal{L}^{p}(y)\left( 1\otimes \sigma
^{x}\right) &=&(1\otimes y^{\frac{\sigma ^{z}}{2}})\left( 1\otimes \sigma
^{x}\right) L^{p}(y^{2})\left( 1\otimes \sigma ^{x}\right) (1\otimes y^{-%
\frac{\sigma ^{z}}{2}}) \\
&=&\mathbf{z}(\phi \otimes 1)\mathcal{L}^{\frak{R}p}(y)\left( \phi
^{-1}\otimes 1\right) \;.
\end{eqnarray*}
This result shows that the additional spectral parameter dependence in (\ref
{RL0}) originates in the choice of gradation imposed when defining the
evaluation homomorphism (\ref{ev}). For the subsequent sections let us
return to the homogeneous gradation unless stated otherwise.

\section{Decomposing the tensor product $\protect\pi _{w}^{p}\otimes \protect%
\pi _{z}^{0}$}

In this section the decomposition the tensor product $\pi _{w}^{p}\otimes
\pi _{z}^{0}$ via the exact sequence (\ref{seq}) will be described providing
the basis for deriving the functional equation (\ref{TQ0}). As the
calculations are straightforward but quite lengthy only the results are
presented here. Throughout this section the coordinate map (\ref{phi}) is
applied and all formulas are to be understood with respect to the convention
(\ref{piphi}). The notation will be simplified by writing simply $\pi
_{w},\pi _{w^{\prime }}^{\prime },\pi _{w^{\prime \prime }}^{\prime \prime }$
instead of $\pi _{w}^{p},\pi _{w^{\prime }}^{p^{\prime }},\pi _{w^{\prime
\prime }}^{p^{\prime \prime }}$ with $p=\varphi (\xi ,\zeta ,\lambda
),p^{\prime }=\varphi (\xi ^{\prime },\zeta ^{\prime },\lambda ^{\prime })$
and $p^{\prime \prime }=\varphi (\xi ^{\prime \prime },\zeta ^{\prime \prime
},\lambda ^{\prime \prime })$.

Our strategy to determine the exact sequence (\ref{seq}) is as follows:
First one considers the action of the quantum loop algebra $U_{q}(\widetilde{%
sl}_{2})$ on the tensor product $\pi _{w}\otimes \pi _{z}^{0}$. Under the
assumption that both of the representations $\pi _{w^{\prime }}^{\prime }$
and $\pi _{w^{\prime \prime }}^{\prime \prime }$ are of the form (\ref{pia})
this allows us to set up a set of equations for the coefficients of the
vectors in the tensor space $\pi _{w}\otimes \pi _{z}^{0}$. They determine
the parameters $(w,\xi ,\zeta ,\lambda ),$ $(w^{\prime },\xi ^{\prime
},\zeta ^{\prime },\lambda ^{\prime })$ and $(w^{\prime \prime },\xi
^{\prime \prime },\zeta ^{\prime \prime },\lambda ^{\prime \prime })$
labeling the respective representations $\pi _{w},\pi _{w^{\prime }}^{\prime
}$ and $\pi _{w^{\prime \prime }}^{\prime \prime }$. In order to solve the
equations for the coefficients one has to guarantee the vanishing of a
determinant. This fixes the value of the ratio $\mu =z/w$ for which the
tensor product $\pi _{w}\otimes \pi _{z}^{0}$ is decomposable. One derives
that $\mu $ is given in terms of the Casimir element (\ref{cpi}) through the
following quadratic equation, 
\begin{equation}
\mu +\mu ^{-1}=\xi \zeta (q-q^{-1})^{2}+q\lambda +q^{-1}\lambda ^{-1}=\pi
_{N}^{\xi ,\zeta ,\lambda }(\mathbf{c}),\quad \mu =z/w\;.  \label{mu}
\end{equation}
The ambiguity in $\mu $ is removed by choosing the branch of the square root
such that the limit to semi-cyclic or nilpotent representations is
consistent (see (\ref{mu0}) below), 
\begin{equation}
\lim_{\xi \rightarrow 0}\mu =\lim_{\zeta \rightarrow 0}\mu =\lim_{\xi ,\zeta
\rightarrow 0}\mu =\lambda ^{-1}q^{-1}\;.  \label{mucon}
\end{equation}
In the following we regard $(z,\xi ,\zeta ,\lambda )$ as the independent
variables while we have to tune the evaluation parameter $w$ such that $\mu $
equals the solution (\ref{mucon}) of (\ref{mu}).

\subsection{The inclusion $\protect\pi _{w^{\prime }}^{\prime }\subset 
\protect\pi _{w}\otimes \protect\pi _{z}^{0}$}

Denote by $\{\uparrow ,\downarrow \}$ (spin up, spin down) the standard
basis for the two-dimensional evaluation representation (\ref{pi0}). The
explicit form of the inclusion map defining the subrepresentation $\pi
_{w^{\prime }}^{\prime }$ in the exact sequence (\ref{seq}) is 
\begin{equation}
\imath :\pi _{w^{\prime }}^{\prime }\hookrightarrow \pi _{w}\otimes \pi
_{z}^{0},\quad w_{n}^{\prime }\hookrightarrow X_{n}=\alpha
_{n}v_{n+1}\otimes \uparrow +\beta _{n}v_{n}\otimes \downarrow ,\;n\in 
\mathbb{Z}_{N^{\prime }}\;.  \label{i}
\end{equation}
Here the coefficients are 
\begin{equation}
\alpha _{n}=\zeta ^{\delta _{N^{\prime }-1,n}}q^{-n}\alpha _{0}\quad \text{%
and\quad }\beta _{n}=\frac{\mu q\lambda ^{-1}\cdot q^{n}-q^{-n}}{q-q^{-1}}%
\,\alpha _{0}  \label{ab}
\end{equation}
The parameter $\alpha _{0}$ is arbitrary unless a specific normalization is
chosen. The representation defined by the inclusion (\ref{i}) is indeed of
the form (\ref{pia}) with 
\begin{equation}
\pi _{w^{\prime }}^{\prime }=\pi _{N^{\prime }}^{\xi ^{\prime },\zeta
^{\prime },\lambda ^{\prime }}\circ \text{ev}_{w^{\prime }}
\end{equation}
and parameters 
\begin{equation}
\xi ^{\prime }\zeta ^{\prime }=\xi \zeta \frac{q\mu -\lambda }{\mu -q\lambda 
},\quad \zeta ^{\prime }=q^{N^{\prime }}\zeta ,\quad \lambda ^{\prime
}=\lambda q^{-1},\quad w^{\prime }=wq\;.  \label{cycp}
\end{equation}
The subrepresentation $\pi _{N^{\prime }}^{\xi ^{\prime },\zeta ^{\prime
},\lambda ^{\prime }}$ of the finite quantum group $U_{q}(sl_{2})$ assigns
the following values to the central elements (\ref{Z0}), 
\begin{equation}
\pi _{N^{\prime }}^{\xi ^{\prime },\zeta ^{\prime },\lambda ^{\prime }}(%
\mathbf{x})=(q-q^{-1})^{N^{\prime }}\eta ,\;\pi _{N^{\prime }}^{\xi ^{\prime
},\zeta ^{\prime },\lambda ^{\prime }}(\mathbf{y})=(q^{2}-1)^{N^{\prime
}}\zeta ,\;\pi _{N^{\prime }}^{\xi ^{\prime },\zeta ^{\prime },\lambda
^{\prime }}(\mathbf{z})=q^{N^{\prime }}\lambda ^{N^{\prime }}  \label{Z0p}
\end{equation}
and for the Casimir element one obtains 
\begin{equation}
\pi _{N^{\prime }}^{\xi ^{\prime },\zeta ^{\prime },\lambda ^{\prime }}(%
\mathbf{c})=q\mu +q^{-1}\mu ^{-1}\;.  \label{cp}
\end{equation}
Note that the first identity in (\ref{Z0p}) is not obvious as one needs to
verify the identity 
\begin{equation}
\frac{\xi ^{\prime }}{\xi }\prod_{n=1}^{N^{\prime }-1}\frac{[\lambda
^{\prime };n-1][n]+\xi ^{\prime }\zeta ^{\prime }}{[\lambda ;n-1][n]+\xi
\zeta }=\frac{\eta ^{\prime }}{\eta }=1\;.  \label{etap}
\end{equation}
From (\ref{mu}) and (\ref{cp}) one infers that the representations $\pi
_{N}^{\xi ,\zeta ,\lambda }$ and $\pi _{N}^{\xi ^{\prime },\zeta ^{\prime
},\lambda ^{\prime }}$ are in general not isomorphic as they belong to
different points in the hypersurface $\limfunc{Spec}\,Z$. As $q^{N^{\prime
}}=1$ for $N$ odd, the derived expressions (\ref{cycp}) imply, however, that
the representations $\pi _{N^{\prime }}^{\xi ,\zeta ,\lambda }$ and $\pi
_{N^{\prime }}^{\xi ^{\prime },\zeta ^{\prime },\lambda ^{\prime }}$
correspond to the same point in $\limfunc{Spec}\,Z_{0}$. For even roots of
unity this ceases to be valid as $q^{N^{\prime }}=-1$.

Most of the above formulas stay valid in the limiting case of semi-cyclic ($%
\xi $ or $\zeta \rightarrow 0$) and nilpotent representations ($\xi ,\zeta
\rightarrow 0$). As already mentioned above the tensor product now becomes
reducible at 
\begin{equation}
\mu =z/w=1/q\lambda \;.  \label{mu0}
\end{equation}
The coefficients determining the inclusion map remain unchanged with the
only exception that 
\begin{equation}
\zeta =0\;\Rightarrow \;\alpha _{N^{\prime }-1}=0\;.
\end{equation}
Also, if $\zeta =0$ and $\xi \neq 0$ the parameters change from (\ref{cycp})
to 
\begin{equation}
\xi ^{\prime }=\xi q^{N^{\prime }}\frac{\lambda -\lambda ^{-1}}{\lambda
q-\lambda ^{-1}q^{-1}},\quad \zeta =\zeta ^{\prime }=0,\quad \lambda
^{\prime }=\lambda q^{-1}
\end{equation}
with the central value $\eta $ being unchanged, 
\begin{equation}
1=\frac{\xi ^{\prime }}{\xi }\prod_{n=1}^{N^{\prime }-1}\frac{[\lambda ;n][n]%
}{[\lambda ;n-1][n]}=\frac{\eta ^{\prime }}{\eta }\;.
\end{equation}
The remaining cases $\xi =0,\zeta \neq 0$ and $\xi ,\zeta =0$ are obtained
by taking the appropriate limit of the previous equations for the cyclic
case.

\subsection{The quotient representation $\protect\pi _{w^{\prime \prime
}}^{\prime \prime }=\protect\pi _{w}\otimes \protect\pi _{z}^{0}/\protect\pi
_{w^{\prime }}^{\prime }$}

In order to determine the representation $\pi _{w}\otimes \pi _{z}^{0}/\pi
_{w^{\prime }}^{\prime }$ one needs to consider the action of the quantum
group generators on vectors in the tensor product which do not lie in the
representation space $\pi _{w^{\prime }}^{\prime }$ and identify the latter
with the null-space. This is implemented by defining the following
projection onto the quotient space via linear extension, 
\begin{equation}
\tau :\pi _{w}\rightarrow \pi _{w^{\prime \prime }}^{\prime \prime }=\pi
_{w}\otimes \pi _{z}^{0}/\pi _{w^{\prime }}^{\prime },\text{\quad }\tau
(X_{n})=0,\quad \tau (Y_{n})=w_{n}^{\prime \prime }  \label{pro}
\end{equation}
with the vectors $X_{n}$ given by equations (\ref{i}), (\ref{ab}) and 
\begin{equation}
Y_{n}=\gamma _{n}\,v_{n}\otimes \uparrow ,\;\gamma _{n}=\prod_{m=1}^{n}\frac{%
[\lambda ^{\prime \prime };m-1][m]+\xi ^{\prime \prime }\zeta ^{\prime
\prime }}{[\lambda ;m-1][m]+\xi \zeta }\,\gamma _{0},\quad n\in \mathbb{Z}%
_{N^{\prime }}\;.  \label{Y}
\end{equation}
The parameters $(\xi ^{\prime \prime },\zeta ^{\prime \prime },\lambda
^{\prime \prime })$ entering the coefficients $\gamma _{n}$ fix the
evaluation representation 
\begin{equation}
\pi _{w^{\prime \prime }}^{\prime \prime }=\pi _{N}^{\xi ^{\prime \prime
},\zeta ^{\prime \prime },\lambda ^{\prime \prime }}\circ \text{ev}%
_{w^{\prime \prime }}
\end{equation}
and read explicitly 
\begin{equation}
\lambda ^{\prime \prime }=\lambda q,\quad \xi ^{\prime \prime }\zeta
^{\prime \prime }=\xi \zeta \frac{\mu q^{-1}-\lambda q^{2}}{\mu -q\lambda }%
,\quad \zeta ^{\prime \prime }=q^{N^{\prime }}\zeta ,\quad w^{\prime \prime
}=wq^{-1}\;.  \label{cycpp}
\end{equation}
Viewing $\pi _{w}$ as a representation of the finite quantum group $%
U_{q}(sl_{2})$ one calculates from these identities the corresponding values
of the central elements giving the points in $\limfunc{Spec}\,Z_{0}$ and $%
\limfunc{Spec}\,Z$. These are 
\begin{equation}
\pi _{N}^{\xi ^{\prime \prime },\zeta ^{\prime \prime },\lambda ^{\prime
\prime }}(\mathbf{x})=(q-q^{-1})^{N^{\prime }}\eta ,\quad \pi _{N}^{\xi
^{\prime \prime },\zeta ^{\prime \prime },\lambda ^{\prime \prime }}(\mathbf{%
y})=(q^{2}-1)^{N^{\prime }}\zeta ,\quad \pi _{N}^{\xi ^{\prime \prime
},\zeta ^{\prime \prime },\lambda ^{\prime \prime }}(\mathbf{z}%
)=q^{N^{\prime }}\lambda ^{N^{\prime }}  \label{Z0pp}
\end{equation}
and 
\begin{equation}
\pi _{N}^{\xi ^{\prime \prime },\zeta ^{\prime \prime },\lambda ^{\prime
\prime }}(\mathbf{c})=\mu q^{-1}+\mu ^{-1}q\;.  \label{cpp}
\end{equation}
Here the identity 
\begin{equation}
1=\eta ^{\prime \prime }/\eta =\xi ^{\prime \prime }/\xi
\prod_{n=1}^{N^{\prime }-1}\frac{[\lambda ^{\prime \prime };n-1][n]+\xi
^{\prime \prime }\zeta ^{\prime \prime }}{[\lambda ;n-1][n]+\xi \zeta }
\label{etapp}
\end{equation}
has been employed. For $N$ odd ($q^{N^{\prime }}=1$) one now immediately
verifies from (\ref{Z0pp}) and (\ref{cpp}) that $[\pi _{N}^{\xi ^{\prime
\prime },\zeta ^{\prime \prime },\lambda ^{\prime \prime }}]$ shares the
same point in the variety $\limfunc{Spec}\,Z_{0}$ as $[\pi _{N}^{\xi ,\zeta
,\lambda }]$ and $[\pi _{N}^{\xi ^{\prime },\zeta ^{\prime },\lambda
^{\prime }}]$, but all three representations belong to different points in $%
\limfunc{Spec}\,Z$, i.e. they are in general not isomorphic. For even roots
of unity ($q^{N^{\prime }}=-1$) only $[\pi _{N}^{\xi ^{\prime \prime },\zeta
^{\prime \prime },\lambda ^{\prime \prime }}]$ and $[\pi _{N}^{\xi ^{\prime
},\zeta ^{\prime },\lambda ^{\prime }}]$ are mapped onto the same location
in $\limfunc{Spec}\,Z_{0}$.

There are certain simplifications in the limit of semi-cyclic and nilpotent
representations, namely one finds for the coefficients in (\ref{Y}) that 
\begin{equation*}
\gamma _{n}=\frac{[\lambda ;-1]}{[\lambda ;n-1]}\,\gamma _{0}\;.
\end{equation*}
The parameters in the case of semi-cyclic representations are now 
\begin{equation*}
\xi =0:\quad \quad \xi ^{\prime \prime }=0,\quad \zeta ^{\prime \prime
}=q^{N^{\prime }}\zeta ,\quad \lambda ^{\prime \prime }=\lambda q,\quad
w^{\prime \prime }=wq^{-1}
\end{equation*}
and 
\begin{equation*}
\zeta =0:\quad \quad \xi ^{\prime \prime }=\xi \,\frac{[\lambda ;N^{\prime
}-2]}{[\lambda ;-1]},\quad \zeta ^{\prime \prime }=0,\quad \lambda ^{\prime
\prime }=\lambda q,\quad w^{\prime \prime }=wq^{-1}\;.
\end{equation*}
The remaining possibility of nilpotent representations follows from the
above by setting $\xi ,\zeta =0$.

\section{The $T$-$Q$ functional equation}

We are now in the position to derive the functional equation (\ref{TQ0}) by
exploiting the previous results on the decomposition of the tensor product $%
\pi _{w}^{p}\otimes \pi _{z}^{0}$ and the explicit construction of the
intertwiner (\ref{L}). Let us start by defining the following family of
auxiliary matrices labelled by points on the hypersurface (\ref{SZ}), 
\begin{equation}
Q_{p}(z):=\limfunc{tr}\limits_{V_{0}=\pi _{w}^{p}}L_{0M}^{p}(z/\mu
_{p})\cdots L_{01}^{p}(z/\mu _{p}),\;\pi _{w}^{p}=\pi ^{p}\circ \text{ev}%
_{w=z/\mu _{p}}\;.  \label{Q0}
\end{equation}
For odd roots of unity we employ the solution (\ref{Lodd}) and for even
roots of unity (\ref{Lsol}). Note that the ambiguity in choosing a
representative $\pi ^{p}$ in the equivalence class $[\pi ^{p}]=\frak{X}%
^{-1}(p)$ only manifests itself in the gauge transformation (\ref{gauge}).
As the trace is taken over the auxiliary space $V_{0}=\pi _{w}^{p}$ in (\ref
{Q0}) the matrix elements of the operator $Q_{p}(z)$ are therefore functions
on the hypersurface (\ref{SZ}). Identifying $p=\varphi (\xi ,\zeta ,\lambda
) $ gives an explicit prescription how to construct the auxiliary matrix via
the representations (\ref{basis}) and (\ref{piphi}). In the following the
``coordinate free'' notation (\ref{Q0}) will be used but for explicit
calculations it will be understood that we invoke the coordinate map (\ref
{phi}). The parameter $\mu _{p}$ corresponds to the solution of (\ref{mu})
satisfying (\ref{mucon}) in order to make contact with the exact sequence (%
\ref{seq}). It can be defined solely in terms of the point $p\in \limfunc{%
Spec}$\thinspace $Z$ and is ``coordinate independent'', 
\begin{equation}
p=(\mathbf{x},\mathbf{y},\mathbf{z},\mathbf{c}=\mu _{p}+\mu _{p}^{-1}),\quad
\lim_{\mathbf{x},\mathbf{y}\rightarrow 0}\mu _{p}^{-N^{\prime
}}=q^{N^{\prime }}\mathbf{z}\;.  \label{mup}
\end{equation}
Finally, note that for $N$ even we have to set $\mathbf{x}=\mathbf{y}=0$ in
order to ensure that the intertwiner (\ref{L}) exists, hence (\ref{Q0})
reduces in this case to a one-parameter family $Q_{p},\;p=(0,0,\mathbf{z},-%
\mathbf{z}-\mathbf{z}^{-1})$.

In order to unburden the notation for the following calculations the
explicit dependence of the auxiliary matrix on the point in the hypersurface
(\ref{SZ}) or that of the parameters $(\xi ,\zeta ,\lambda )$ will be
temporarily dropped. That is, for fixed but arbitrary $p\in \limfunc{Spec}%
\,Z $ set $Q(z)\equiv Q_{p}(z)=Q_{\varphi (\xi ,\zeta ,\lambda )}(z)$ and $%
\pi _{w}^{p}\equiv \pi _{w}$. Similarly, the matrices $Q_{p^{\prime
}}(z),Q_{p^{\prime \prime }}(z)$ with the parameters $\varphi ^{-1}(p)=(\xi
^{\prime },\zeta ^{\prime },\lambda ^{\prime }),\varphi ^{-1}(p^{\prime
\prime })=(\xi ^{\prime \prime },\zeta ^{\prime \prime },\lambda ^{\prime
\prime })$ given in equations (\ref{cycp}) and (\ref{cycpp}) will simply be
written as $Q^{\prime }(z),Q^{\prime \prime }(z)$ and $\pi _{w}^{p^{\prime
}}\equiv \pi _{w}^{\prime },\pi _{w}^{p^{\prime \prime }}\equiv \pi
_{w}^{\prime \prime }$. Furthermore, we set $\mu ^{\prime }=\mu _{\xi
^{\prime },\zeta ^{\prime },\lambda ^{\prime }}$ and $\mu ^{\prime \prime
}=\mu _{\xi ^{\prime \prime },\zeta ^{\prime \prime },\lambda ^{\prime
\prime }}$ which according to (\ref{cp}), (\ref{cpp}) and (\ref{mucon}) are
given by 
\begin{equation}
\mu ^{\prime }=q\mu \quad \text{and\quad }\mu ^{\prime \prime }=q^{-1}\mu \;.
\end{equation}

We start the derivation of the functional equation (\ref{TQ0}) by
considering the operator product of the auxiliary and transfer matrix which
can be written as, 
\begin{equation}
Q(z)T(z)=\limfunc{tr}\limits_{\pi _{w}\otimes \pi _{z}^{0}}L_{\pi
_{w},M}(z/\mu )R_{\pi _{z}^{0},M}(z)\cdots L_{\pi _{w},1}(z/\mu )R_{\pi
_{z}^{0},1}(z)\;.  \label{tq1}
\end{equation}
As the $L$-matrix is an intertwiner, it has a non-trivial kernel only when $%
w=\mu ^{\pm 1}$, i.e. when the tensor product becomes reducible. In
particular, one has 
\begin{equation}
L(\mu ^{-1})|_{\imath \,\pi _{wq}^{\prime }}\equiv 0\;.
\end{equation}
Here $\imath :\pi _{wq}^{\prime }\hookrightarrow \pi _{w}\otimes \pi
_{z}^{0} $ is the inclusion map (\ref{i}). The above relation can be
calculated explicitly. (The other solution $w=\mu $ is related by
spin-reversal, cf equation (\ref{Rpi}) and (\ref{muR}).) As a consequence of
this observation and from the identity (\ref{RLL}) for the three-fold tensor
product 
\begin{equation*}
\underset{1}{\pi _{w}}\otimes \underset{2}{\pi _{z}^{0}}\otimes \underset{3}{%
\pi _{1}^{0}}\;,
\end{equation*}
we can conclude that the operator products 
\begin{equation*}
L_{13}(w=z/\mu )R_{23}(z)
\end{equation*}
in expression (\ref{tq1}) leave the image of the representation space $\pi
_{wq}^{\prime }$ under the inclusion map (\ref{i}) invariant. Suppose the
following equations are satisfied 
\begin{eqnarray}
L_{13}(w)R_{23}(z)(\imath \otimes 1) &=&\phi _{1}(z,q)\,(\imath \otimes
1)L^{\prime }(wq)\;,  \label{phi1} \\
(\tau \otimes 1)L_{13}(w)R_{23}(z) &=&\phi _{2}(z,q)\,L^{\prime \prime
}(w/q)(\tau \otimes 1)\;.  \label{phi2}
\end{eqnarray}
Here $\imath :\pi _{wq}^{\prime }\hookrightarrow \pi _{w}\otimes \pi
_{z}^{0} $ is again the inclusion map (\ref{i})and $\tau :\pi _{w}\otimes
\pi _{z}^{0}\rightarrow \pi _{w/q}^{\prime \prime }$ the projection (\ref
{pro}). Then a functional equation of the following form must hold, 
\begin{eqnarray}
Q(z)T(z) &=&\limfunc{tr}\limits_{\pi _{w}\otimes \pi _{z}^{0}}L_{\pi
_{w},M}(z/\mu )R_{\pi _{z}^{0},M}(z)\cdots L_{\pi _{w},1}(z/\mu )R_{\pi
_{z}^{0},1}(z)  \notag \\
&=&\phi _{1}(z,q)Q^{\prime }(zq^{2})+\phi _{2}(z,q)Q^{\prime \prime
}(zq^{-2}),\quad \mu ^{\prime }=q\mu ,\;\mu ^{\prime \prime }=q^{-1}\mu .
\label{TQ}
\end{eqnarray}
It remains to prove (\ref{phi1}) and (\ref{phi2}) and to derive the explicit
form of the coefficients functions $\phi _{1},\phi _{2}$. From the
definition of the $R$-matrix (\ref{R}) one explicitly calculates the
following identities, 
\begin{multline}
L_{13}R_{23}\,X_{n}\otimes \uparrow =\left\{ (a\,\alpha
_{n}\,Av_{n+1}+c\,\beta _{n}\,Bv_{n})\otimes \uparrow +b\,\beta
_{n}\,Av_{n}\otimes \downarrow \right\} \otimes \uparrow  \notag \\
\quad +\left\{ (a\,\alpha _{n}\,Cv_{n+1}+c\,\beta _{n}\,Dv_{n})\otimes
\uparrow +b\,\beta _{n}\,Cv_{n}\otimes \downarrow \right\} \otimes \downarrow
\notag \\
=\phi _{1}\,(\imath \otimes 1)(A^{\prime }v_{n}^{\prime }\otimes \uparrow
+C^{\prime }v_{n}^{\prime }\otimes \downarrow )
\end{multline}
\begin{multline}
L_{13}R_{23}\,X_{n}\otimes \downarrow =\left\{ b\,\alpha
_{n}\,Bv_{n+1}\otimes \uparrow +(c^{\prime }\alpha _{n}\,Av_{n+1}+a\,\beta
_{n}\,Bv_{n})\otimes \downarrow \right\} \otimes \uparrow  \notag \\
\quad +\left\{ b\,\alpha _{n}\,Dv_{n+1}\otimes \uparrow +(c^{\prime }\alpha
_{n}\,Cv_{n+1}+a\frak{\,}\beta _{n}\,Dv_{n})\otimes \downarrow \right\}
\otimes \downarrow  \notag \\
=\phi _{1}\,(\imath \otimes 1)(B^{\prime }v_{n}^{\prime }\otimes \uparrow
+D^{\prime }v_{n}^{\prime }\otimes \downarrow )
\end{multline}
and 
\begin{multline}
(\tau \otimes 1)L_{13}R_{23}\,Y_{n}\otimes \uparrow =(\tau \otimes
1)(a\,\gamma _{n}\,Av_{n}\otimes \uparrow \otimes \uparrow +a\,\gamma
_{n}\,Cv_{n}\otimes \uparrow \otimes \downarrow )  \notag \\
=\phi _{2}\,(A^{\prime \prime }v_{n}^{\prime \prime }\otimes \uparrow
+C^{\prime \prime }v_{n}^{\prime \prime }\otimes \downarrow )
\end{multline}
\begin{multline}
(\tau \otimes 1)L_{13}R_{23}\,Y_{n}\otimes \downarrow =(\tau \otimes
1)\left( b\,\gamma _{n}\,Bv_{n}\otimes \uparrow +c^{\prime }\gamma
_{n}\,Av_{n}\otimes \downarrow \right) \otimes \uparrow  \notag \\
+(\tau \otimes 1)\left( b\,\gamma _{n}\,Dv_{n}\otimes \uparrow +c^{\prime
}\gamma _{n}\,Cv_{n}\otimes \downarrow \right) \otimes \downarrow  \notag \\
=\phi _{2}\,(B^{\prime \prime }v_{n}^{\prime \prime }\otimes \uparrow
+D^{\prime \prime }v_{n}^{\prime \prime }\otimes \downarrow )
\end{multline}
These relations translate into equations involving the coefficients in (\ref
{ab}) and (\ref{Y}). The solution consistent with all the equations is for
the intertwiner (\ref{Lsol}) 
\begin{equation}
N\;\text{even}:\phi _{1}(z,q)=b(z,q)q^{-\frac{1}{2}}\rho _{-}/\rho
_{-}^{\prime },\text{\quad }\phi _{2}(z,q)=a(z,q)q^{\frac{1}{2}}\rho
_{-}/\rho _{-}^{\prime \prime }  \label{Phiev}
\end{equation}
and for the alternative $L$-matrix (\ref{Lodd}) 
\begin{equation}
N\;\text{odd}:\phi _{1}(z,q)=b(z,q)q^{\frac{N-1}{2}}\rho _{-}/\rho
_{-}^{\prime },\text{\quad }\phi _{2}(z,q)=a(z,q)q^{\frac{1-N}{2}}\rho
_{-}/\rho _{-}^{\prime \prime }\;.  \label{Phodd}
\end{equation}
Here $a,b$ are the Boltzmann weights (\ref{h}) of the six-vertex $R$-matrix
and $\rho _{-},\rho _{-}^{\prime },\rho _{-}^{\prime \prime }$ the
normalization functions (\ref{rho}) of the intertwiners associated with the
tensor products $\pi _{w}\otimes \pi _{1}^{0},\pi _{wq}^{\prime }\otimes \pi
_{1}^{0}$ and $\pi _{w/q}^{\prime \prime }\otimes \pi _{1}^{0}$. If one
wants to eliminate the powers of $q$ the normalization functions should be
set to 
\begin{equation}
\rho _{\pm }(w,q)=q^{\frac{1\pm 1}{2}}w^{\pm \frac{1}{2}}\;\Rightarrow
\;\phi _{1}(z,q)=b(z,q),\;\phi _{2}(z,q)=a(z,q)\;.  \label{Phab}
\end{equation}
For $N$ odd the correct square root has to be chosen such that $w^{\prime 
\frac{1}{2}}=q^{\frac{1-N}{2}}w^{\frac{1}{2}}$ and $w^{\prime \prime \frac{1%
}{2}}=q^{\frac{N-1}{2}}w^{\frac{1}{2}}$. This gives the desired functional
equation (\ref{TQ0}) between the transfer matrix (\ref{T}) and the auxiliary
matrix (\ref{Q0}).

Again it needs to be emphasized that the three auxiliary matrices appearing
in the functional equation are not equivalent in general. In fact, for $N$
odd the values of the central elements (\ref{Z0}) agree for all three
representations $\pi _{w},\pi _{wq}^{\prime }$ and $\pi _{w/q}^{\prime
\prime }$ but the values of the Casimir element are different according to
the identities (\ref{mu}), (\ref{cp}) and (\ref{cpp}). Setting all three
Casimir elements equal leads to the condition 
\begin{equation*}
\pi _{N}^{\xi ,\zeta ,\lambda }(\mathbf{c})=\pi _{N}^{\xi ^{\prime },\zeta
^{\prime },\lambda ^{\prime }}(\mathbf{c})=\pi _{N}^{\xi ^{\prime \prime
},\zeta ^{\prime \prime },\lambda ^{\prime \prime }}(\mathbf{c})\Rightarrow
\mu ^{2}=-q=-q^{-1}
\end{equation*}
implying that $q=\pm 1$ which is excluded from our construction. Thus, there
is a genuine difference between the functional equation (\ref{BTQ}) and the
one derived here on the basis of representation theory. Nonetheless, we may
construct now in an obvious manner solutions to Baxter's functional equation
(\ref{BTQ}). Recall from (\ref{fiber}) that for a given point $p\in \limfunc{%
Spec}Z$ the auxiliary matrices associated to the fiber over $p_{\ast }$ are
given by 
\begin{equation}
Q_{\ell }(z)\equiv Q_{p_{\ell }}(z),\quad p_{\ell }=(\mathbf{x},\mathbf{y},%
\mathbf{z},q^{\ell }\mu +\mu ^{-1}q^{-\ell }),\;\ell \in \mathbb{Z}%
_{N},\;p_{0}\equiv p\;.  \label{Qel}
\end{equation}
The functional equation (\ref{TQ0}) is now rewritten as 
\begin{equation}
Q_{\ell }(z)T(z)=\phi _{1}(z)^{M}Q_{\ell +1}(zq^{2})+\phi _{2}(z)^{M}Q_{\ell
-1}(zq^{-2})\;.  \label{TQel}
\end{equation}
If we now sum over all the points in the fiber one immediately deduces that
the operator 
\begin{equation}
Q_{p_{\ast }}^{(s)}(z)\equiv \sum_{\ell \in \mathbb{Z}_{N}}q^{-s\ell
}Q_{\ell }(z),\quad p_{\ast }=(\mathbf{x},\mathbf{y},\mathbf{z})\in \limfunc{%
Spec}Z_{0},\;s\in \mathbb{Z}_{N}  \label{Qfiber}
\end{equation}
provides a valid solution to Baxter's functional equation (\ref{BTQ}) with a
slight modification of the coefficient functions $\phi _{1},\phi _{2}$, 
\begin{equation}
Q_{p_{\ast }}^{(s)}(z)T(z)=\phi _{1}(z)^{M}q^{s}Q_{p_{\ast
}}^{(s)}(zq^{2})+\phi _{2}(z)^{M}q^{-s}Q_{p_{\ast }}^{(s)}(zq^{-2})\;.
\end{equation}
As indicated this solution lives on $\limfunc{Spec}Z_{0}$ rather than $%
\limfunc{Spec}Z$, cf (\ref{Zseq}).

For $N$ even one has to set $\mathbf{x}=\mathbf{y}=0$ and the functional
equation (\ref{TQ0}) in terms of the coordinate map (\ref{phi}) now reads 
\begin{equation}
N\;\text{even}:Q_{\lambda }(z)T(z)=\phi _{1}(z)^{M}Q_{\lambda
q^{-1}}(zq^{2})+\phi _{2}(z)^{M}Q_{\lambda q}(zq^{-2}),\quad Q_{\lambda
}\equiv Q_{\varphi (0,0,\lambda )}\;.  \label{TQe}
\end{equation}
The auxiliary matrices are again not equivalent as they belong to different
root of unity representations. This time not only the values of the Casimir
elements are different but also the values of the central element $\mathbf{z}
$ (cf equations (\ref{Z0p}) and (\ref{Z0pp})). We therefore have to sum over
two fibers, i.e. $\mathbb{Z}_{N}$ instead of $\mathbb{Z}_{N^{\prime }}$, to
obtain a solution to (\ref{BTQ}). Thus, the expression (\ref{Qfiber})
applies to all roots of unity when setting $Q_{\ell }(z)\equiv Q_{\varphi
(0,0,\lambda q^{\ell })}(z)$.\smallskip

The following cautious remark with regard to the solutions (\ref{Qfiber}) of
Baxter's functional equation (\ref{BTQ}) must be made. Depending on the
length of the spin-chain as well as the spin-sector it can happen that these
solutions are trivial, i.e. they might sum up to zero. We will investigate
this below for two examples. There we verify that the solutions (\ref{Qfiber}%
) are non-trivial in the spin-zero sector of the four-chain, where we
compare them with Baxter's expression (\ref{BQ}). This shows that at roots
of unity one can find (at least in certain sectors) solutions to (\ref{BTQ})
which are of a simpler form than (\ref{QB}), namely finite sums of the
expression (\ref{Q}). However, from the construction it is clear that the
auxiliary matrices (\ref{Q0}) defined on $\limfunc{Spec}Z$ should be
regarded as the fundamental objects in the present setting.

\subsection{Transformation properties of the auxiliary matrix}

In this section the transformation properties of the auxiliary matrix
related with the symmetries of the six-vertex transfer matrix (\ref{T}) are
investigated. The first transformation law involves the total spin (\ref{Sz}%
) and is a direct consequence from the intertwining relation (\ref{L}). As
the coproduct and opposite coproduct coincide for the Cartan elements one
has 
\begin{equation*}
L^{p}(w)\,\pi ^{p}(k)\otimes q^{\sigma ^{z}}=\pi ^{p}(k)\otimes q^{\sigma
^{z}}\,L^{p}(w)
\end{equation*}
which in turn implies 
\begin{equation}
\lbrack Q_{p}(z),\,q^{\sigma ^{z}}\otimes q^{\sigma ^{z}}\cdots \otimes
q^{\sigma ^{z}}]=0\;.
\end{equation}
Hence, the auxiliary matrix flips at most multiples of $N^{\prime }$ spins.
This property is due to non-vanishing contributions in the trace in (\ref{Q0}%
) containing the matrices $B^{N^{\prime }},C^{N^{\prime }}$ which depend on
the central elements $\mathbf{x},\mathbf{y}$. The only other non-vanishing
contributions contain the operators $B,C$ in pairs and thus do not change
the total spin. Consequently, we have the transformation law 
\begin{equation}
e^{tS^{z}}Q_{p}(z)e^{-tS^{z}}=Q_{e^{tS^{z}}p}(z),  \label{QSz}
\end{equation}
where the following map on the hypersurface (\ref{SZ}) has been introduced, 
\begin{equation}
\text{$\limfunc{Spec}$}\,Z\ni p=(\mathbf{x},\mathbf{y},\mathbf{z},\mathbf{c}%
)\rightarrow e^{tS^{z}}p:=(e^{-tN}\mathbf{x},e^{tN}\mathbf{y},\mathbf{z},%
\mathbf{c})\;.  \label{ZSz}
\end{equation}
For nilpotent representations, $\mathbf{x}=\mathbf{y}=0$, the auxiliary
matrices obviously commute with the total spin operator.

Next let us investigate the transformation of the auxiliary matrices under
the action of $\frak{S}$ defined in (\ref{R&S}). From the following simple
transformation of the $L$-matrix, 
\begin{equation}
(1\otimes \sigma ^{z})L(1\otimes \sigma ^{z})=\left( 
\begin{array}{cc}
A & -B \\ 
-C & D
\end{array}
\right)
\end{equation}
and the fact that the only non-vanishing terms in the trace (\ref{Q0})
contain the step operators $B,C$ either in pairs or to the power $N$ when $N$
is odd, one deduces the transformation property 
\begin{equation}
\frak{S\,}Q_{p}(z)\frak{S}=Q_{\frak{S}p}(z)  \label{SQ}
\end{equation}
with 
\begin{equation}
\text{$\limfunc{Spec}$}\,Z\ni p=(\mathbf{x},\mathbf{y},\mathbf{z},\mathbf{c}%
)\rightarrow \frak{S}p:=(-\mathbf{x},-\mathbf{y},\mathbf{z},\mathbf{c})\;.
\label{ZS}
\end{equation}
This includes the case $N$ even where $\mathbf{x}=\mathbf{y}=0$. The
transformation behaviour (\ref{SQ}) together with the second identity in (%
\ref{RST}) allows us to discuss the case of even primitive roots of unity $%
q^{2N^{\prime }}=1$ with $N^{\prime }$ odd and cyclic representations.
Performing the replacement $q\rightarrow -q$ we obviously recover the case
of odd roots of unity and can therefore conclude for even $M$, 
\begin{equation}
Q_{p}(z,-q)\frak{S}T(z,q)=b(z,q)^{M}Q_{p^{\prime
}}(zq^{2},-q)+a(z,q)^{M}Q_{p^{\prime \prime }}(zq^{-2},-q)\;.  \label{TQS}
\end{equation}
Here the trivial relations $b(z,-q)=-b(z,q),a(z,-q)=a(z,q)$ have been used.
The operators in this functional equation do not in general commute with
each other due to (\ref{SQ}). Nevertheless, this functional equation might
be useful to gain insight in the different structure encountered for even
and odd roots of unity.

Finally, we investigate the behaviour under spin-reversal employing the
previous investigations of Section 3.1. There we already saw that the
spin-reversal transformation induces the mapping (cf (\ref{Rp})) 
\begin{equation*}
\text{$\limfunc{Spec}$}\,Z\ni p=(\mathbf{x},\mathbf{y},\mathbf{z},\mathbf{c}%
)\rightarrow \frak{R}\,p:=(\mathbf{y},\mathbf{x},\mathbf{z}^{-1},\mathbf{c})
\end{equation*}
According to the definition (\ref{Q0}) we have $w=z/\mu _{p}$ where $\mu
_{p} $ is given by the quadratic equation (\ref{mup}). While the value of
the Casimir element stays invariant under spin reversal we now need to
switch to the other solution $\mu _{p}^{-1}$ satisfying 
\begin{equation}
\lim_{\mathbf{x},\mathbf{y}\rightarrow 0}\mu _{\frak{R}p}^{-N^{\prime
}}=\lim_{\mathbf{x},\mathbf{y}\rightarrow 0}(\mu _{p}^{-1})^{-N^{\prime }}=%
\mathbf{z}^{-1}\;.  \label{muR}
\end{equation}
This follows directly from (\ref{Rp}). Hence, we obtain the following
transformation law for the auxiliary matrix, 
\begin{eqnarray}
\frak{R\,}Q_{p}(z)\frak{R} &=&\limfunc{tr}\limits_{V_{0}=\pi _{w}^{p}}\sigma
_{M}^{x}L_{0M}^{p}(w)\sigma _{M}^{x}\cdots \sigma
_{1}^{x}L_{01}^{p}(w)\sigma _{1}^{x},\;  \notag \\
&=&\mathbf{z}^{M}\limfunc{tr}\limits_{V_{0}=\pi _{w}^{\frak{R}p}}w^{-\sigma
_{M}^{z}/2}L_{0M}^{\frak{R}p}(w)w^{\sigma _{M}^{z}/2}\cdots w^{-\sigma
_{1}^{z}/2}L_{01}^{\frak{R}p}(w)w^{\sigma _{1}^{z}/2},\;  \notag \\
&=&\mathbf{z}^{M}\,w^{-S^{z}}Q_{\frak{R}p}(z\mu
_{p}^{-2})w^{S^{z}},\;w=z/\mu _{p}\;.  \label{QR}
\end{eqnarray}
This transformation law simplifies for nilpotent representations, $\mathbf{x}%
,\mathbf{y}=0$, using the coordinate map (\ref{phi}) to 
\begin{equation}
\frak{R\,}Q_{\lambda }(z)\frak{R}=\lambda ^{N^{\prime }M}Q_{\lambda
^{-1}q^{-2}}(z\lambda ^{2}q^{2})\;.  \label{QR0}
\end{equation}
For even roots of unity and the solution (\ref{Lsol}) the factor $\lambda
^{N^{\prime }M}=\mathbf{z}^{M}$ has to be omitted. Note that spin-reversal
symmetry is only broken for spin-chains which are sufficiently long.

The last transformation law we are going to derive involves auxiliary
matrices with inverted arguments. Setting for simplicity $\rho
_{+}=wq,\;\rho _{-}=1$ in (\ref{Lsol}) respectively (\ref{Lodd}) it is
straightforward to verify the identity 
\begin{equation}
(1\otimes \sigma ^{x})L^{p}(w)(1\otimes \sigma ^{x})=-wq(1\otimes (-wq)^{-%
\frac{\sigma ^{z}}{2}})L^{p}(w^{-1}q^{-2})^{t_{2}}(1\otimes (-wq)^{\frac{%
\sigma ^{z}}{2}})\;.
\end{equation}
Here the superscript $t_{2}$ denotes transposition with respect to the
second factor. Consequently, the auxiliary matrix obeys the relation 
\begin{equation}
\frak{R\,}Q_{p}(z)\frak{R}=(-wq)^{M}\,(-w)^{-S^{z}}Q_{p}(z^{-1}q^{-2}\mu
_{p}^{2})^{t}(-w)^{S^{z}},\;w=z/\mu _{p}\;.
\end{equation}
This quite complicated looking transformation behaviour again simplifies for
nilpotent representations where spin-conservation is restored.

\subsubsection{Principal gradation}

For completeness let us now consider the principal gradation (\ref{evp}).
According to the discussion in Section 3.2, cf (\ref{Lp}), the corresponding
auxiliary matrix is calculated to be 
\begin{eqnarray}
\mathcal{Q}_{p}(x) &=&\limfunc{tr}\limits_{0}\mathcal{L}_{0M}(x/\mu ^{\frac{1%
}{2}})\cdots \mathcal{L}_{01}(x/\mu ^{\frac{1}{2}})  \notag \\
&=&(x/\mu ^{\frac{1}{2}})^{-S^{z}}\limfunc{tr}\limits_{0}L_{0M}(x^{2}/\mu
)\cdots L_{01}(x^{2}/\mu )\,(x/\mu ^{\frac{1}{2}})^{S^{z}}  \notag \\
&=&(x/\mu ^{\frac{1}{2}})^{-S^{z}}Q_{p}(x^{2})(x/\mu ^{\frac{1}{2}%
})^{S^{z}}\;.  \label{Qp}
\end{eqnarray}
Exploiting the transformation law (\ref{QSz}) this allows us to rewrite all
the previous relations for the auxiliary matrix (\ref{Q0}) in terms of the
principal gradation. For example, exploiting (\ref{Sz}) the functional
equation with the transfer matrix now reads 
\begin{equation}
\mathcal{Q}_{p}(x)T(x^{2})=\phi _{1}(x^{2},q)^{M}\mathcal{Q}_{p^{\prime
}}(xq)+\phi _{2}(x^{2},q)^{M}\mathcal{Q}_{p^{\prime \prime }}(xq^{-1}),\quad
z=x^{2}\;.  \label{TQp}
\end{equation}
For nilpotent representations both gradations are obviously equivalent as
the auxiliary matrix then conserves the total spin.

\subsection{Commutation relations}

In order to make contact between the functional equation (\ref{TQ0}) and to
derive the Bethe ansatz equations (\ref{BE}) for the six-vertex model one
needs to ensure that all the operators in (\ref{TQ0}) commute with each
other. The commutation of the transfer matrix with the different auxiliary
matrices is immediate from the existence of the intertwiner (\ref{L}) and (%
\ref{RLL}). We have already seen that there are no restrictions on $p\in 
\limfunc{Spec}Z\backslash D$ when $N$ is odd. For $N$ even we have to set $%
\mathbf{x}=\mathbf{y}=0$ as mentioned before.

Employing the same argument to guarantee the commutation of the auxiliary
matrices in (\ref{TQ0}) let us verify whether the necessary existence
criteria (\ref{crit}) are satisfied for the corresponding intertwiners. It
is worthwhile doing this for two arbitrary points $p=(\mathbf{x},\mathbf{y},%
\mathbf{z},\mathbf{c})$ and $\bar{p}=(\mathbf{\bar{x}},\mathbf{\bar{y}},%
\mathbf{\bar{z}},\mathbf{\bar{c}})$. We are looking for an operator such
that 
\begin{equation}
S_{p\bar{p}}(w,\bar{w})(\pi _{w}^{p}\otimes \pi _{\bar{w}}^{\bar{p}})\Delta
(x)=\left[ (\pi _{w}^{p}\otimes \pi _{\bar{w}}^{\bar{p}})\Delta ^{\text{op}%
}(x)\right] S_{p\bar{p}}(w,\bar{w}),\;x\in U_{q}(\widetilde{sl}_{2})\;.
\label{Spp}
\end{equation}
The case of even roots of unity is trivial as only nilpotent representations
occur and one finds that the necessary requirements (\ref{crit}) are met for
arbitrary values of the spectral parameters $w,\bar{w}$. For $N$ odd cyclic
representations are allowed and one finds 
\begin{gather}
\mathbf{x}+\mathbf{z}\frak{\,}\mathbf{\bar{x}}=\mathbf{\bar{x}}+\mathbf{x}%
\frak{\,}\mathbf{\bar{z}},\quad \mathbf{y}\frak{\,}\mathbf{\bar{z}}^{-1}+%
\mathbf{\bar{y}}\frak{\,}=\mathbf{z}^{-1}\mathbf{\bar{y}}+\mathbf{y},  \notag
\\
\mathbf{x}\frak{\,}\mathbf{\bar{z}}+\mathbf{\bar{x}}\frak{\,}(w/\bar{w})^{N}=%
\mathbf{z\,\bar{x}}(w/\bar{w})^{N}+\mathbf{x}\,,\quad (w/\bar{w})^{N}\mathbf{%
y}+\mathbf{\bar{y}}\frak{\,}\mathbf{z}^{-1}=\mathbf{\bar{y}}+\mathbf{y}\frak{%
\,}\mathbf{\bar{z}}^{-1}\,(w/\bar{w})^{N}\;.  \label{critpp}
\end{gather}
From these equations one deduces that $(w/\bar{w})^{N}=1$ needs to hold
unless $\mathbf{z}=\mathbf{\bar{z}}=1$. According to (\ref{Z0p}) and (\ref
{Z0pp}) the values of the central elements (\ref{Z0}) coincide for all three
representations in the exact sequence (\ref{seq}). Hence, the criteria are
met for all three auxiliary matrices in (\ref{TQ0}) respectively (\ref{TQel}%
). Strictly speaking these requirements are necessary for the existence of
an intertwiner but not sufficient. However, employing the results of the
important paper \cite{BS90}, the intertwiner (\ref{Spp}) can be shown to
exist for $q^{N}=1$ with $N$ odd.

\subsubsection{Connection with the chiral Potts model}

Bazhanov and Stroganov pointed out \cite{BS90} that the Boltzmann weights of
the chiral Potts model \cite{AYMCPTY,BPAY} solve the Yang-Baxter algebra of
the $L$-operators at roots of unity. As the solution in \cite{BS90} is given
in a particular root-of-unity representation different from (\ref{basis}) it
is helpful to briefly review the results and make the connection with
evaluation representations of $U_{q}(\widetilde{sl}_{2})$ explicit.

Denote by $\{v_{n}\}_{n\in \mathbb{Z}_{N}}$ as before the standard basis in $%
\mathbb{C}^{N}$ and define the operators 
\begin{equation}
Zv_{n}=q^{-n}v_{n},\;Xv_{n}=v_{n+1},\quad n\in \mathbb{Z}_{N}\;.
\end{equation}
Then the solution to the Yang-Baxter algebra (\ref{RLL}) found in \cite{BS90}
is given by (setting $A=Z,\;B=Z^{-1},\;C=X$ with respect to the notation
used in equations (2.12) and (2.13) in \cite{BS90}) 
\begin{equation}
\tilde{L}(w)=\left( 
\begin{array}{cc}
w^{\frac{1}{2}}d_{+}Z+w^{-\frac{1}{2}}d_{-}Z^{-1} & w^{\frac{1}{2}}\left(
g_{+}Z^{-1}+g_{-}Z\right) X \\ 
w^{-\frac{1}{2}}\left( h_{+}Z^{-1}+h_{-}Z\right) X^{-1} & w^{\frac{1}{2}%
}f_{+}Z^{-1}+w^{-\frac{1}{2}}f_{-}Z
\end{array}
\right) ,  \label{BSL}
\end{equation}
where the six coefficients $\chi =\{d_{+},d_{-},f_{+},f_{-},g_{+},g_{-}\}$
are independent complex parameters and $h_{\pm }=f_{\pm }d_{\mp }/g_{\pm }$.
Under a suitable choice of the parameters this solution can be identified as
an intertwiner of the Hopf algebra $\breve{U}_{q}(\widetilde{sl}_{2})$.
Given a representation $\pi $ of the finite subalgebra $\breve{U}%
_{q}(sl_{2}) $ the solution to the intertwiner equation (\ref{L}) for the
associated evaluation representation now is 
\begin{equation}
\breve{L}(w)=\left( 
\begin{array}{cc}
\breve{\rho}_{+}\pi (t)-\breve{\rho}_{-}\pi (t)^{-1} & \breve{\rho}%
_{+}(q-q^{-1})q^{-\frac{1}{2}}\pi (\breve{f}) \\ 
\breve{\rho}_{-}(q-q^{-1})q^{\frac{1}{2}}\pi (\breve{e}) & \breve{\rho}%
_{+}\pi (t)^{-1}-\breve{\rho}_{-}\pi (t)
\end{array}
\right) ,\;\breve{\rho}_{+}/\breve{\rho}_{-}=wq\;.  \label{Lbar}
\end{equation}
The additional factors of $q^{\pm \frac{1}{2}}$ in the off-diagonal matrix
elements in comparison to (\ref{Lsol}) are explained by choosing (\ref{pi0})
as representation of $\breve{U}_{q}(\widetilde{sl}_{2})$ instead of $U_{q}(%
\widetilde{sl}_{2})$. Invoking the following cyclic representation 
\begin{equation}
\pi (\breve{e})=s_{0}^{-1}\,\frac{s_{1}Z^{-1}-s_{1}^{-1}Z}{q-q^{-1}}%
\,X^{-1},\;\pi (\breve{f})=s_{0}\,\frac{s_{2}Z-s_{2}^{-1}Z^{-1}}{q-q^{-1}}%
\,X,\;\pi (t)=\left( \frac{s_{1}}{s_{2}}\right) ^{-\frac{1}{2}}Z
\label{piBS}
\end{equation}
with $s_{0},s_{1},s_{2}\in \mathbb{C}^{\times }$ and central values 
\begin{eqnarray}
\pi (\mathbf{x}) &=&\left( \frac{s_{1}}{s_{2}}\right) ^{-\frac{N}{2}%
}(s_{1}^{N}-s_{1}^{-N})s_{0}^{-N},\;\pi (\mathbf{y})=\left( \frac{s_{1}}{%
s_{2}}\right) ^{\frac{N}{2}}(s_{2}^{N}-s_{2}^{-N})s_{0}^{N},\;  \notag \\
\pi (\mathbf{z}) &=&(s_{1}/s_{2})^{-N},\;\pi (\mathbf{c}%
)=q^{-1}s_{1}s_{2}+q(s_{1}s_{2})^{-1}
\end{eqnarray}
the solution (\ref{BSL}) can be identified as the intertwiner (\ref{Lbar})
upon setting 
\begin{equation}
\breve{\rho}_{\pm }=q^{\pm \frac{1}{2}}w^{\pm \frac{1}{2}},\;d_{\pm
}=-q^{\pm 1}f_{\mp }=\pm q^{\pm \frac{1}{2}}(s_{1}/s_{2})^{\mp \frac{1}{2}%
},\;g_{\pm }=\mp s_{0}s_{2}^{\mp 1},\;h_{\pm }=\pm s_{0}^{-1}s_{1}^{\pm 1}\;.
\label{BSid}
\end{equation}

The three additional parameters of (\ref{BSL}) in comparison to the solution
(\ref{Lsol}) can be accounted for as follows. First note that the ratio of
the coefficient functions (\ref{rho}) is arbitrary if one requires the $L$%
-matrix only to satisfy (\ref{RLL}). The remaining two parameters can be
understood in terms of Drinfel'd's quantum double construction \cite{Drin}.
They fix the value of two additional central elements which arise when one
considers the quantum double of the upper Borel subalgebra, see e.g. \cite
{DJMM90} for an explanation in the context of the chiral Potts model. The
solution (\ref{Lsol}) is obtained when setting these central elements to one.

In \cite{BS90} the authors showed by construction that the operator products 
\begin{equation*}
\tilde{L}_{13}(w_{1},\chi _{1})\tilde{L}_{23}(w_{2},\chi _{2})\text{\quad
and\quad }\tilde{L}_{23}(w_{2},\chi _{2})\tilde{L}_{13}(w_{1},\chi _{1})
\end{equation*}
are equivalent provided the parameter sets $\chi _{1},\chi _{2}$ share three
common invariants (cf equations (4.4) and (4.5) on page 809 in \cite{BS90}).
Employing the identification (\ref{BSid}) these invariants can be expressed
in terms of the central elements (\ref{Z0}) in the representation (\ref{piBS}%
), 
\begin{eqnarray*}
\Gamma _{1} &=&\frac{(d_{+}^{N}-f_{+}^{N})(d_{-}^{N}-f_{-}^{N})}{%
(g_{+}^{N}+g_{-}^{N})(h_{+}^{N}+h_{-}^{N})}=\frac{%
(1-s_{2}^{N}/s_{1}^{N})(1-s_{1}^{N}/s_{2}^{N})}{%
(s_{1}^{N}-s_{1}^{-N})(s_{2}^{N}-s_{2}^{-N})}=\frac{(1-\pi (\mathbf{z}%
)^{-1})(1-\pi (\mathbf{z}))}{\pi (\mathbf{x})\pi (\mathbf{y})},\quad \\
\Gamma _{2} &=&w^{-N}\frac{d_{-}^{N}-f_{-}^{N}}{d_{+}^{N}-f_{+}^{N}}=w^{-N}%
\frac{(s_{1}^{N}/s_{2}^{N}-1)}{(s_{1}^{N}/s_{2}^{N}-1)}=w^{-N}, \\
\Gamma _{3} &=&w^{-N}\frac{h_{+}^{N}+h_{-}^{N}}{g_{+}^{N}+g_{-}^{N}}=-w^{-N}%
\frac{s_{0}^{-N}(s_{1}^{N}-s_{1}^{-N})}{s_{0}^{N}(s_{2}^{N}-s_{2}^{-N})}%
=-w^{-N}\pi (\mathbf{z}^{-1})\pi (\mathbf{x})/\pi (\mathbf{y})\;.
\end{eqnarray*}
Comparing with (\ref{critpp}) one now deduces that the intertwiner (\ref{Spp}%
) indeed exists and is a special case of the chiral Potts model. As
demonstrated the auxiliary matrix (\ref{Q0}) is independent of the choice of
the root of unity representation, hence we can conclude that the following
auxiliary matrices commute 
\begin{equation}
\lbrack Q_{p}(z),Q_{\bar{p}}(\bar{z})]=0,\text{\quad }\frac{\mathbf{x}}{1-%
\mathbf{z}}=\frac{\mathbf{\bar{x}}}{1-\mathbf{\bar{z}}},\;\frac{\mathbf{y}}{%
1-\mathbf{z}^{-1}}=\frac{\mathbf{\bar{y}}}{1-\mathbf{\bar{z}}^{-1}}%
,\;z^{N}=(z\mu /\bar{\mu})^{N}\;.  \label{Qcom}
\end{equation}
Clearly for any two points $p=p_{\ell },\bar{p}=p_{k},\;k,\ell \in \mathbb{Z}
$ in the fiber (\ref{fiber}) the above conditions are satisfied and all the
auxiliary matrices (\ref{Qel}) evaluated at the same spectral value $z$
commute with each other. In particular, this allows us to write the
functional equation (\ref{TQ0}) in terms of eigenvalues and to derive the
Bethe ansatz equations (\ref{BE}).

\subsubsection{Comment on previous solutions to Baxter's functional equation}

Note that auxiliary matrices for the functional equation (\ref{BTQ}) and not
(\ref{TQ0}) have been considered in \cite{BS90} and \cite{BOU01}. In \cite
{BS90} (page 805), see also Section 3 in \cite{BOU01}, it is stated that one
is then forced to make the specific choice 
\begin{equation*}
\{d_{+},d_{-},f_{+},f_{-},g_{+},g_{-},h_{+},h_{-}\}\rightarrow \{a,b,cw^{-%
\frac{1}{2}},dw^{\frac{1}{2}},\lambda b,\lambda a,\lambda cw^{-\frac{1}{2}%
},\lambda dw^{\frac{1}{2}}\}
\end{equation*}
of the parameters in (\ref{BSL}). The corresponding five-parameter auxiliary
matrices are built from the operators 
\begin{equation}
\tilde{L}(w)=\left( 
\begin{array}{cc}
w^{\frac{1}{2}}aZ+w^{-\frac{1}{2}}bZ^{-1} & w^{\frac{1}{2}}\lambda \left(
bZ^{-1}+aZ\right) X \\ 
w^{-1}\lambda \left( cZ^{-1}+wdZ\right) X^{-1} & cZ^{-1}+dZ
\end{array}
\right) \;.  \label{BSL2}
\end{equation}
Note, that due to the additional spectral variable dependence in the
parameters this $L$-matrix does not satisfy the Yang-Baxter equation (\ref
{RLL}). (Because $Q_{R,L}$ in Baxter's construction need not commute with
the transfer matrix, this poses no problem.) Moreover, the above solution (%
\ref{BSL2}) cannot be interpreted in terms of evaluation representations of
the quantum loop algebra. The auxiliary matrices considered in \cite
{BS90,BOU01} are assumed to be of the form (\ref{Q}) and to solve (\ref{BTQ}%
). This corresponds to Baxter's construction procedure for the eight \cite
{Bx72,Bx73a,Bx73b,Bx73c,BxBook} and six-vertex-model \cite{BxBook} which is
different from the one applied here. We already saw that the solution to (%
\ref{BTQ}) in the present framework is given as a sum of the expressions (%
\ref{Q}), cf (\ref{Qfiber}).

\section{Two simple examples: $N=3,\;M=3,4$}

In this section the specific examples $N=3,\;M=3,4$ are considered in order
to illustrate the construction procedure of the auxiliary matrix (\ref{Q0})
and to demonstrate the working of the functional equation (\ref{TQ0}). The
matrices in (\ref{TQ0}) are diagonalized and it is shown explicitly that the
additional parameter dependence of the auxiliary matrices then drops out of
the functional equation. This must be the case as the eigenvalues of the
transfer matrix (\ref{T}) and the Bethe ansatz equations (\ref{BE}) only
depend on the variables $z$ and $q$. In particular, it is shown that the
center of the complete $N$-strings (\ref{Nstring}) describing the degenerate
eigenstates of the six-vertex model are given in terms of the central
elements of the quantum group. Furthermore, we will compare the auxiliary matrix (%
\ref{Q0}) for the four chain with Baxter's expression (\ref{BQ}) in the spin
zero sector.

Invoking the representation (\ref{basis}) the Chevalley generators of the
quantum group are, 
\begin{gather}
\pi _{3}^{\xi ,\zeta ,\lambda }(k)=\lambda \left( 
\begin{array}{ccc}
1 &  &  \\ 
& q^{-2} &  \\ 
&  & q^{-4}
\end{array}
\right) ,\;\pi _{3}^{\xi ,\zeta ,\lambda }(f)=\left( 
\begin{array}{ccc}
0 & 0 & \zeta \\ 
1 & 0 & 0 \\ 
0 & 1 & 0
\end{array}
\right) \;,  \notag \\
\pi _{3}^{\xi ,\zeta ,\lambda }(e)=\left( 
\begin{array}{ccc}
0 & \xi \zeta +\frac{\lambda -\lambda ^{-1}}{q-q^{-1}} & 0 \\ 
0 & 0 & \xi \zeta -q^{N^{\prime }}\frac{\lambda q^{-1}-\lambda ^{-1}q}{%
q-q^{-1}} \\ 
\xi & 0 & 0
\end{array}
\right) \;.
\end{gather}
Using these expression one can now explicitly write down the intertwiner (%
\ref{Lodd}) and the auxiliary matrix (\ref{Q0}). Nonetheless, all
expressions will be given in terms of the values of the central elements,
independent of the representation used. This emphasizes the aforementioned
fact that the matrix elements of (\ref{Q0}) are functions on the
hypersurface (\ref{SZ}).

\subsection{The $M=3$ spin chain}

We start with $M=3$ as it is the minimum length of the spin-chain required
for the breaking of spin-reversal symmetry and spin conservation when $N=3$.
One finds the following non-vanishing matrix elements of (\ref{Q0}), 
\begin{eqnarray}
2S^{z} &=&\pm 3:Q_{\uparrow \uparrow \uparrow }^{\uparrow \uparrow \uparrow
}=\limfunc{tr}A^{3},\;Q_{\downarrow \downarrow \downarrow }^{\uparrow
\uparrow \uparrow }=\limfunc{tr}B^{3},\;Q_{\uparrow \uparrow \uparrow
}^{\downarrow \downarrow \downarrow }=\limfunc{tr}C^{3},\;Q_{\downarrow
\downarrow \downarrow }^{\downarrow \downarrow \downarrow }=\limfunc{tr}%
D^{3},  \notag \\
2S^{z} &=&\pm 1:Q_{\downarrow \uparrow \uparrow }^{\downarrow \uparrow
\uparrow }=Q_{\uparrow \downarrow \uparrow }^{\uparrow \downarrow \uparrow
}=Q_{\uparrow \uparrow \downarrow }^{\uparrow \uparrow \downarrow }=\limfunc{%
tr}A^{2}D,\;Q_{\uparrow \downarrow \downarrow }^{\uparrow \downarrow
\downarrow }=Q_{\downarrow \uparrow \downarrow }^{\downarrow \uparrow
\downarrow }=Q_{\downarrow \downarrow \uparrow }^{\downarrow \downarrow
\uparrow }=\limfunc{tr}D^{2}A,  \notag \\
2S^{z} &=&1:Q_{\downarrow \uparrow \uparrow }^{\uparrow \downarrow \uparrow
}=Q_{\uparrow \downarrow \uparrow }^{\uparrow \uparrow \downarrow
}=Q_{\uparrow \uparrow \downarrow }^{\downarrow \uparrow \uparrow }=\limfunc{%
tr}ACB,\;Q_{\downarrow \uparrow \uparrow }^{\uparrow \uparrow \downarrow
}=Q_{\uparrow \downarrow \uparrow }^{\downarrow \uparrow \uparrow
}=Q_{\uparrow \uparrow \downarrow }^{\uparrow \downarrow \uparrow }=\limfunc{%
tr}ABC,  \notag \\
2S^{z} &=&-1:Q_{\uparrow \downarrow \downarrow }^{\downarrow \downarrow
\uparrow }=Q_{\downarrow \uparrow \downarrow }^{\downarrow \downarrow
\uparrow }=Q_{\downarrow \downarrow \uparrow }^{\uparrow \downarrow
\downarrow }=\limfunc{tr}DBC,\;Q_{\uparrow \downarrow \downarrow
}^{\downarrow \downarrow \uparrow }=Q_{\downarrow \uparrow \downarrow
}^{\uparrow \downarrow \downarrow }=Q_{\downarrow \downarrow \uparrow
}^{\downarrow \uparrow \downarrow }=\limfunc{tr}DCB\;.
\end{eqnarray}
To simplify the notation we have dropped the dependence on the point in the
hypersurface and the spectral variable. The matrix elements are defined
according to the convention $Q(z)\left| \underline{\alpha }\right\rangle
=\sum_{\underline{\beta }}Q(z)_{\underline{\alpha }}^{\underline{\beta }%
}\,\left| \underline{\beta }\right\rangle \;.$ From the matrix elements we
infer that each of the sectors $S^{z}=\pm 1/2$ is mapped into itself. Thus
we need only diagonalize the matrices 
\begin{equation}
Q|_{S^{z}=\pm 3/2}=\left( 
\begin{array}{cc}
\limfunc{tr}A^{3} & \limfunc{tr}B^{3} \\ 
\limfunc{tr}C^{3} & \limfunc{tr}D^{3}
\end{array}
\right) \quad \text{and\quad }Q|_{S^{z}=1/2}=\left( 
\begin{array}{ccc}
\limfunc{tr}A^{2}D & \limfunc{tr}ABC & \limfunc{tr}ACB \\ 
\limfunc{tr}ACB & \limfunc{tr}A^{2}D & \limfunc{tr}ABC \\ 
\limfunc{tr}ABC & \limfunc{tr}ACB & \limfunc{tr}A^{2}D
\end{array}
\right) \;.
\end{equation}
The matrix $Q|_{S^{z}=-1/2}$ is obtained by exploiting the transformation
law (\ref{QR}). The matrix elements turn out to take an algebraically
simpler form when the following choice of the normalization functions (\ref
{rho}) is made, 
\begin{equation*}
\rho _{\pm }=3^{-\frac{1}{3}}(wq)^{\frac{1\pm 1}{2}}\;.
\end{equation*}
As $M=N=3$ this does not change the form of the functional equation (\ref
{TQ0}) (cf (\ref{TQ}) and (\ref{Phodd})). The matrix elements are calculated
to 
\begin{eqnarray}
\limfunc{tr}A^{3} &=&w^{3}\mathbf{z}^{2}-\mathbf{z},\frak{\;}\limfunc{tr}%
B^{3}=w^{3}\mathbf{y}\frak{\,}\mathbf{z}^{2},\;\limfunc{tr}C^{3}=\mathbf{x}%
\frak{\,}\mathbf{z},\;\limfunc{tr}D^{3}=\,w^{3}\mathbf{z}-\mathbf{z}^{2} 
\notag \\
\limfunc{tr}A^{2}D &=&wq\,\mathbf{z}(1-wq\,\mathbf{z}),\;\limfunc{tr}ABC=w\,%
\mathbf{z}(1-w\,\mathbf{z}),\;\limfunc{tr}ACB=wq\,\mathbf{z}(q-w\,\mathbf{z}%
)\;.
\end{eqnarray}
Here the spectral variable is set to $w=z/\mu $ with $\mu $ given by (\ref
{mu}) and the central elements take the values detailed in (\ref{eta}). The
eigenvalues of the auxiliary matrix in the sector $S^{z}=\pm 3/2$ are found
to be 
\begin{eqnarray}
Q_{\pm }(w) &=&\tfrac{1}{2}(\limfunc{tr}A^{3}+\limfunc{tr}D^{3}\pm \sqrt{(%
\limfunc{tr}A^{3}-\limfunc{tr}D^{3})^{2}+4\limfunc{tr}B^{3}\limfunc{tr}C^{3}}%
)  \notag \\
&=&\frac{\mathbf{z}}{2}\left( (w^{3}-1)(\mathbf{z}+1)\pm \sqrt{(\mathbf{z}%
-1)^{2}(w^{3}+1)^{2}+4\,w^{3}\mathbf{xyz})}\right) \;.
\end{eqnarray}
The zeroes of these two eigenvalues form two-complete 3-strings, 
\begin{equation}
Q_{\pm }(w_{n})=0,\quad w_{n}=q^{n}\left\{ \mathbf{xy}+\mathbf{z}+\mathbf{z}%
^{-1}\pm \sqrt{\left( \mathbf{xy}+\mathbf{z}+\mathbf{z}^{-1}\right) ^{2}-4}%
\right\} ^{\frac{1}{3}},\;n=0,1,2\;.
\end{equation}
Note that the string center is determined by the central elements (\ref{Z0})
and one has 
\begin{equation*}
\mathbf{xy}+\mathbf{z}+\mathbf{z}^{-1}=F_{3}(\mathbf{c})=\mu ^{3}+\mu
^{-3}\;.
\end{equation*}
Thus, the above complete $3$-strings simplify to $w_{n}=q^{n}\mu ^{\pm 1}$.
If the nilpotent limit is taken the simple form (\ref{Nstring}) for the
eigenvalues is recovered. The complete $3$-string contributions cancel on
both sides of the functional equation (\ref{TQ0}) due to the $q$-periodicity
of $Q_{\pm }$. Thus, one obtains the corresponding eigenvalues $T_{\pm
}(z)=a(z)^{3}+b(z)^{3}$ of the transfer matrix (\ref{T}) as required.

For the sector $S^{z}=1/2$ one finds the eigenvalues 
\begin{eqnarray}
Q_{0}(w) &=&\limfunc{tr}A^{2}D+\limfunc{tr}ABC+\limfunc{tr}ACB\equiv 0,\; \\
Q_{1}(w) &=&\limfunc{tr}A^{2}D+q\limfunc{tr}ABC+q^{2}\limfunc{tr}ACB=3qw%
\mathbf{z}, \\
Q_{2}(w) &=&\limfunc{tr}A^{2}D+q^{2}\limfunc{tr}ABC+q\limfunc{tr}%
ACB\}=-3q^{2}w^{2}\mathbf{z}^{2}\;.
\end{eqnarray}
From the first expression we infer that the auxiliary matrix turns out to be
singular and not all eigenvalues of the transfer matrix can be calculated. The solution (\ref{Qfiber}) for $s=0$
vanishes completely. If we set $s=1,2$ either $Q_{1}^{(s)}$ or $Q_{2}^{(s)}$
becomes zero showing that the solution (\ref{Qfiber}) has a nullspace of
higher rank than (\ref{Q0}).

For the remaining two eigenvalues in the sector $S^{z}=1/2$ one reads off a
simple and a double zero at $w=z=0$. These correspond to ``Bethe roots at
infinity'' when the parametrization $z=e^{u}q^{-1}$ is used in the Bethe
ansatz equations (\ref{BE}). That such ``Bethe roots'' can occur is a known
phenomenon, see e.g. the discussion in \cite{Bx02} and references therein.
Note that also here the dependence on the central elements drops out of the
equation (\ref{TQ0}) showing as expected that the corresponding eigenvalues
of the transfer matrix 
\begin{equation*}
T_{1}(z)=b(z)^{3}q+a(z)^{3}q^{2},\quad T_{2}(z)=b(z)^{3}q^{2}+a(z)^{3}q
\end{equation*}
are independent of the point $p=\varphi (\xi ,\zeta ,\lambda )$ in the
hypersurface (\ref{SZ}).

\subsection{The $M=4$ spin chain}

We now consider the spin-sectors $S^{z}=0,-1$ in the four-chain. As all of
the following matrix elements vanish, 
\begin{eqnarray}
Q_{\uparrow \uparrow \uparrow \uparrow }^{\uparrow \downarrow \downarrow
\downarrow } &=&Q_{\uparrow \uparrow \uparrow \uparrow }^{\downarrow
\uparrow \downarrow \downarrow }=Q_{\uparrow \uparrow \uparrow \uparrow
}^{\downarrow \downarrow \uparrow \downarrow }=Q_{\uparrow \uparrow \uparrow
\uparrow }^{\downarrow \downarrow \downarrow \uparrow }=\limfunc{tr}%
AC^{3}=0,\; \\
Q_{\uparrow \downarrow \downarrow \downarrow }^{\uparrow \uparrow \uparrow
\uparrow } &=&Q_{\downarrow \uparrow \downarrow \downarrow }^{\uparrow
\uparrow \uparrow \uparrow }=Q_{\downarrow \downarrow \uparrow \downarrow
}^{\uparrow \uparrow \uparrow \uparrow }=Q_{\downarrow \downarrow \downarrow
\uparrow }^{\uparrow \uparrow \uparrow \uparrow }=\limfunc{tr}AB^{3}=0,
\end{eqnarray}
the remaining sectors are either trivial ($S^{z}=\pm 2$) or related by
spin-reversal ($S^{z}=1$). Let us start with the spin-sector of smaller
dimension, i.e. $S^{z}=-1$.

\subsubsection{$S^{z}=-1$}

The auxiliary matrix in this sector is computed to be 
\begin{equation}
Q|_{S^{z}=-1}=\left( 
\begin{array}{cccc}
\limfunc{tr}AD^{3} & \limfunc{tr}CBD^{2} & \limfunc{tr}BDCD & \limfunc{tr}%
BCD^{2} \\ 
\limfunc{tr}BCD^{2} & \limfunc{tr}AD^{3} & \limfunc{tr}CBD^{2} & \limfunc{tr}%
BDCD \\ 
\limfunc{tr}BDCD & \limfunc{tr}BCD^{2} & \limfunc{tr}AD^{3} & \limfunc{tr}%
CBD^{2} \\ 
\limfunc{tr}CBD^{2} & \limfunc{tr}BDCD & \limfunc{tr}BCD^{2} & \limfunc{tr}%
AD^{3}
\end{array}
\right)
\end{equation}
where the basis vectors in the spin-sector have been chosen such that the
first column vector is given by 
\begin{equation}
Q_{\uparrow \downarrow \downarrow \downarrow }^{\uparrow \downarrow
\downarrow \downarrow }=\limfunc{tr}AD^{3},\;Q_{\uparrow \downarrow
\downarrow \downarrow }^{\downarrow \uparrow \downarrow \downarrow }=%
\limfunc{tr}BCD^{2},\;Q_{\uparrow \downarrow \downarrow \downarrow
}^{\downarrow \downarrow \uparrow \downarrow }=\limfunc{tr}%
BDCD,\;Q_{\uparrow \downarrow \downarrow \downarrow }^{\downarrow \downarrow
\downarrow \uparrow }=\limfunc{tr}CBD^{2}\;.
\end{equation}
Choosing as before the conventions 
\begin{equation*}
\rho _{\pm }=(wq)^{\frac{1\pm 1}{2}},\;\phi _{1}=bq^{\frac{N-1}{2}},\;\phi
_{2}=aq^{\frac{1-N}{2}},\;\mathbf{c}=\mu +\mu ^{-1},\;w=z/\mu
\end{equation*}
one finds after some algebra the expressions 
\begin{eqnarray}
Q_{\uparrow \downarrow \downarrow \downarrow }^{\uparrow \downarrow
\downarrow \downarrow } &=&\limfunc{tr}AD^{3}=-9w\mathbf{z}^{2}(w^{2}+q), 
\notag \\
Q_{\uparrow \downarrow \downarrow \downarrow }^{\downarrow \uparrow
\downarrow \downarrow } &=&\limfunc{tr}BCD^{2}=-3w\mathbf{z}%
^{2}(1+qw^{2}+2wq^{2}\mathbf{c}), \\
Q_{\uparrow \downarrow \downarrow \downarrow }^{\downarrow \downarrow
\uparrow \downarrow } &=&\limfunc{tr}BDCD=-3w\mathbf{z}^{2}(w^{2}+q-wq^{2}%
\mathbf{c}),  \notag \\
Q_{\uparrow \downarrow \downarrow \downarrow }^{\downarrow \downarrow
\downarrow \uparrow } &=&\limfunc{tr}CBD^{2}=-3w\mathbf{z}%
^{2}(q^{2}+q^{2}w^{2}+2wq^{2}\mathbf{c})\;.
\end{eqnarray}
Diagonalizing the above matrix then yields the eigenvalues 
\begin{eqnarray}
Q_{1} &=&Q_{\uparrow \downarrow \downarrow \downarrow }^{\uparrow \downarrow
\downarrow \downarrow }+Q_{\uparrow \downarrow \downarrow \downarrow
}^{\downarrow \uparrow \downarrow \downarrow }+Q_{\uparrow \downarrow
\downarrow \downarrow }^{\downarrow \downarrow \uparrow \downarrow
}+Q_{\uparrow \downarrow \downarrow \downarrow }^{\downarrow \downarrow
\downarrow \uparrow }  \notag \\
&=&-9w\mathbf{z}^{2}(w^{2}+wq^{2}\mathbf{c}+q)=-9\mathbf{z}%
^{2}\,w(w+q^{2}\mu ^{-1})(w+q^{2}\mu ),  \notag \\
Q_{2} &=&Q_{\uparrow \downarrow \downarrow \downarrow }^{\uparrow \downarrow
\downarrow \downarrow }+Q_{\uparrow \downarrow \downarrow \downarrow
}^{\downarrow \downarrow \uparrow \downarrow }-Q_{\uparrow \downarrow
\downarrow \downarrow }^{\downarrow \uparrow \downarrow \downarrow
}-Q_{\uparrow \downarrow \downarrow \downarrow }^{\downarrow \downarrow
\downarrow \uparrow }  \notag \\
&=&-15\mathbf{z}^{2}w(w^{2}-wq^{2}\mathbf{c}+q)=-15\mathbf{z}%
^{2}w(w-q^{2}\mu ^{-1})(w-q^{2}\mu ),  \notag \\
Q_{3,4} &=&Q_{\uparrow \downarrow \downarrow \downarrow }^{\uparrow
\downarrow \downarrow \downarrow }-Q_{\uparrow \downarrow \downarrow
\downarrow }^{\downarrow \downarrow \uparrow \downarrow }\pm i(Q_{\uparrow
\downarrow \downarrow \downarrow }^{\downarrow \uparrow \downarrow
\downarrow }-Q_{\uparrow \downarrow \downarrow \downarrow }^{\downarrow
\downarrow \downarrow \uparrow })  \label{q1} \\
&=&-3\mathbf{z}^{2}w\{w^{2}(2\mp \sqrt{3})+wq^{2}\mathbf{c}+q(2\pm \sqrt{3}%
)\}\;.  \notag
\end{eqnarray}
In order to verify the functional equation (\ref{TQ0}) we also need to
compute the corresponding eigenvalues of the transfer matrix (\ref{T}). The
relevant matrix elements are 
\begin{equation}
T_{\uparrow \downarrow \downarrow \downarrow }^{\uparrow \downarrow
\downarrow \downarrow }=a^{3}b+ab^{3},\;T_{\uparrow \downarrow \downarrow
\downarrow }^{\downarrow \uparrow \downarrow \downarrow }=b^{2}cc^{\prime
},\;T_{\uparrow \downarrow \downarrow \downarrow }^{\downarrow \downarrow
\uparrow \downarrow }=abcc^{\prime },\;T_{\uparrow \downarrow \downarrow
\downarrow }^{\downarrow \downarrow \downarrow \uparrow }=a^{2}cc^{\prime }
\end{equation}
from which one calculates the eigenvalues 
\begin{eqnarray}
T_{1} &=&a^{3}b+ab^{3}+(a^{2}+ab+b^{2})cc^{\prime
}=b(b^{2}+1)+(b^{2}+b+1)cc^{\prime },  \notag \\
T_{2} &=&a^{3}b+ab^{3}-(a^{2}-ab+b^{2})cc^{\prime
}=b(b^{2}+1)-(b^{2}-b+1)cc^{\prime },  \notag \\
T_{3,4} &=&a^{3}b+ab^{3}\mp (ia^{2}\pm ab-ib^{2})cc^{\prime }=b(b^{2}+1)\pm
i(b^{2}\pm ib-1)cc^{\prime }\;.  \label{T1}
\end{eqnarray}
One now verifies for this example that the functional equation 
\begin{equation*}
Q(z)T(z)=\phi _{1}(z)^{4}Q^{\prime }(zq^{2})+\phi _{2}(z)^{4}Q^{\prime
\prime }(zq^{-2})
\end{equation*}
with $\phi _{1}=bq,\;\phi _{2}=aq^{-1},\;\mu ^{\prime }=q\mu ,\;\mu ^{\prime
\prime }=\mu q^{-1}$ is valid. Let us do this explicitly for the first
eigenvalues $Q_{1},T_{1}$. Using (\ref{q1}) we write down the functional
relation 
\begin{eqnarray}
z(z+q^{2})(z+q^{2}\mu ^{2})T_{1}(z) &=&\phi
_{1}(z)^{4}\,zq^{2}(zq^{2}+q^{2})(zq^{2}+q^{4}\mu ^{2})  \notag \\
&&+\phi _{2}(z)^{4}\,zq^{-2}(zq^{-2}+q^{2})(zq^{-2}+\mu ^{2})  \label{tqex1}
\end{eqnarray}
which yields in accordance with (\ref{T1}) the eigenvalue 
\begin{equation*}
T_{1}(z)=b(z)^{4}q\,\frac{z+1}{z+q^{2}}+a(z)^{4}q\,\frac{z+q}{z+q^{2}}\;.
\end{equation*}
Note that the factors in (\ref{tqex1}) which contain zeroes depending on $%
\mu $ cancel on both sides of the equation. We are left with one Bethe root $%
z^{B}=-q^{2}$ which upon the identification $z=e^{u}q^{-1},q=e^{i\gamma }$
is seen to trivially fulfill the Bethe ansatz equations (\ref{BE}), 
\begin{equation*}
\left( \frac{a(z^{B})}{b(z^{B})}\right) ^{4}=\left( \frac{1-z^{B}q^{2}}{%
q-z^{B}q}\right) ^{4}=1\;.
\end{equation*}
The remaining eigenvalues work out in a similar manner. We now turn to the
spin-sector $S^{z}=0$ where we can compare our expressions for the auxiliary
matrices with the one found by Baxter, see (\ref{BQ}).

\subsubsection{$S^{z}=0$}

The spin-sector is six-dimensional and the non-vanishing matrix elements of (%
\ref{Q0}) are computed to 
\begin{eqnarray}
m_{1} &=&Q_{\downarrow \downarrow \uparrow \uparrow }^{\downarrow \downarrow
\uparrow \uparrow }=Q_{\downarrow \uparrow \downarrow \uparrow }^{\downarrow
\uparrow \downarrow \uparrow }=Q_{\downarrow \uparrow \uparrow \downarrow
}^{\downarrow \uparrow \uparrow \downarrow }=Q_{\uparrow \downarrow
\downarrow \uparrow }^{\uparrow \downarrow \downarrow \uparrow }=Q_{\uparrow
\downarrow \uparrow \downarrow }^{\uparrow \downarrow \uparrow \downarrow
}=Q_{\uparrow \uparrow \downarrow \downarrow }^{\uparrow \uparrow \downarrow
\downarrow }=\limfunc{tr}A^{2}D^{2}  \notag \\
&=&3\mathbf{z}^{2}(qw^{4}+4q^{2}w^{2}+1),  \notag \\
m_{2} &=&Q_{\downarrow \downarrow \uparrow \uparrow }^{\downarrow \uparrow
\downarrow \uparrow }=Q_{\downarrow \uparrow \downarrow \uparrow
}^{\downarrow \uparrow \uparrow \downarrow }=Q_{\downarrow \uparrow
\downarrow \uparrow }^{\uparrow \downarrow \downarrow \uparrow
}=Q_{\downarrow \uparrow \uparrow \downarrow }^{\uparrow \downarrow \uparrow
\downarrow }=Q_{\uparrow \uparrow \downarrow \downarrow }^{\downarrow
\uparrow \downarrow \uparrow }=Q_{\uparrow \downarrow \downarrow \uparrow
}^{\uparrow \downarrow \uparrow \downarrow }=Q_{\uparrow \downarrow \uparrow
\downarrow }^{\uparrow \uparrow \downarrow \downarrow }=Q_{\uparrow
\downarrow \uparrow \downarrow }^{\downarrow \downarrow \uparrow \uparrow }=%
\limfunc{tr}ADCB  \notag \\
&=&3w\mathbf{z}^{2}q^{2}((q+w^{2})q\mathbf{c}-w)=3\mathbf{z}^{2}(\mathbf{c}%
\frak{\,}w^{3}-q^{2}w^{2}+\mathbf{c}q\,w),  \notag \\
m_{3} &=&Q_{\downarrow \downarrow \uparrow \uparrow }^{\downarrow \uparrow
\uparrow \downarrow }=Q_{\downarrow \uparrow \uparrow \downarrow }^{\uparrow
\uparrow \downarrow \downarrow }=Q_{\uparrow \uparrow \downarrow \downarrow
}^{\uparrow \downarrow \downarrow \uparrow }=Q_{\uparrow \downarrow
\downarrow \uparrow }^{\downarrow \downarrow \uparrow \uparrow }=\limfunc{tr}%
ABDC  \notag \\
&=&3\mathbf{z}^{2}q^{2}w((1+w^{2}q)q\mathbf{c}+2w)=3\mathbf{z}^{2}(\mathbf{c}%
q\frak{\,}w^{3}+2q^{2}w^{2}+\mathbf{c}\,w),  \notag \\
m_{4} &=&Q_{\downarrow \downarrow \uparrow \uparrow }^{\uparrow \downarrow
\downarrow \uparrow }=Q_{\downarrow \uparrow \uparrow \downarrow
}^{\downarrow \downarrow \uparrow \uparrow }=Q_{\uparrow \uparrow \downarrow
\downarrow }^{\downarrow \uparrow \uparrow \downarrow }=Q_{\uparrow
\downarrow \downarrow \uparrow }^{\uparrow \uparrow \downarrow \downarrow }=%
\limfunc{tr}ACDB  \notag \\
&=&3\mathbf{z}^{2}q^{2}w((1+w^{2})\mathbf{c}+2w)=3\mathbf{z}^{2}(\mathbf{c}%
q^{2}\frak{\,}w^{3}+2q^{2}w^{2}+\mathbf{c}q^{2}\,w),  \notag \\
m_{5} &=&Q_{\downarrow \downarrow \uparrow \uparrow }^{\uparrow \downarrow
\uparrow \downarrow }=Q_{\downarrow \uparrow \downarrow \uparrow
}^{\downarrow \downarrow \uparrow \uparrow }=Q_{\downarrow \uparrow
\downarrow \uparrow }^{\uparrow \uparrow \downarrow \downarrow
}=Q_{\downarrow \uparrow \uparrow \downarrow }^{\downarrow \uparrow
\downarrow \uparrow }=Q_{\uparrow \uparrow \downarrow \downarrow }^{\uparrow
\downarrow \uparrow \downarrow }=Q_{\uparrow \downarrow \uparrow \downarrow
}^{\uparrow \downarrow \downarrow \uparrow }=Q_{\uparrow \downarrow \uparrow
\downarrow }^{\downarrow \uparrow \uparrow \downarrow }=Q_{\uparrow
\downarrow \downarrow \uparrow }^{\downarrow \uparrow \downarrow \uparrow }=%
\limfunc{tr}ADBC  \notag \\
&=&m_{2}  \notag \\
m_{6} &=&Q_{\downarrow \downarrow \uparrow \uparrow }^{\uparrow \uparrow
\downarrow \downarrow }=Q_{\downarrow \uparrow \uparrow \downarrow
}^{\uparrow \downarrow \downarrow \uparrow }=Q_{\uparrow \uparrow \downarrow
\downarrow }^{\downarrow \downarrow \uparrow \uparrow }=Q_{\uparrow
\downarrow \downarrow \uparrow }^{\downarrow \uparrow \uparrow \downarrow }=%
\limfunc{tr}B^{2}C^{2}  \notag \\
&=&3\mathbf{z}^{2}q^{2}w^{2}(\mathbf{c}^{2}-2-q-q^{-1}),  \notag \\
m_{7} &=&Q_{\downarrow \uparrow \downarrow \uparrow }^{\uparrow \downarrow
\uparrow \downarrow }=Q_{\uparrow \downarrow \uparrow \downarrow
}^{\downarrow \uparrow \downarrow \uparrow }=\limfunc{tr}BCBC  \notag \\
&=&3\mathbf{z}^{2}q^{2}w^{2}(\mathbf{c}^{2}+2)\;.
\end{eqnarray}
Under the appropriate choice of basis the auxiliary matrix is 
\begin{equation*}
Q|_{S^{z}=0}=\left( 
\begin{array}{cccccc}
m_{1} & m_{2} & m_{3} & m_{4} & m_{5} & m_{6} \\ 
m_{5} & m_{1} & m_{2} & m_{2} & m_{7} & m_{5} \\ 
m_{4} & m_{5} & m_{1} & m_{6} & m_{2} & m_{3} \\ 
m_{3} & m_{5} & m_{6} & m_{1} & m_{2} & m_{4} \\ 
m_{2} & m_{7} & m_{5} & m_{5} & m_{1} & m_{2} \\ 
m_{6} & m_{2} & m_{4} & m_{3} & m_{5} & m_{1}
\end{array}
\right)
\end{equation*}
and has the six eigenvalues 
\begin{eqnarray}
Q_{1} &=&m_{1}-m_{7}=3\mathbf{z}^{2}\{q\,w^{4}-(\mathbf{c}%
^{2}-2)q^{2}\,w^{2}+1\}=3\mathbf{z}^{2}q(w^{2}-q\mu ^{2})(w^{2}-q\mu ^{-2}) 
\notag \\
Q_{2} &=&m_{1}+m_{6}-m_{3}-m_{4}  \notag \\
&=&3\mathbf{z}^{2}\{q\,w^{4}-q\mathbf{c}(1+q)\,w^{3}+(\mathbf{c}%
^{2}-2-q-q^{2})q^{2}\,w^{2}-\mathbf{c}(1+q^{2})\,w+1\}  \notag \\
Q_{3,4} &=&m_{1}-m_{6}\pm i(m_{3}-m_{4})  \notag \\
&=&3\mathbf{z}^{2}\{q\,w^{4}\pm iq\mathbf{c}(1-q)\,w^{3}+(6-\mathbf{c}%
^{2}+q+q^{2})q^{2}\,w^{2}\pm i\mathbf{c}(1-q^{2})\,w+1\}  \notag \\
Q_{5,6} &=&\tfrac{1}{2}\left( 2m_{1}+m_{3}+m_{4}+m_{6}+m_{7}\pm \sqrt{%
32m_{2}^{2}+(m_{3}+m_{4}+m_{6}-m_{7})^{2}}\right)  \label{q40}
\end{eqnarray}
Again we see that the matrix elements as well as the eigenvalues of the
auxiliary matrix only depend on the central elements of the quantum group.

In comparison we obtain from Baxter's formula (\ref{BQ}) the matrix elements 
\begin{equation*}
m_{1}^{Bax}=1/m_{7}^{Bax}=1/m_{6}^{Bax}=zq,\;m_{2}^{Bax}=m_{5}^{Bax}=1,%
\;m_{3}^{Bax}=1/m_{4}^{Bax}=q^{-1}\;.
\end{equation*}
In order to match the different conventions in the choice of Boltzmann
weights one has to multiply by the additional normalization factor $\tilde{%
\rho}=(zq)^{\frac{M}{4}}$ , see \cite{RW02}. One then finds the eigenvalues 
\begin{eqnarray}
Q_{1}^{Bax} &=&\tilde{\rho}(z q-1/(zq))=z^2 q^2-1,\;  \notag \\
Q_{2}^{Bax} &=&\tilde{\rho}\left( zq+z^{-1}q^{2}-q^{2}-q\right)
=q^{2}(z^{2}-(1+q)z+q),  \notag \\
Q_{3,4}^{Bax} &=&\tilde{\rho}\left( zq-z^{-1}q^{2}\pm i(q^{2}-q)\right)
=q^{2}(z^{2}\mp i(1-q)z-q),  \notag \\
Q_{5,6}^{Bax} &=&\tfrac{1}{2}\left( 2z^2 q^2+(1+q^2)z+2\pm z 
\sqrt{32 q^2+(1+q^2)^{2}}\right) \;.  \label{qe40B}
\end{eqnarray}
In Section 5 we derived from (\ref{Q0})
solutions to Baxter's functional equation (\ref{BTQ}) by summing over all
the points in one fiber of the hypersurface, cf (\ref{Qfiber}). Let us check
for $s=0$ whether those solutions match (\ref{BQ}) up to a possible
normalization factor. As all the auxiliary matrices in the same fiber
commute with each other, we can simply sum up the eigenvalues. Let us do
this for the first four eigenvalues, 
\begin{eqnarray}
Q_{1}^{(s)} &=&\sum_{\ell =0}^{2}Q_{1,\ell }=3\mathbf{z}^{2}q(z^{2}-q)\sum_{%
\ell =0}^{2}q^{-\ell s}\{q^{-\ell }\mu ^{-4}z^{2}-q\}\underset{s=0}{=}-9%
\mathbf{z}^{2}q^{2}(z^{2}-q)  \notag \\
Q_{2}^{(s=0)} &=&\sum_{\ell =0}^{2}Q_{2,\ell }=9\mathbf{z}%
^{2}q^{2}\{z^{2}-(1+q)z+q\},\quad  \notag \\
Q_{3,4}^{(s=0)} &=&\sum_{\ell =0}^{2}Q_{3,4,\ell }=-9\mathbf{z}%
^{2}q^{2}\{z^{2}\pm i(1-q)z-q\}\;.
\end{eqnarray}
For the fifth and sixth eigenvalue it has been checked numerically for
several values that the eigenvalues $Q_{5,6}^{(s=0)}$ are nonzero as well.
Thus, in this case the summed expression (\ref{Qfiber}) turns out to be proportional to the expression (\ref{BQ}), whence we obtain the same Bethe
roots and eigenvalues for the transfer matrix.

As before we can now check that the eigenvalues of the transfer matrix are
correctly obtained from the functional relation. Let us perform this
consistency check for the first eigenvalue in (\ref{q40}). The non-vanishing
matrix elements of the transfer matrix are 
\begin{eqnarray}
T_{\downarrow \downarrow \uparrow \uparrow }^{\downarrow \downarrow \uparrow
\uparrow } &=&T_{\downarrow \uparrow \downarrow \uparrow }^{\downarrow
\uparrow \downarrow \uparrow }=T_{\downarrow \uparrow \uparrow \downarrow
}^{\downarrow \uparrow \uparrow \downarrow }=T_{\uparrow \downarrow
\downarrow \uparrow }^{\uparrow \downarrow \downarrow \uparrow }=T_{\uparrow
\downarrow \uparrow \downarrow }^{\uparrow \downarrow \uparrow \downarrow
}=T_{\uparrow \uparrow \downarrow \downarrow }^{\uparrow \uparrow \downarrow
\downarrow }=2a^{2}b^{2},  \notag \\
T_{\downarrow \downarrow \uparrow \uparrow }^{\downarrow \uparrow \downarrow
\uparrow } &=&T_{\downarrow \uparrow \downarrow \uparrow }^{\downarrow
\uparrow \uparrow \downarrow }=T_{\downarrow \uparrow \downarrow \uparrow
}^{\uparrow \downarrow \downarrow \uparrow }=T_{\downarrow \uparrow \uparrow
\downarrow }^{\uparrow \downarrow \uparrow \downarrow }=T_{\uparrow \uparrow
\downarrow \downarrow }^{\downarrow \uparrow \downarrow \uparrow
}=T_{\uparrow \downarrow \downarrow \uparrow }^{\uparrow \downarrow \uparrow
\downarrow }=T_{\uparrow \downarrow \uparrow \downarrow }^{\uparrow \uparrow
\downarrow \downarrow }=T_{\uparrow \downarrow \uparrow \downarrow
}^{\downarrow \downarrow \uparrow \uparrow }=abcc^{\prime }  \notag \\
T_{\downarrow \downarrow \uparrow \uparrow }^{\downarrow \uparrow \uparrow
\downarrow } &=&T_{\downarrow \uparrow \uparrow \downarrow }^{\uparrow
\uparrow \downarrow \downarrow }=T_{\uparrow \uparrow \downarrow \downarrow
}^{\uparrow \downarrow \downarrow \uparrow }=T_{\uparrow \downarrow
\downarrow \uparrow }^{\downarrow \downarrow \uparrow \uparrow
}=a^{2}cc^{\prime }  \notag \\
T_{\downarrow \downarrow \uparrow \uparrow }^{\uparrow \downarrow \downarrow
\uparrow } &=&T_{\downarrow \uparrow \uparrow \downarrow }^{\downarrow
\downarrow \uparrow \uparrow }=T_{\uparrow \uparrow \downarrow \downarrow
}^{\downarrow \uparrow \uparrow \downarrow }=T_{\uparrow \downarrow
\downarrow \uparrow }^{\uparrow \uparrow \downarrow \downarrow
}=b^{2}cc^{\prime }  \notag \\
T_{\downarrow \downarrow \uparrow \uparrow }^{\uparrow \downarrow \uparrow
\downarrow } &=&T_{\downarrow \uparrow \downarrow \uparrow }^{\downarrow
\downarrow \uparrow \uparrow }=T_{\downarrow \uparrow \downarrow \uparrow
}^{\uparrow \uparrow \downarrow \downarrow }=T_{\downarrow \uparrow \uparrow
\downarrow }^{\downarrow \uparrow \downarrow \uparrow }=T_{\uparrow \uparrow
\downarrow \downarrow }^{\uparrow \downarrow \uparrow \downarrow
}=T_{\uparrow \downarrow \uparrow \downarrow }^{\uparrow \downarrow
\downarrow \uparrow }=T_{\uparrow \downarrow \uparrow \downarrow
}^{\downarrow \uparrow \uparrow \downarrow }=T_{\uparrow \downarrow
\downarrow \uparrow }^{\downarrow \uparrow \downarrow \uparrow
}=abcc^{\prime }  \notag \\
T_{\downarrow \uparrow \downarrow \uparrow }^{\uparrow \downarrow \uparrow
\downarrow } &=&T_{\uparrow \downarrow \uparrow \downarrow }^{\downarrow
\uparrow \downarrow \uparrow }=(cc^{\prime })^{2}\;.
\end{eqnarray}
From these identities one calculates the eigenvalues 
\begin{eqnarray}
T_{1} &=&2(ab)^{2}-(cc^{\prime })^{2}  \notag \\
T_{2} &=&2(ab)^{2}-(a^{2}+b^{2})cc^{\prime }  \notag \\
T_{3,4} &=&2(ab)^{2}\pm i(a^{2}-b^{2})cc^{\prime }  \notag \\
T_{5,6} &=&\tfrac{1}{2}\left( 4(ab)^{2}+cc^{\prime }(a^{2}+b^{2}+cc^{\prime
})\pm cc^{\prime }\sqrt{32(ab)^{2}+(a^{2}+b^{2}-cc^{\prime })^{2}}\right) \;.
\end{eqnarray}
The functional relation 
\begin{equation*}
Q(z)T(z)=\phi _{1}(z)^{4}Q^{\prime }(zq^{2})+\phi _{2}(z)^{4}Q^{\prime
\prime }(zq^{-2})
\end{equation*}
with $\phi _{1}=bq,\;\phi _{2}=aq^{-1},\;\mu ^{\prime }=q\mu ,\;\mu ^{\prime
\prime }=\mu q^{-1}$ implies for the eigenvalues $T_{1},Q_{1}$ the identity
(setting $\rho =1$ in (\ref{h})), 
\begin{equation*}
T_{1}(z)=2b(z)^{2}-c(z)^{2}c^{\prime }(z)^{2}=\phi _{1}(z)^{4}\frac{z^{2}q-q%
}{z^{2}-q}+\phi _{2}(z)^{4}\frac{z^{2}q^{2}-q}{z^{2}-q}\;.
\end{equation*}
After a short calculation the above equation is shown to be true. The
corresponding Bethe roots of the eigenvalue are easily deduced from (\ref
{q40}) to be $z_{\pm }^{B}=\pm q^{2}$ which are easily seen to solve the
Bethe ansatz equations (\ref{BE}) when setting $z=e^{u}q^{-1},\;q=e^{i\gamma
}$, 
\begin{equation*}
\left( \frac{a(z_{\pm })}{b(z_{\pm })}\right) ^{4}=\left( \frac{1-z_{\pm
}q^{2}}{q-z_{\pm }q}\right) ^{4}=\frac{z_{\mp }/z_{\pm }-q^{2}}{z_{\mp
}/z_{\pm }q^{2}-1}=1\;.
\end{equation*}
For the remaining eigenvalues the functional relation has been checked
numerically.

\section{Conclusions}

Starting from evaluation representations of the quantum loop algebra $U_{q}(%
\widetilde{sl}_{2})$ families of auxiliary matrices for the six-vertex model
at roots of unity have been explicitly constructed. See the definitions (\ref
{Q0}), (\ref{Lsol}), (\ref{Lodd}) and apply the representation (\ref{basis}%
). For odd roots of unity the auxiliary matrices depend on three, for even
roots of unity on one additional parameter besides the spectral variable $z$
and the deformation parameter $q$. In comparison to earlier results in the
literature the auxiliary matrices (\ref{Q0}) have several advantages. They
extend to all spin-sectors and do not contain formal power series since the
auxiliary space can be kept finite-dimensional at roots of unity. They have
been demonstrated to be of the simple form (\ref{Q}) and to satisfy the
functional equation (\ref{TQ0}) which can be interpreted in terms of
representation theory, cf (\ref{seq}). All operators in this functional
equation have been shown to commute with each other (cf (\ref{Qcom})),
whence the eigenvalues of the transfer matrix (\ref{T}) and the Bethe ansatz
equations (\ref{BE}) can be derived. In the present article this has been
done for the two simple examples $N=3,M=3,4$. In a forthcoming paper \cite
{KP} the eigenvalues of the constructed auxiliary matrices will be
investigated for general $N,\;M$.

\subsection{The geometric interpretation}

Applying the concept of evaluation representations of $U_{q}(\widetilde{sl}%
_{2})$ allowed for a simple classification of the auxiliary matrices and a
geometric interpretation of their parameters. Regardless which root-of-unity
representation is used to write down solutions for the $L$-operator (\ref
{Lodd}) the final auxiliary matrix (\ref{Q0}) only depends on the values of
the central elements of the quantum algebra. Employing the results of \cite
{CK,CKP} these values were shown to specify points on a three-dimensional
complex hypersurface $\limfunc{Spec}Z$ defined in (\ref{SZ}).

In fact, the presented construction of auxiliary matrices can be interpreted
as the definition of the following map from the direct product of the
complex numbers with $\limfunc{Spec}Z$ into the operator space over the
spin-chain, 
\begin{equation}
Q:\mathbb{C}\times \left( \text{$\limfunc{Spec}$\thinspace }Z\backslash
D\right) \rightarrow \text{End}(\pi _{1}^{0\,\otimes M}),\quad p=(\mathbf{x},%
\mathbf{y},\mathbf{z},\mathbf{c}=\mu +\mu ^{-1})\rightarrow Q_{p}(z)
\label{map}
\end{equation}
with $Q_{p}(z)$ given by (\ref{Q0}). The hypersurface $\limfunc{Spec}$%
\thinspace $Z\backslash D$ is an $N$-fold fibration and under multiplication
with the transfer matrix the auxiliary matrix is shifted to the neighbouring
points 
\begin{equation*}
p^{\prime }=(\mathbf{x},\mathbf{y},\mathbf{z},\mathbf{c}^{\prime }=\mu
q+q^{-1}\mu ^{-1})\quad \text{and\quad }p^{\prime \prime }=(\mathbf{x},%
\mathbf{y},\mathbf{z},\mathbf{c}^{\prime \prime }=\mu q^{-1}+q\mu ^{-1})
\end{equation*}
in the same fiber. If we would have allowed for points $p$ in the
discriminant set $D$, see (\ref{D}), the result would have been the known
functional relations between transfer matrices of higher spin $1\leq 2s\leq
N-1$, see e.g. \cite{KNS94}. The constructed map (\ref{map}) allows us to
carry important mathematical structures of the hypersurface $\limfunc{Spec}Z$
over to the family of auxiliary matrices $\{Q_{p}(z)\}_{p\in \limfunc{Spec}%
Z} $.

\subsubsection{The quantum coadjoint action on auxiliary matrices}

The quantum coadjoint action given by the automorphisms (\ref{G}) on the
hypersurface (\ref{SZ}) can be extended in a natural manner to the family of
auxiliary matrices (\ref{Q0}) setting (for $N$ odd), 
\begin{equation}
G\times \limfunc{Spec}Z\ni (g,p)\rightarrow g\cdot Q_{p}(z):=Q_{gp}(z)\;.
\label{GQ}
\end{equation}
As the points $p,p^{\prime },p^{\prime \prime }$ in (\ref{TQ0}) belong to
the same fiber over $\limfunc{Spec}Z_{0}$ and the Casimir element remains
invariant under the group action, the functional equation is preserved. The
definition (\ref{GQ}) is a manifestation of the infinite-dimensional
symmetry of the six-vertex model at roots of unity. Since the six-vertex
transfer matrix does not depend on the point $p$ in the hypersurface one
might choose any element in the family of auxiliary matrices to solve the
eigenvalue problem of (\ref{T}) respectively (\ref{H}). Therefore, one is
lead to look for transformations in the parameters $p$ which leave the set
of auxiliary matrices invariant. These transformations are precisely given
by the infinite-dimensional group $G$. Another way of expressing the
symmetry is by observing that for any group element $g\in G$ one has 
\begin{equation}
\left[ T(z),Q_{gp}(w)\right] =0,\quad \forall g\in G,\;z,w\in \mathbb{C}\;.
\end{equation}
Since the auxiliary matrices belonging to different fibers in the
hypersurface do not commute in general, the vanishing of the above
commutator exhibits infinitely many non-abelian conserved quantities defined
for all spin-sectors.

Mathematically the quantum coadjoint action is interesting since it allows
to explore the structure of the base space $\limfunc{Spec}Z_{0}$ which can
be given the structure of a Poisson-Lie group \cite{CK,CKP}. In order to
connect these mathematical structures with the eigenvalue problem of the
transfer matrix further investigations are needed. In particular, it would
be helpful to have a direct implementation of the group action (\ref{GQ}) on
the spin-chain. This would allow one to obtain further insight how the
earlier observed loop symmetry in the commensurate sectors $S^{z}=0\func{mod}%
N$ \cite{DFM,KM01,KR02} extends to the non-commensurate ones. It is an
intriguing observation that the generators of the quantum coadjoint action
are closely related with the restricted quantum group expressing the loop
algebra symmetry. This aspect will be subject to future investigations.

\subsection{Broken symmetries}

Besides the infinite-dimensional symmetry, the auxiliary matrices also break
the finite symmetries (\ref{R&S}) and (\ref{Sz}) of the six-vertex model.
(Note that all three of these symmetries have also been implemented as
mappings on the hypersurface $\limfunc{Spec}Z$.) This shows that the present
setting is more general than the coordinate space Bethe ansatz.
Spin-conservation, which is essential for the application of the coordinate
space Bethe ansatz, can be restored by taking the nilpotent limit. That is,
for the following subvariety of auxiliary matrices one has 
\begin{equation}
\lbrack Q_{p_{\mu }}(z),S^{z}]=0,\quad p_{\mu }=(0,0,\mu ^{N^{\prime }},\mu
+\mu ^{-1}),\;\mu \in \mathbb{C}^{\times }\;.
\end{equation}
However, the action (\ref{GQ}) of the infinite-dimensional automorphism
group $G$ on this one-parameter family forces one to consider also cyclic
representations which do not preserve the total spin and violate the
conditions (\ref{BQC1}) and (\ref{BQC2}). This explicitly shows the
statement made in the introduction that the full symmetry present at odd
roots of unity becomes manifest when the coordinate space Bethe ansatz
ceases to be applicable.\smallskip

For even roots of unity cyclic representations had to be excluded since an
intertwiner does not exist. Nevertheless, the hypersurface, the
decomposition of the tensor product via the exact sequence (\ref{seq}) and
the quantum coadjoint action are equally well defined. It would be
interesting to find the construction for cyclic representations also in this
case. The functional relation (\ref{TQS}) derived for even roots of unity
with $N^{\prime }$ odd might serve as a starting point. Another way to
proceed is to exploit the quantum coadjoint action (\ref{GQ}) for even roots
of unity. Again this subject is left to future investigations.\smallskip

Within the framework of the coordinate space Bethe ansatz spin-reversal
symmetry is broken in the sectors $S^{z}\neq 0$. The auxiliary matrices
constructed in this work also break spin-reversal symmetry, cf (\ref{QR}).
For example one has for the above one-parameter family $Q_{\mu }\equiv
Q_{p_{\mu }}$ the transformation, 
\begin{equation}
\frak{R\,}Q_{\mu }(z)\frak{R}=Q_{\mu ^{-1}}(z\mu ^{-2})\,.
\end{equation}
From this transformation law one infers that spin-reversal symmetry is
restored when $\mu \rightarrow 1$. In the construction presented here this
amounts to choosing a reducible representation in the definition of (\ref{L}%
). This transformation behaviour which has been derived from the algebra
automorphism (\ref{om}) is different from previous constructed auxiliary
matrices for the six-vertex model. Baxter's expression (\ref{BQ}) only
applies to the spin-zero sectors where it does not break spin-reversal
invariance in accordance with the Bethe ansatz. The auxiliary matrix
considered in \cite{RW02} breaks spin-reversal symmetry outside the sectors $%
S^{z}=0$ due to an additional factor $s_{0}^{S^{z}},s_{0}\in \mathbb{C}$ in
the ``constant'' (\ref{N8}).

\subsection{Outlook}

The difference between the two construction procedures for auxiliary
matrices in the literature needs to be further clarified. We explicitly
verified for the $S^{z}=0$ sector of the four-chain that the auxiliary
matrices constructed here are different form Baxter's expression (\ref{BQ}).

For the auxiliary matrices (\ref{Q0}) this is to be expected as Baxter's
functional equation (\ref{BTQ}) is formulated in terms of a single auxiliary
matrix, while (\ref{TQ0}) obtained from representation theory involves three
different ones.

This discrepancy between the functional equations is removed when summing
the auxiliary matrices (\ref{Q0}) over the points in a single fiber. (In the
case of even roots of unity one has to sum over two fibers). One then
obtains the solutions (\ref{Qfiber}) to Baxter's functional equation (\ref
{BTQ}) which are defined on the base manifold $\limfunc{Spec}Z_{0}$, cf (\ref
{Zseq}).

The solutions (\ref{Qfiber}) might be singular matrices in general, although
for the spin-zero sector of the four-chain this was not the case. Singular
matrices were also encountered for the auxiliary matrix (\ref
{Q0}) in the case of the three-chain. Whether these singularities persist
for longer spin-chains needs to be numerically investigated. This is of
particular importance in order to clarify whether all eigenvalues of the
transfer matrix can be obtained using the method of auxiliary matrices. It
would also shed further light on the differences between the two different
construction methods.

In this context it would also be of particular interest to make contact with
the results in \cite{RW02} for $q^{N}\neq 1$ and investigate further the
implications of the formal power series (\ref{N8}). This would be another
step forward to understand the representation theoretic meaning of the Bethe
ansatz solutions for all values of the deformation parameter $q$.

The construction procedure for auxiliary matrices at roots of unity
presented here can be generalized to higher spin and higher rank. While the
analysis of the irreducible representations at roots of unity has been
carried out for all simple quantum Lie algebras \cite{CK,CKP} the crucial
input needed is the existence of an evaluation homomorphism (\ref{ev}). The
latter allowed us to make contact with the corresponding quantum loop
algebra underlying the respective trigonometric vertex model. Such
evaluation homomorphisms only exist for $sl_{n}$ \cite{Jimbo2}. In the case
of the other algebras it might depend on the specific nature of the
evaluation representation whether one finds analogous results.\smallskip

The other generalization which comes to mind is the connection with elliptic
models, most of all the eight-vertex model which historically has been the
starting point for the observation of extra symmetries at roots of unity.
Fabricius and McCoy observed \cite{FM8v} that the eigenvalues of the
eight-vertex auxiliary matrix constructed by Baxter \cite{Bx72} satisfy a
functional equation at roots of unity which does not involve the transfer
matrix. This functional equation in conjunction with sum rules for the Bethe
roots allowed them to calculate the dimension of the degenerate eigenspaces
of the transfer matrix. The formulation of an analogous functional equation
for the six-vertex model is an open problem.{\small \medskip }

\textbf{Acknowledgments}. It is a pleasure to thank Harry Braden, Iain
Gordon, Tom Lenagan, Barry McCoy and Robert Weston for interesting
discussions and comments. The work on this article started at the C.N. Yang
Institute, State University of New York at Stony Brook and was completed at
the School of Mathematics, University of Edinburgh. The financial support by
the Research Foundation Stony Brook, NSF Grants DMR-0073058 and PHY-9988566,
and the EPSRC Grant GR/R93773/01 is gratefully acknowledged.


\begin{thebibliography}{99}
\bibitem{Lb67a}  {\small Lieb E H 1967 \emph{Phys. Rev.} \textbf{162}
162--172}

\bibitem{Lb67b}  {\small Lieb E H 1967 \emph{Phys. Rev. Lett.} \textbf{18}
1046--1048}

\bibitem{Lb67c}  {\small Lieb E H 1967 \emph{Phys. Rev. Lett.} \textbf{19}
108--110}

\bibitem{St67}  {\small Sutherland B 1967 \emph{Phys. Rev. Lett.} \textbf{19}
103--104}

\bibitem{Bethe}  {\small Bethe H A 1931\ \emph{Z. Physik} \textbf{71} 205-226%
}

\bibitem{DFM}  {\small Deguchi T, Fabricius K and McCoy B M 2001 \emph{J.
Stat. Phys.} \textbf{102} 701-736}

\bibitem{FM01a}  {\small Fabricius K and McCoy B M 2001 \emph{J. Stat. Phys.}
\textbf{103} 647-678}

\bibitem{BA}  {\small Braak D and Andrei N 2001 \emph{J. Stat. Phys.} 
\textbf{105} 677-709}

\bibitem{FM01b}  {\small Fabricius K and McCoy B M 2002 \emph{MathPhys
Odyssey 2001} (\emph{Progress in Mathematical Physics 23}) ed Kashiwara M
and Miwa T (Boston: Birkhauser) pp 119}

\bibitem{Bx02}  {\small Baxter R J 2002 \emph{J. Stat. Phys.} \textbf{108}
1-48}

\bibitem{D02}  {\small Deguchi T 2002} {\small \emph{XXZ Bethe states as the
highest weight vector of the} }$sl_{2}${\small \ \emph{loop algebra at roots
of unity} cond-mat/0212217}

\bibitem{KM01}  {\small Korff C and McCoy B M 2001 \emph{Nucl. Phys}. B 
\textbf{618} [FS] 551-569}

\bibitem{KR02}  {\small Korff C and Roditi I 2002 \emph{J. Phys. A: Math. Gen%
}. \textbf{35} 5115-5137}

\bibitem{Bx71}  {\small Baxter R J 1971 \emph{Phys. Rev. Lett}.\ \textbf{26}
193-228}

\bibitem{Bx72}  {\small Baxter R J 1972 \emph{Ann. Phys., NY} \textbf{70}
193--228}

\bibitem{Bx73a}  {\small Baxter R J 1973 \emph{Ann. Phys., NY} \textbf{76}
1--24}

\bibitem{Bx73b}  {\small Baxter R J 1973 \emph{Ann. Phys., NY} \textbf{76}
25--47}

\bibitem{Bx73c}  {\small Baxter R J 1973 \emph{Ann. Phys., NY} \textbf{76}
48--71}

\bibitem{BxBook}  {\small Baxter R J 1982 \emph{Exactly Solved Models in
Statistical Mechanics} (London: Academic Press)}

\bibitem{CK}  {\small De Concini C and Kac V 1990~\emph{Operator algebras,
unitary representations, enveloping algebras, and invariant theory\
(Progress in Mathematical Physics 92)\ }ed Connes A et al (Boston: Birkh\"{a}%
user) pp 471}

\bibitem{CKP}  {\small De Concini~C, Kac~V and Procesi C 1992 \emph{J. Amer.
Math. Soc.} \textbf{5} 151-189}

\bibitem{BK}  {\small Beck~J and Kac~V 1996 \emph{J. Amer. Math. Soc}. 
\textbf{9} 391-423}

\bibitem{KP}  {\small Korff C, in preparation.}

\bibitem{BS90}  {\small Bazhanov V V and~Stroganov Yu G~1990 \emph{J. Stat.
Phys.\ }\textbf{59} 799-817}

\bibitem{AF97}  {\small Antonov A and Feigin B 1997 \emph{Phys. Lett.} B 
\textbf{392} 115--122}

\bibitem{BOU01}  {\small Belavin A A, Odesskii A V and Usmanov R A 2002 
\emph{Theor. Math. Phys.} \textbf{130} 323-350}

\bibitem{RW02}  {\small Rossi M and~Weston R~2002 \emph{J. Phys. A: Math.
Gen.} \textbf{35} 10015-10032}

\bibitem{Y}  {\small Yang C N~1967 \emph{Phys. Rev. Lett.} \textbf{19} 1312-4%
}

\bibitem{Bx70}  {\small Baxter R J 1972~\emph{Ann. Phys., NY} \textbf{70}
323-37}

\bibitem{Drin}  {\small Drinfel'd V G~1987 \emph{Proceedings of the 1986
International Congress of Mathematics, Berkeley} ed Gleason A M (Providence,
RI: American Mathematical Society) pp 798-820}

\bibitem{Jimbo}  {\small Jimbo M 1985 \emph{Lett. Math. Phys}.\emph{\ }%
\textbf{10} 63-69}

\bibitem{QISM}  {\small Fadeev L D, Sklyanin E K and Takhtajan L A 1979 
\emph{Theor. Math. Phys.} \textbf{40} 194-220}

\bibitem{QISM1}  {\small Fadeev L D and Takhtajan LA 1979 \emph{Russian.
Math. Surveys} \textbf{34}:5 11-68}

\bibitem{FRT}  {\small Fadeev L D, Reshetikhin N Yu and Takhtajan L A 1988 
\emph{Algebraic analysis} \textbf{1} ed Kashiwara M and Kawai T (New York:
Academic Press) pp 129-39}

\bibitem{DJMM90}  {\small Date E,~Jimbo M, Miki K and Miwa T 1990 \emph{%
Phys. Lett.} A \textbf{148} 45-49}

\bibitem{T92}  {\small Tarasov V O 1992~\emph{Int. J. Mod. Phys}. A\emph{\ }%
\textbf{7,} Suppl. 1B 963-975}

\bibitem{AYMCPTY}  {\small Au-Yang H, McCoy B M, Perk J H H, Tang S and Yan
M L 1987 \emph{Phys. Lett.} A \textbf{123} 219-223}

\bibitem{BPAY}  {\small Baxter R J, Perk J H H and Au-Yang H 1988 \emph{%
Phys. Lett}. A \textbf{128}, 138-142}

\bibitem{CP}  {\small Chari V and~Pressley A 1994 \emph{A Guide to Quantum
Groups} (Cambridge: Cambridge Univ. Press)}

\bibitem{JmbNotes}  {\small Jimbo M~\emph{Topics from Representations of} }$%
U_{q}(g)${\small \ -- \emph{An Introductory Guide to Physicists} (Kyoto:
Faculty of Science, Kyoto University, Kyoto 606)}

\bibitem{A94}  {\small Arnaudon D 1994 \emph{Comm. Math. Phys}. \textbf{159}
175-194}

\bibitem{RA89}  {\small Roche P and Arnaudon D 1989 \emph{Lett. Math. Phys}. 
\textbf{17} 295-300}

\bibitem{Jimbo2}  {\small Jimbo M 1986~\emph{Lett. Math. Phys.\ }\textbf{11}
247-252}

\bibitem{KNS94}  {\small Kuniba A,~Nakanishi T and~Suzuki J~1994 \emph{Int. J.
Mod. Phys}. A \textbf{9} 5215--5266}

\bibitem{FM8v}  {\small Fabricius K and McCoy B M 2002 \emph{New
Developments in the Eight Vertex Model} cond-mat/0207177}
\end{thebibliography}
\end{document}